\documentclass[cits]{JINST}
\let\ifpdf\relax
\pdfoutput=1

\usepackage[squaren,textstyle]{SIunits}
\usepackage{subfigure}
\usepackage{multirow}
\usepackage{url}

\title{Prototype ATLAS IBL Modules using the FE-I4A Front-End Readout Chip \\ }

\author{\Large The ATLAS IBL Collaboration}

\abstract{
The ATLAS Collaboration will upgrade its semiconductor pixel tracking detector with a new Insertable B-layer (IBL) between the existing pixel detector and the vacuum pipe of the Large Hadron Collider. The extreme operating conditions at this location have necessitated the development of new radiation hard pixel sensor technologies and a new front-end readout chip, called the FE-I4. Planar pixel sensors and 3D pixel sensors have been investigated to equip this new pixel layer, and prototype modules using the FE-I4A have been fabricated and characterized using 120 GeV pions at the CERN SPS and 4 GeV positrons at DESY, before and after module irradiation. Beam test results are presented, including charge collection efficiency, tracking efficiency and charge sharing.
}

\keywords{ATLAS, upgrade, tracker, silicon, pixel, FE-I4, planar sensors, 3D sensors, test beam}

\clearpage

\begin{document}

\section{Introduction}{\label{section:introduction}}

The Insertable B-Layer (IBL) \cite{Capeans:1291633} is a fourth pixel layer added to the present pixel detector of the ATLAS experiment \cite{Aad:2008zz} at the Large Hadron Collider (LHC), between a new vacuum pipe and the current inner pixel layer. The principal motivation of the IBL is to provide increased tracking robustness as the instantaneous luminosity of the LHC increases beyond the design luminosity of $10^{34}  \rm{cm^{-2}s^{-1}}$ and the integrated radiation deteriorates the performance of inner pixel layers. The  existing tracker performance will be significantly improved in several aspects: the track pattern recognition capability; the reconstructed track accuracy; and the primary and displaced vertex identification performance in high-luminosity conditions. The IBL will be installed in the ATLAS experiment during the LHC shut-down in 2013-14. It is designed to operate at least until a full tracker upgrade, that is planned for high luminosity LHC (HL-LHC) operation from approximately  2023, is completed. 
 
The constraints of the IBL project are stringent, and have influenced the mechanical, sensor and electronics technologies developed for the detector. The small radius of the IBL requires a more radiation hard technology for both the sensors and the front-end electronics, with a required radiation tolerance for fluences of up to 6$\times10^{15}\,\mathrm{n_{eq}/cm^{2}}$ NIEL and 250 Mrad TID \cite{Capeans:1291633}\footnote{The non-ionising energy loss (NIEL) is normally quoted as the equivalent damage of a fluence of 1 MeV neutrons ($\mathrm{n_{eq}/cm^{2}}$). This is an important measure of the radiation dose for silicon sensors. The Total Ionising Dose measured in silicon (TID) is a more relevant measure of the radiation dose for the front-end electronics.}. Because of the high occupancy of the individual pixel elements, a more efficient front-end readout is required. The available space does not allow module overlaps in the longitudinal direction (along the beam) and sensors with either an active edge or a slim edge guard ring have been developed to reduce geometrical inefficiencies. Radiation induced displacement damage of silicon sensors causes an increased sensor leakage current (I$\rm{_{l}}$), resulting in an increase of noise in the analog front-end, and an increased detector bias voltage (V$\rm{_{b}}$) needed for full depletion (V$\rm{_{d}}$). Depending on the sensor technology, operating voltages of up to 1000~V are required for the sensor after heavy irradiation. The power dissipation at these operating voltages must be controlled, and this places strong requirements on the cooling to prevent thermal run-away. Following irradiation, sensor temperatures of below $-15^{\circ}\ \mathrm{C}$ must be maintained, allowing a power consumption of order 0.65 $\mathrm{W/cm^{2}}$.  Minimizing the material is very important to optimize the tracking and vertexing performance, and the average IBL radiation length target is 0.015 $X_0$ for perpendicular traversing tracks, as opposed to more than 0.03 $X_0$ for the existing inner pixel layer. This is achieved using aggressive technology solutions, including a new module based on optimized sensor and front-end chip designs, local support structures (staves) made of recently developed low density, thermally conducting carbon foam, the use of $\rm{CO_{2}}$ evaporative cooling which is more efficient in terms of mass flow and pipe size, and electrical power services using aluminum conductors. 
  
\begin{figure}[ht!] 
 \centering
  \includegraphics[height=3.2in]{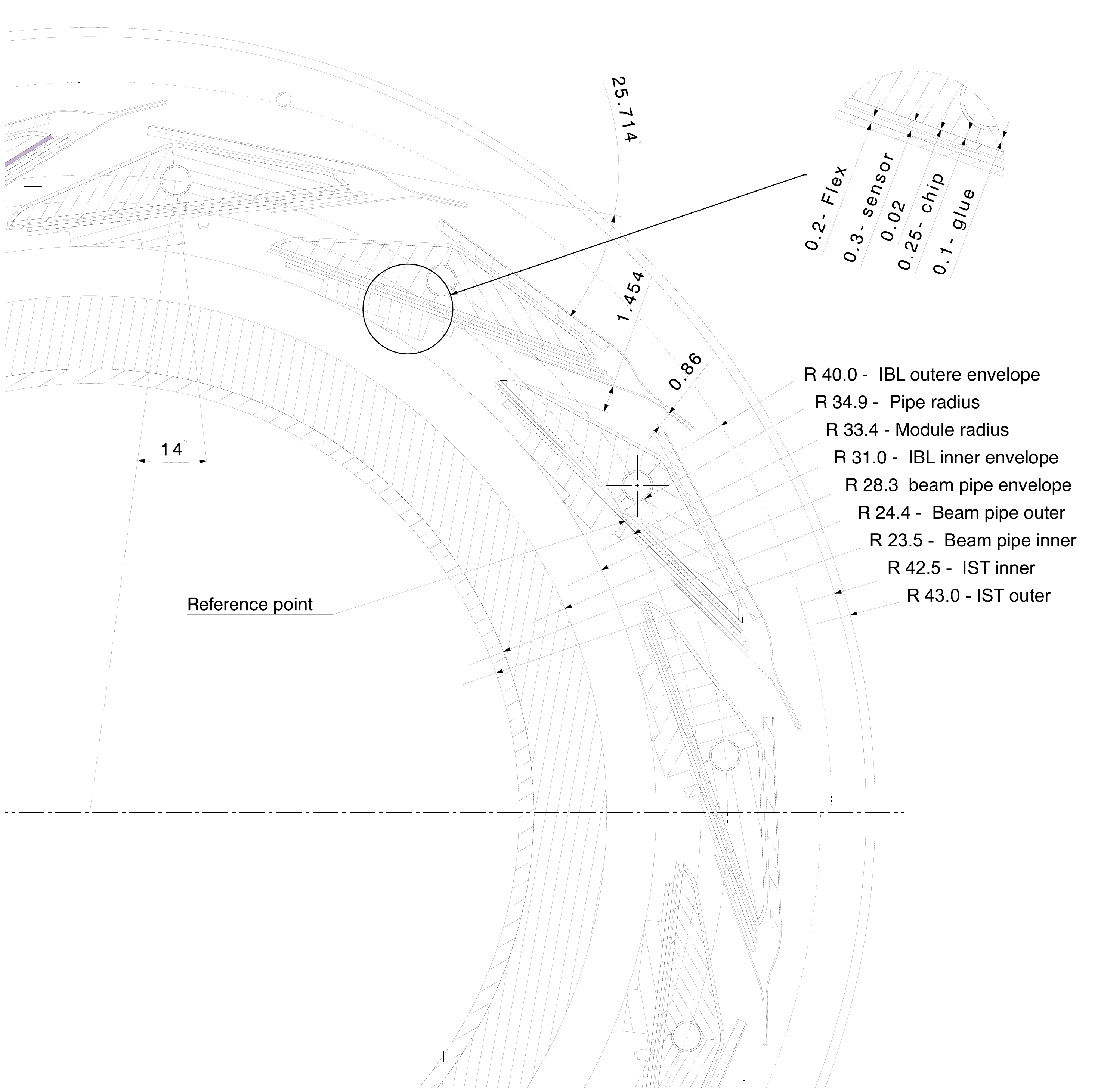}
  \caption{Drawing of the IBL pixel layer in the ATLAS experiment.}
  \label{IBL_layout_A19}
\end{figure}

\begin{figure}[ht!] 
 \centering
  \includegraphics[height=3.2in]{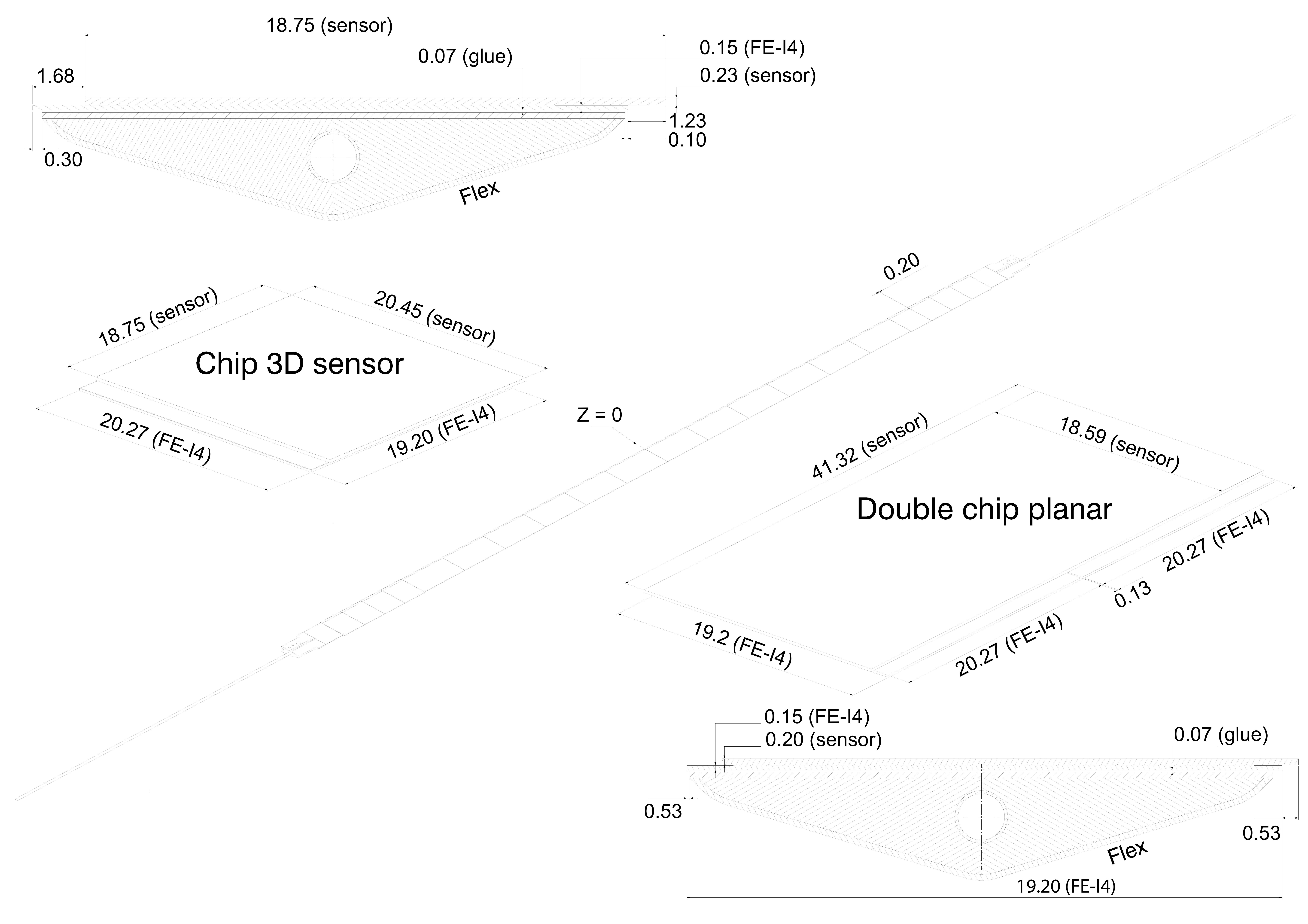}
  \caption{Sketch of individual pixel staves, indicating the placement of planar and 3D pixel modules, on the IBL pixel detector.}
  \label{IBL_stave}
\end{figure}

Figure~\ref{IBL_layout_A19} shows a drawing of the new IBL detector. It consists of 14 staves, located at a mean active geometric radius of 33.4 mm, each loaded with silicon sensors bump-bonded to the newly developed front-end integrated circuit (IC) FE-I4 \cite{Karagounis:2008zz}. Figure~\ref{IBL_stave} shows an individual stave. Figure~\ref{IBL_stave} also shows the module design using the FE-I4 IC and either  a planar \cite{Muenstermann:2009iq} or 3D sensors \cite{daVia:2011iq}. The planar sensors will be bump-bonded to 2 ICs, while the 3D sensors will be bump-bonded to a single IC. The ATLAS collaboration intends to build modules using the FE-I4 IC and both planar and 3D sensor technologies for the IBL. This paper describes the development and test of IBL modules using prototype sensors and the FE-I4A prototype IC. Measurements of the module performance have been made before and after irradiation to the fluence levels expected during IBL operation.  

Section \ref{section:FE-I4A} describes the FE-I4A prototype IC, an array of pixel cells arranged in 80 columns and 336 rows of size 250$\times$50~$\mathrm{\mu m^{2}}$. After a discussion of the required FE-I4 specifications, the analog and digital architectures of the IC are described, and test results of the FE-I4A IC are presented. On the basis of prototype studies, minor design changes have been implemented for the production IC iteration (FE-I4 B).

Section {\ref{section:sensor-design-performance}  describes the requirements, the resulting operational specifications and the subsequent technical designs for both the planar and 3D sensor technologies. Measurements are shown for bare sensors of each type. In Section \ref{section:IBL-module}, prototype modules that have been constructed for each sensor type are described, and their performances are evaluated and compared with the specification for each type, before and after irradiation to fluences of up to 6$\times10^{15 }\mathrm{n_{eq}/cm^{2}}$.

Finally, in Section Section \ref{section:IBL-module-test-beam}, the performance of individual prototype modules in a test beam is assessed before and after irradiation. Section \ref{section:conclusions} provides some conclusions. 

\section{The FE-I4 Front-End Readout Chip} {\label{section:FE-I4A}}

\subsection{Requirements and Specifications for the FE-I4 chip} {\label{section:FE-I4A-specifications}}
    
The present 3-layer ATLAS pixel detector is based on 16-chip modules using the $7.6\times10.8\, \rm{mm^{2}}$, 2880 pixel FE-I3 readout IC \cite{Aad:2008zz}, with a pixel granularity $400\times50\,\rm{\mu m^{2}}$ in a 250 nm feature size bulk CMOS process. The limitations of the FE-I3, in particular its radiation hardness for the fluences expected at the IBL radius and its ability to cope with high hit rates, make the FE-I3 unusable at the IBL for the expected LHC luminosities of up to 3$\times10^{34}\rm{\,cm^{-2}s^{-1}}$. The new FE-I4 IC has been developed with a pixel size of $250\times50\,\rm{\mu m^{2}}$ in a 130 nm feature size bulk CMOS process, in view of future ATLAS high luminosity pixel applications, and is well matched to the IBL requirements. Using a smaller feature size presents advantages. Firstly, the increased digital circuitry density means that more complexity can be implemented despite the smaller available pixel area, enabling a new readout architecture able to cope with much higher hit rates. Secondly, the 130 nm feature size is advantageous in terms of TID hardness, the chip being rated for 250~Mrad and therefore able to cope with the IBL environment \cite{GarciaSciveres:2011vd}.

A major innovation of the FE-I4 IC is a pixel matrix architecture that is radically different from the existing pixel column drain readout followed by peripheral data storage and trigger logic. The data storage is now made locally at the pixel level until triggering and subsequent propagation of the trigger inside the pixel array. This organization overcomes the efficiency limitation of the FE-I3 architecture. Demonstrations using physics based simulations have shown that the FE-I4 readout will remain efficient up to luminosities of 3$\times10^{34}~\,\rm{cm^{-2}s^{-1}}$ at the IBL radius \cite{Arutinov:2008a,Barbero:2009zz}.

The 130 nm CMOS process core transistors have a gate-oxide thickness of about 2 nm. The trapping of positive charges in the gate oxide is therefore reduced compared with previous generation processes (250 nm processes typically have a 5 nm gate-oxide thickness), and threshold shifts after irradiation as well as radiation induced leakage current paths are better controlled \cite{Gonella:2007zz}. The use of specific hardening techniques (e.g. enclosed layout transistors) is no longer necessary for all digital and even for most analog transistors. Guard rings are used for sensitive analog transistors, and minimal sized transistors are in general avoided. However, it is noted that processes with smaller feature sizes are not intrinsically more Single Event Effect (SEE) hard, and the resistance of the design in particular to Single Event Upset (SEU) needs to be assessed (see Section \ref{section:test-results}).

Hybrid pixel readout IC's are based on analog sections interleaved with digital sections. In the FE-I4 IC, the digital sections are based on standard synthesized cells. This allows to use the full power of available industry tools developed for digital logic synthesis and verification, and not to rely on custom cells developed for our specific application. Digital logic is synthesized from a high level description language. The digital logic library therefore has at first sight the drawback of directly coupling the local digital and analog substrates. This would normally be a major concern for sensitive analog sections that require a noise-free substrate. This problem is avoided in the FE-I4 by providing a deep n-well option that then allows the isolation of the digital cell local substrate from the global one, and leads to reduced noise coupling to the analog parts.

The FE-I4 IC consists of an array of 26,880 pixels, 80 in the z-direction (beam direction) by 336 in the azimuthal r$\phi$ direction (referring to the ATLAS detector coordinates). The pixel size is 250$\times$50 $\rm{\mu m^{2}}$. The r$\phi$ granularity is chosen to match the established bump-bonding pitch used to build existing pixel modules. The pixel length is then chosen to have sufficient area to embed the more complex digital section, taking into account power routing constraints. A diced FE-I4 IC is 18.8$\times$20.2 mm$^{2}$, with 2 mm in the r$\phi$ direction devoted to the IC periphery. The IC is the largest so far designed for High Energy Physics applications. Going to the large FE-I4 size is beneficial in many respects. It enhances the active over total area ratio, and allows integrated module and stave concepts. As a consequence, it reduces the inert material and therefore the IBL material budget, resulting in a significant enhancement in physics performance, for example the b-tagging efficiency versus light-jet rejection factor \cite{Capeans:1291633}. A large FE-I4 IC also reduces the bump-bonding cost which scales as the number of manipulated IC's (not important for the IBL but an important parameter for large area detectors such as the outer pixel layers at a future HL-LHC). Such a large IC can only be designed if a solid power distribution can be established, and a satisfactory yield model can be achieved. The former point is addressed by the powerful features of the CMOS process used, with 8 metal layers among which 2 are made of thick aluminium (also good to provide effective shielding). The latter point is addressed by an active yield enhancing policy as will be outlined in Section \ref{section:FE-I4A-periphery}. Table \ref{specs} shows the main specifications of the FE-I4.
 
  
\begin{table*}[ht!]
\footnotesize
\begin{center}
\caption{FE-I4 main specifications.}
\medskip
\begin{tabular}{|l|c|c|}
\hline
\hline

         \textbf{Item} & \textbf{Value} & \textbf{Unit (note)} \\
\hline
\hline
 Number of pixels & 80$\times$336 = 26,880 & (column $\times$ row) \\
 Pixel unit size & 250$\times$50 & $\rm{\mu m^{2}}$ (direction $z\times r\phi$) \\
 Last bump to physical edge on bottom  & $\leq2.0$ & mm \\
\hline
 Nominal analog supply voltage & 1.4 & V \\
 Nominal digital supply voltage & 1.2 & V \\
 Nominal analog current & 12 & $\rm{\mu A}$ \\
 Nominal digital current & 6 & $\rm{\mu A}$ \\ 
 DC leakage current tolerance per pixel  & 100 & nA \\
 Normal pixel input capacitance range & 0-500 & fF \\
 Edge pixel input capacitance range & 0-750 & fF \\
 Hit trigger association resolution & 25 & ns \\
 Single channel ENC & $<$ 300 & e$^{-}$ \\
 In-time threshold within 20 ns (400 fF) & $\leq$ 4000 & e$^{-}$ (at discriminator output) \\
 Tuned threshold dispersion & $<$100 & e$^{-}$ \\
 Charge $\rightarrow$ Digital coding method & ToT & (on 4 bits)\\
\hline 
 Radiation tolerance (specs met at dose) & 250 & Mrad \\
 Operating temperature & -40 to +60 & $^{o}$C \\
\hline
 Readout initiation & Trigger & \\
 Maximum number of consecutive triggers & 16 & (internal multiplication) \\
 Minimal time between external triggers & 125 & ns \\
 Maximum trigger latency & 6.4 & $\rm{\mu s}$ \\
 Maximum sustained trigger rate & 200 & kHz \\
\hline
 I/O signals & custom LVDS & \\
 Nominal clock input frequency & 40 & MHz \\
 (design includes 20\% frequency margin) & & \\
 Nominal serial command input rate & 40 & Mb/s \\
 Output data encoding & 8-10 & bits \\
 Nominal data output rate & 160 (up to 320) & Mb/s \\
\hline         
\hline
\end{tabular}
\label{specs}
\end{center}
\end{table*} 

\subsection{Design of the FE-I4A chip} {\label{section:FE-I4A-design}}

Figure~\ref{FE-I4Apict} shows a photograph of the first full scale prototype, the FE-I4A, fabricated in Summer 2010. The superimposed photograph of the FE-I3 IC emphasizes the IC size. It is the basic element of all IBL prototype modules tested during 2011. Figure~\ref{FE-I4Asketch} gives an insight into the IC organization. Each pixel consists of an independent analog section with continuous reset, amplifying the collected charge from the bump-bonded sensor. In the analog section, hits are discriminated at the level of a tunable comparator with an adjustable threshold, and charge is translated to Time over Threshold (ToT) with a proportionality factor that the user can tune by changing the return to baseline behavior of the pixel (see section \ref{section:FE-I4A-analog}). The 26,880 pixel array is organized in columns of analog pixels, each pair of analog columns tied to a shared digital double-column unit centred between them. Inside the double-column, 4 analog pixels communicate to a single so-called ``4-pixel digital region'' (4-PDR). Details of the architecture and the 4-PDR benefits will be described in Section \ref{section:FE-I4A-digital}. Communication is organized inside the digital double-column and coordinated with peripheral logic. Section \ref{section:FE-I4A-periphery} will describe how communication to the FE is established and how the data output is organized, as well as give an insight into the major peripheral blocks. Finally, Section \ref{section:test-results} will give a few test results.

\begin{figure}[ht!] 
\centering
  \includegraphics[height=2.4in]{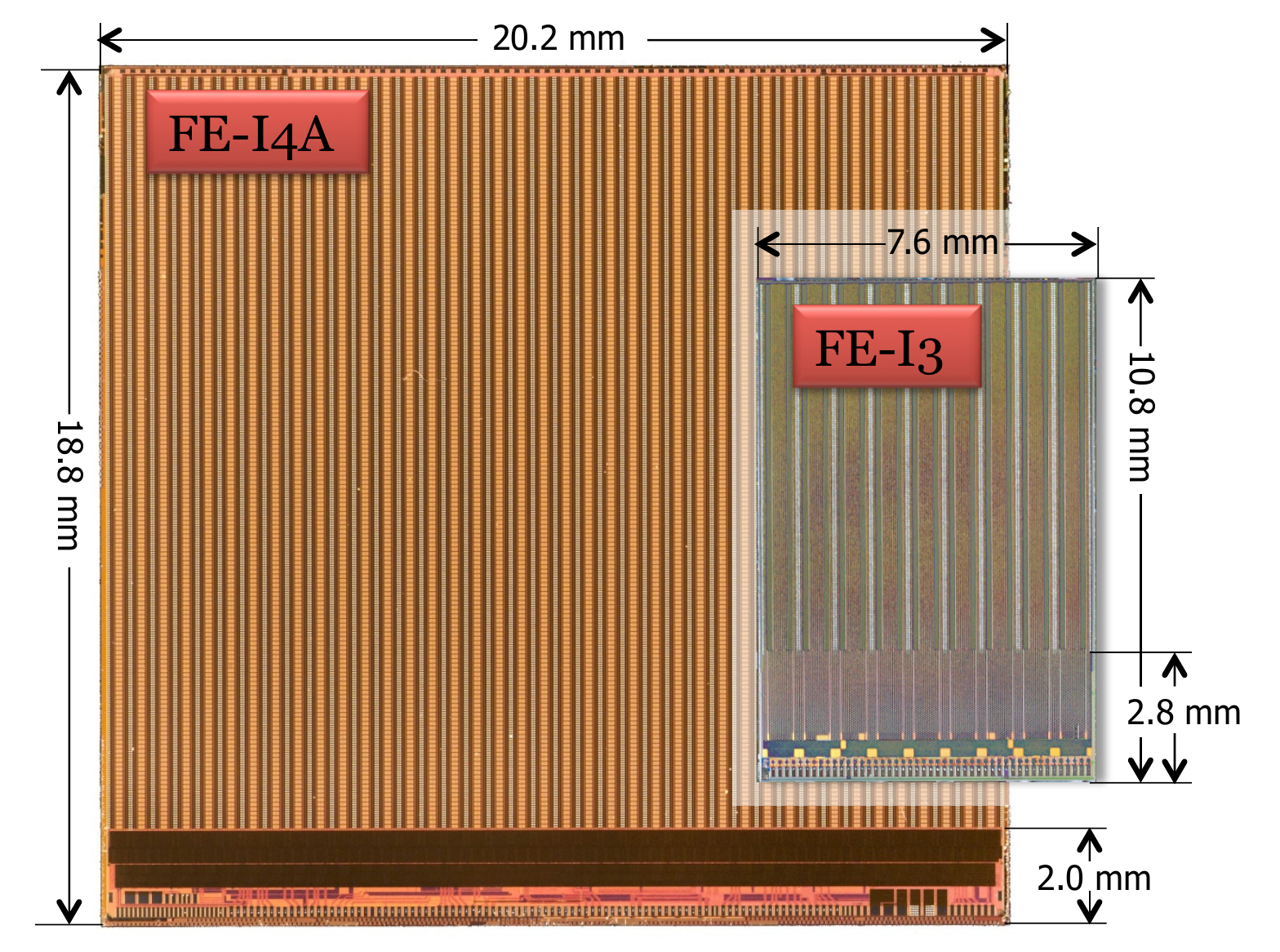}
  \caption{Picture of the FE-I4A IC (the to-scale FE-I3 IC is shown for comparison).}
  \label{FE-I4Apict}
\end{figure}

\begin{figure}[ht!]
\centering
  \includegraphics[height=5.4in]{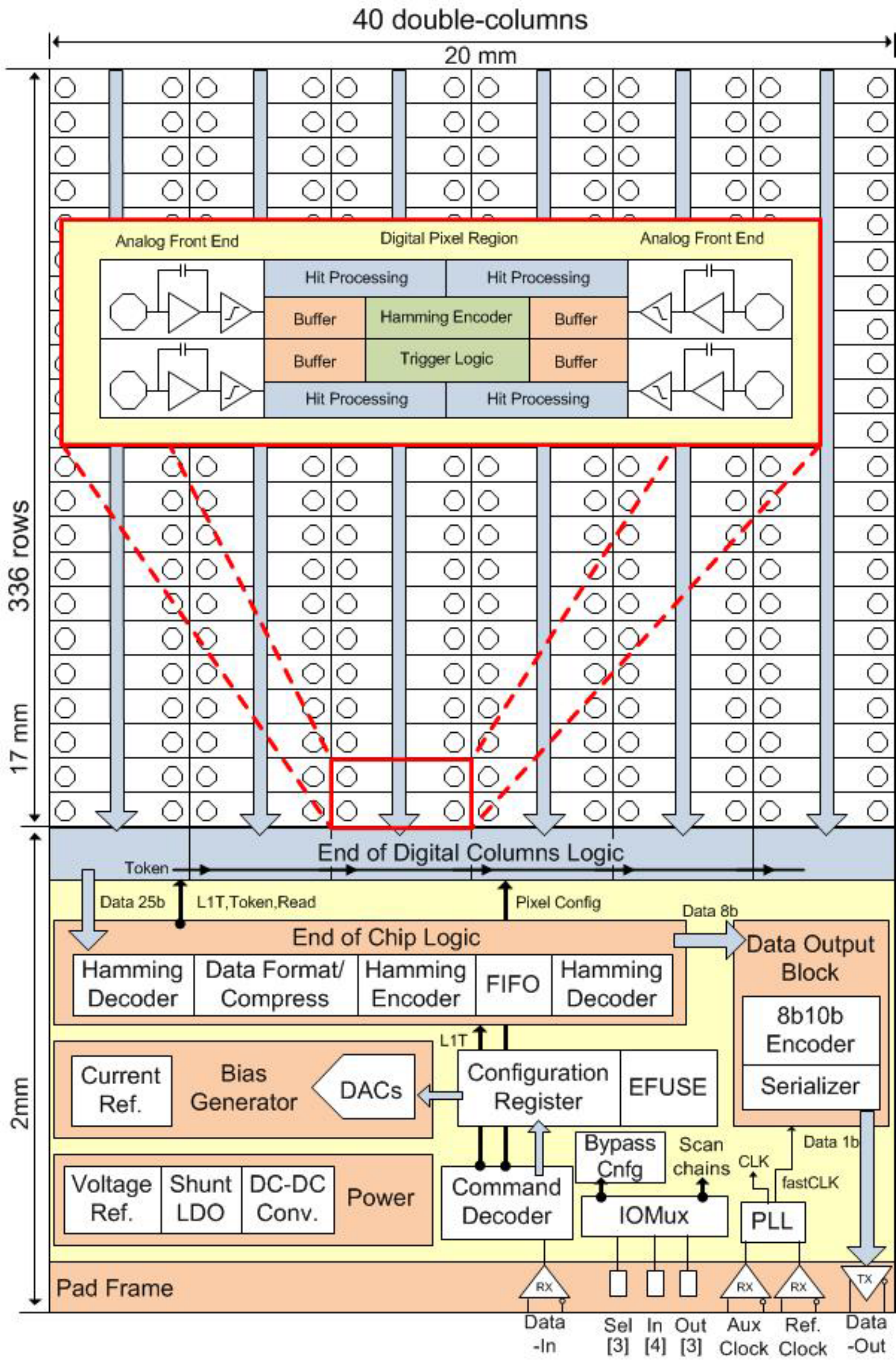}
  \caption{Sketch of the organization of FE-I4A IC.}
  \label{FE-I4Asketch}
\end{figure}

\subsubsection{Analog front-end of the FE-I4A chip} {\label{section:FE-I4A-analog}}

The analog front-end of the FE-I4A is implemented as a 2-stage amplifier optimized for low power, low noise and fast rise time, followed by a discriminator. A schematic of the analog section is shown in figure~\ref{analog_PUC_sketch}: it covers approximately 60\% of the total pixel area. 

\begin{figure}[ht!] 
\centering
  \includegraphics[height=2.2in]{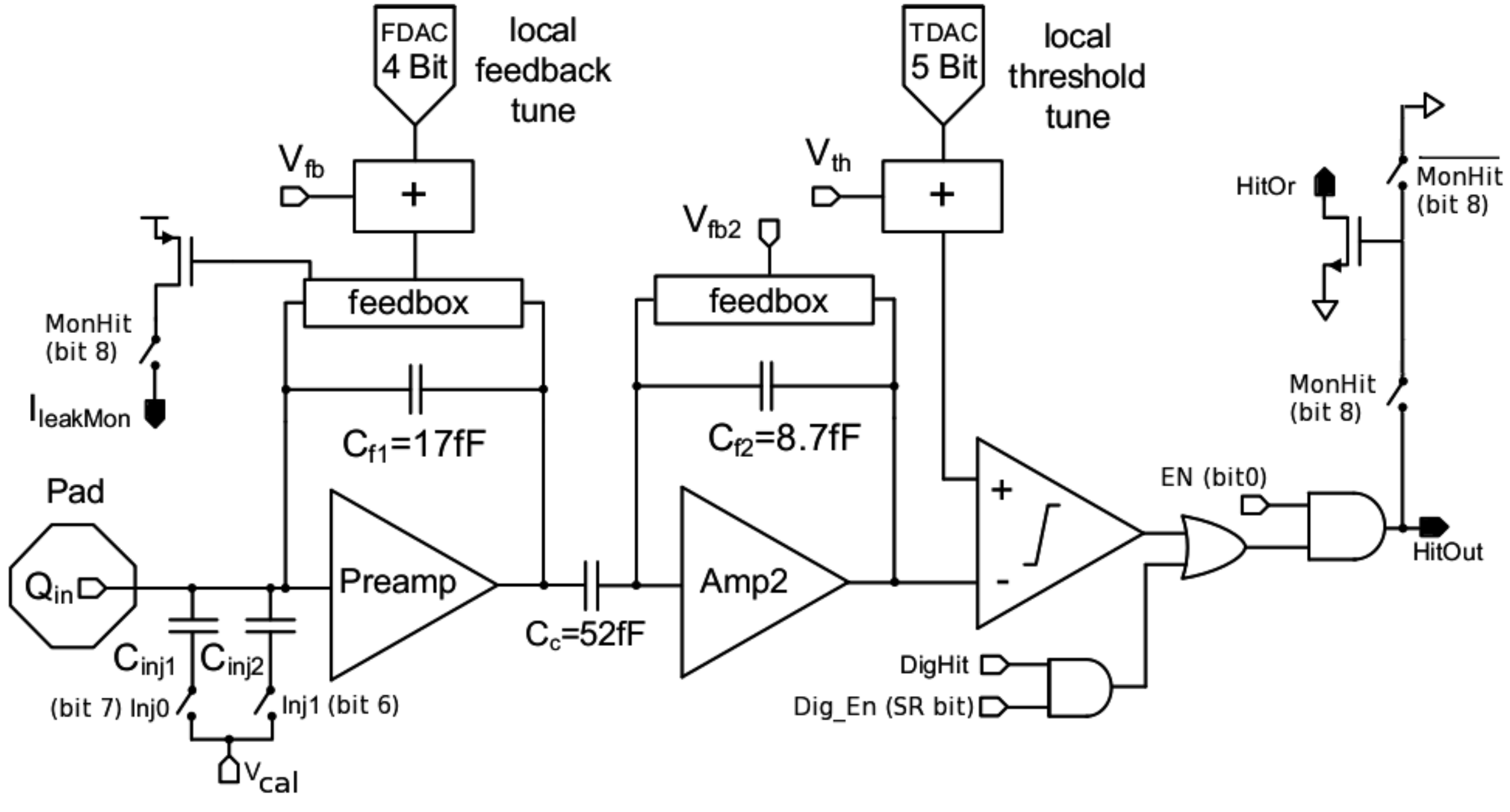}
  \caption{Analog pixel section schematic.}
  \label{analog_PUC_sketch}
\end{figure}

The first stage of amplification consists of a regulated cascode with triple-well NMOS input. The preamplifier uses a continuous current-source reset that gives an approximately linear signal ToT.  It contains an active slow differential pair, tying the first stage input to its output, that is used to compensate radiation induced leakage current coming from the sensor and to provide DC leakage current tolerance above 100 nA. The second stage is AC coupled to the first stage, and implemented as a folded cascode with PMOS input. This AC coupling has the advantage of decoupling the second stage from potential leakage current related DC shifts, and of reducing the pixel-to-pixel threshold mismatch. It also gives an extra gain factor coming from the ratio of the second stage feedback capacitance to the coupling capacitance ($\approx$6 in the actual FE-I4A design). This allows to increase the first stage feedback capacitance with beneficial consequences on charge collection efficiency, signal rise time and power consumption, but with no degradation of the signal amplitude at the discriminator input \cite{Karagounis:2008zz}. Finally, the discriminator consists of a classic two stage architecture. 

For test purposes and calibration, a test charge can be injected at the pre-amplifier input through a set of 2 injection capacitors. Hits can also be injected after the discriminator to test the digital part of the pixel. In total, 13 bits are stored locally in each pixel for tuning of operation: 2 bits for the control of the injection switches, 4 bits for the local tuning of the feedback current of the first stage (they control the return to baseline of the first stage output, hence the charge to ToT conversion factor), 5 bits for the local tuning of the discriminator threshold, 1 bit for switching on the MonHit output (leakage current output) and the HitOR output (global OR of all pixel hits), and the last bit used for masking off the pixel.

\subsubsection{Digital organization of the FE-I4A chip} {\label{section:FE-I4A-digital}}

A critical innovation of the FE-I4 IC is its digital architecture, which allows to accept much higher hit rates than was possible with the FE-I3. With a smaller feature size, the trigger can be propagated inside the array and the hits stored locally at pixel level until triggering or erasing. For each analog pixel there exist 5 buffer memories where ToT information can be stored during the trigger latency. Studies  have shown that an organization with 4 analog pixels tied to a single 4-PDR as shown in figure~\ref{4PDR-sketch} is a very efficient implementation \cite{Arutinov:2008a}.

\begin{figure}[ht!] 
 \centering
  \includegraphics[height=2.6in]{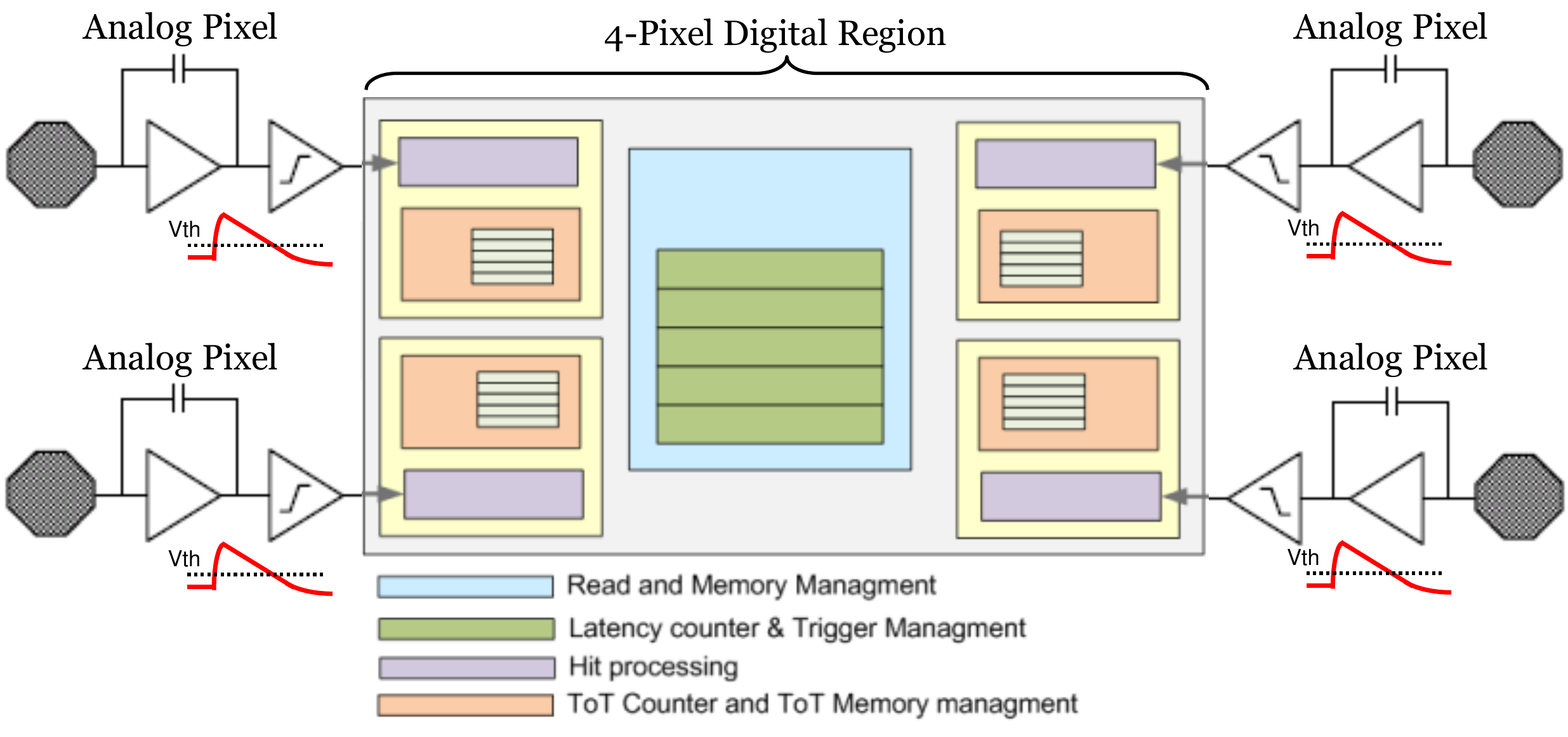}
  \caption{Organization of the 4-pixel region, focusing on a schematic of the digital region.}
  \label{4PDR-sketch}
\end{figure}

In this structure, 4 independently working analog pixels share a common digital block. The outputs of the 4 discriminators are fed to 4 separate hit processing units (in purple in figure~\ref{4PDR-sketch}) that provide time-stamping and compute the ToT. An extra level of digital discrimination can also be programmed, to distinguish large and small recorded charges. When one (or more) of the four hit processing units detects a ``large'' hit, the unit books one of the central latency counters (in green). Regardless of which of the four pixels has initiated the booking of the latency counter, four ToT memories for the four pixels will be associated to the event, thanks to a fixed geographical association (the 1st latency counter corresponds to 4 1st ToT buffer memories, the 2nd latency counter to 4 2nd ToT memories, and so on).

This architecture presents several advantages. First, it makes good use of the fact that the pixels inside the 4-PDR are in geographic proximity, by sharing some resources. Resource sharing is efficient to record hits, as real hits are clustered. Second, it is advantageous in terms of lowering the power used, as the un-triggered hits are not transferred to the periphery, and some logic inside the 4-PDR is common to several hits at a time. Third, it is efficient in terms of time-walk compensation, as one can use the digital discriminator to associate a small hit (below digital discriminator threshold) with a large hit (above digital discriminator threshold) occurring in previous bunch-crossing, by simple geographical association. Finally, it also improves the active fraction of the FE-I4A as the memory is located at the level of the pixels, not in the periphery.
 
\subsubsection{Periphery of the FE-I4A chip} {\label{section:FE-I4A-periphery}}

The periphery schematic of the FE-I4A is shown in figure~\ref{FE-I4Asketch}. It contains blocks that fulfill the following operations: communication and operational tuning of the IC; organization of the data read back and fast data output serialization. Finally some blocks are implemented to provide extra testing capabilities (e.g. redundant memories, low speed multi-purpose multiplexer), or as prototype blocks for the future production IC (e.g. powering section).

Two LVDS inputs are required to communicate to the FE-I4A: the clock (nominally 40 MHz); and the command input Data-In (40 Mb/s). In the FE-I4 command decoder, the command stream is decoded into local pixel configuration, global configuration and trigger commands. It is based on the architecture for the module control chip of the existing ATLAS pixel detector. No separate module control chip is needed for the IBL, further reducing the IBL mass. The decoded pixel configuration is sent to the pixels for storage in the 13 local register bits of the pixel. The global configuration is stored in SEU-hardened  configuration registers using Dual Interlock storage CEII (DICE) latches \cite{Menouni:2008zz,Calin:556880} and triplication logic. The 32 16-bit deep registers are used for global tuning of the operation of the chip. In the bias generator section, based on an internal current reference, DACs convert the stored configuration values to voltages and currents needed to tune the various sections of the IC. The decoded trigger is propagated to the pixels and to the ``End of Chip Logic'' block where the readout is initiated.

When a trigger confirms a hit (the coincidence of a trigger with a latency counter reaching its latency value inside a 4-PDR), data stored in the 4-PDR ToT buffers are sent to the periphery and associated to the bunch-crossing corresponding to the specific trigger. In the double-column, the 4-PDR address as well as the 4 ToTs are propagated to the "End of Chip Logic" (the transmitted signals are Hamming coded for yield enhancement). The data are then re-formatted (for band-width reduction and to facilitate the following data processing steps) and stored in a FIFO to be sent out. In addition to stored pixel data, read back information from pixel and global registers, as well as some diagnostic information (error messages), can be included. The data is then 8b10b-encoded \cite{Widmer:5390392} in the "Data Output Block" and serialized at 160 Mb/s. Fast serialization is made possible by use of a high speed clock provided by a "Phase Lock Loop" clock generator \cite{Kruth:1235880}. The custom LVDS receiver and transmitter have been described elsewhere \cite{Barbero:2011aa}.

In addition, a few structures are used for test purposes, for example the IOMux-based redundant configuration memories that also provide diagnostic access to various signals and the prototype powering section (Shunt-LDO \cite{Gonella:2010zz,Karagounis:5325974} and DC-DC converters \cite{Denes:2009zz}). 

During implementation, design strategies have been followed to SEU-harden the FE-I4A: the test of 2 flavors of DICE latches for the in-pixel memories; DICE structures with interleaved layout and triplication logic for the global configuration; and triplication of counters and logic in the End of Chip Logic block. The yield enhancement for such a large IC has also been an important consideration: the use of Hamming coding with a minimal number of gates for the data transfer in the array; triplication and majority voting for the peripheral Command Decoder; redundant (and user selectable) configuration shift registers in each double-column; triple redundant read token passing inside the double-column and at the level of the End of Digital Column logic; the use of lithography friendly bus widening and bus spacing; and using multi-via digital cells when synthesizing the design.

\subsection{Test results on bare FE-I4A} {\label{section:test-results}}

\subsubsection{The USBpix system} {\label{section:USBpix-test-system}}

Most of the testing of both the bare FE-I4 chips and the subsequent pixel modules has been made with a portable USB data acquisition system called USBpix \cite{Backhaus:201137}. USBpix is a modular test system, developed for lab measurements with FE-I3 and FE-I4 chips. It consists of a Multi-IO board that provides a USB interface with a micro controller, a FPGA and 2~MB of on-board memory. The Multi-IO board is 
connected to an adapter card that is specific to the choice of FE-I3 or FE-I4 chip to be tested. The adapter card provides all signals to the chip using LVDS transmitters or CMOS level shifters. The FE-I4 adapter card allows to either route all power and signal lines via a flat cable to the chip, or to connect power lines and signal lines separately in which case the signals are routed via an RJ45 connector to the chip. 

\subsubsection{Characterization of bare FE-I4A chips} {\label{section:bare-test-results}}

The analog pixel, the 4-PDR implementation, the communication and programming, and the data output logic path, have all been successfully characterized for the prototype FE-I4A IC. 

The two main contributions to the FE-I4A analog power consumption are the biasing of the pre-amplifier and of the discriminator. The digital power is the sum of a static contribution and a contribution that is proportional to the hit rate. Typical values for operation are approximately 16~$\rm{\mu W}$/pixel (11 $\rm{\mu A}$) for the analog power and 7 $\rm{\mu W}$/pixel (6 $\rm{\mu A}$) for the digital part, giving a total power consumption of approximately 160 mW/cm$^{2}$.

The measured threshold and noise values of the FE-I4A IC depend on the operating parameters, for example the global threshold, the global feedback current, and the analog working point of the pixel preamplifier. Figure~\ref{fig:ThreshfeedbackIBL} shows the dependence of the Equivalent Noise Charge (ENC) on the global threshold setting and feedback current for a typical IC. Bare ICs show a noise in the range 110-120 ENC for a 3000 e$^{-}$ threshold. For a typical FE-I4A working point at the LHC, the noise from a pixel module with its attached sensor is of order 150 ENC. Within an IC, the threshold dispersion after tuning is 30-40 e$^{-}$, well within specification.  In general, the ENC increases with reduced threshold, and with increased feedback current. Calibrations of the FE-I4A IC require measurements using a radioactive source or a test beam, and  are difficult to make. The noise results when quoted in units of electron ENC therefore have an intrinsic  10-20 \% normalisation uncertainty.  

\begin{figure}[ht!]
 \centering
 \includegraphics[width=0.95\textwidth]{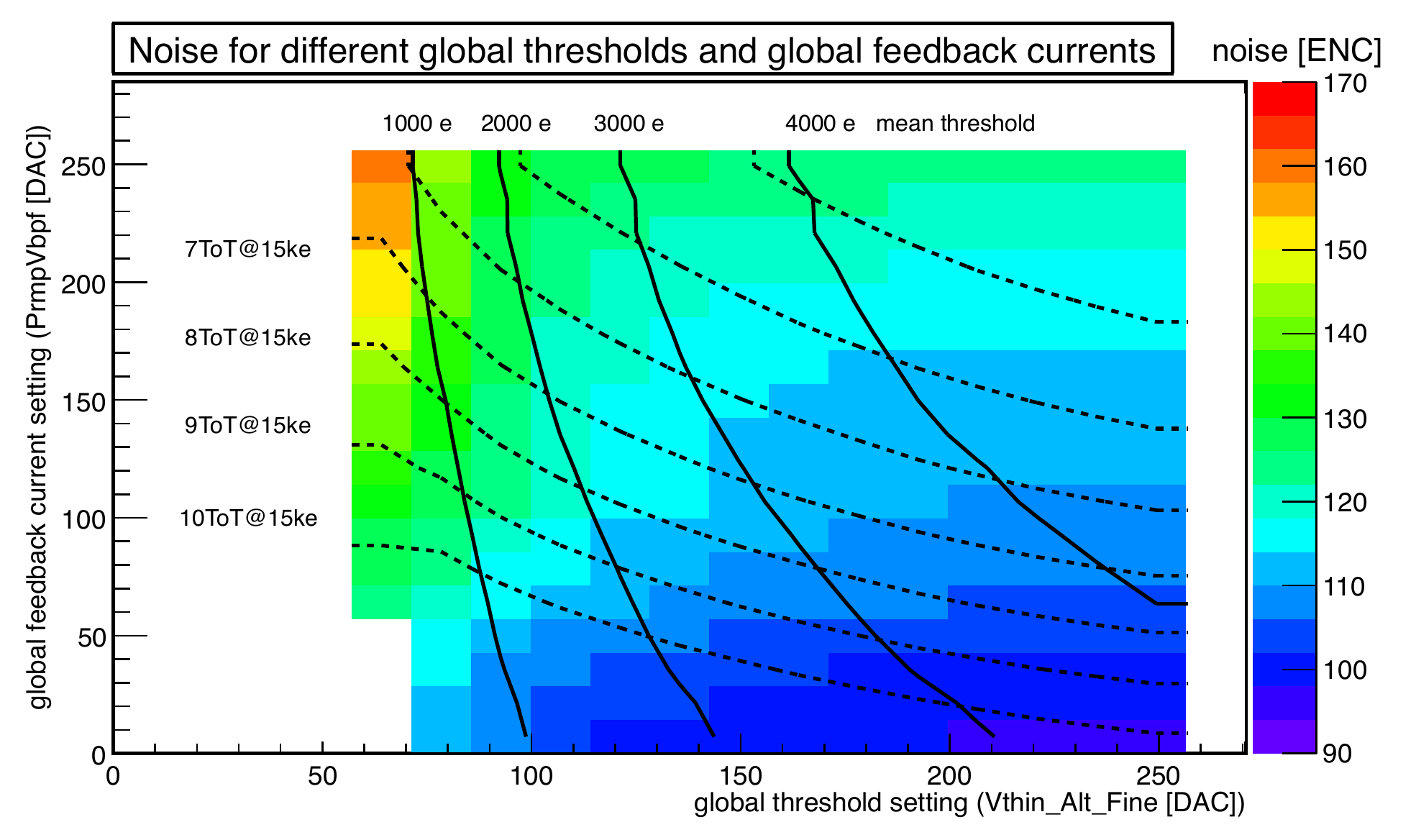}
 \caption{The dependence of the ENC on the global threshold setting (x-axis) and the global feedback current (y-axis) for a typical FE-I4A IC. The solid lines show the mean threshold determined using a gaussian fit. The dashed lines show the mean of the ToT distribution for an input charge of 15 ke$^{-}$.}
 \label{fig:ThreshfeedbackIBL}
\end{figure}

As expected, the 130 nm technology used for the IC design is very radiation tolerant. Three fully powered bare ICs have been exposed to TID doses of respectively 6, 75 and 200 Mrad in an 800 MeV proton beam at the Los Alamos Laboratory. Over this range of irradiations, the measured threshold dispersion was almost unchanged, and the noise increased by 15-25 \% in comparison to pre-irradation values. The FE-I4A front-end IC is rated to be radiation hard until TID values of approximately 250 Mrad. As discussed in Section \ref{section:IBL-module-prepost-rad}, some bump-bonded FE-I4A modules were irradiated in a 25 MeV proton beam to an estimated TID dose of 750 Mrad. Even then, the chip remained operational, although with some dead and noisy pixel cells that could be recovered following a reconfiguration of the IC. 

FE-I4A wafers have been tested in laboratories of the collaboration. The selection criteria included the analog and digital power taken by the ICs in different configuration states, global configuration scans, local pixel configuration scans, the analog and digital pixel maps,  and the threshold and noise pixel maps. The criteria retained were sufficient for the purpose of building high quality prototype modules in the first phase of prototyping. With this custom testing, the wafer yield (on a sample of 21 wafers) reached on average close to 70\%. Nevertheless, to enhance the failure mode coverage of the test primitives, the future wafer probe test list will address more points, for example  enhanced coverage of failure mechanisms in the 4-PDR, cross-talk assessment, current probing as a function of activity and the testing of stuck at bits in the synthesized peripheral digital blocks through scan chain probing.

The command decoder, the configuration register section and the DACs have been extensively characterized and only minor tuning was needed in the IBL production FE-I4B IC. The data output path, the data readout organization in the end of chip logic (EOCHL) block, the 8b10b encoder making use of the PLL-based generated high frequency clock and the LVDS transmission at 160 Mb/s have all been successfully characterized. For the FE-I4B IC iteration, new functionalities have been added to the EOCHL based on DAQ requirements, including increasing the bunch crossing and trigger counters, and the implementation of a user-defined event size limit.  Based on test results, a specific flavour of the FE-I4A pixel implementation was selected and the modifications brought to the peripery were of limited scope to ensure robustness of the FE-I4B production IC. 

The results, together with pre- and post-irradiated test results when bump-bonded to both planar and 3D sensors (see Section \ref{section:IBL-module-prepost-rad}), indicate that the FE-I4 IC is a very solid component for future IBL module developments. 

\section{The IBL Sensor Design and Performance} {\label{section:sensor-design-performance}}

As noted in Section \ref{section:introduction}, the main challenge for the IBL sensor is to retain adequate detection efficiency following fluences up to 6$\times10^{15}\,\mathrm{n_{eq}/cm^{2}}$. The current ATLAS pixel sensor (APS) is specified for a fluence of only $10^{15}\,\mathrm{n_{eq}/cm^{2}}$). Several promising new sensor technologies have been developed, for example advanced silicon designs \cite{Andricek:2011zz,Unno:2011S24,Tsurin:2011140,DaVia:2009505}, p-CVD diamond sensors \cite{Gorisek:2011iq} and pixelised gas detector concepts (GOSSIP) \cite{Hessey:2006iq}). Because of the tight IBL construction schedule, n$^+$-in-n planar and double-sided 3D silicon pixel sensor technologies, driven by the ATLAS Upgrade Planar Pixel Sensor and 3D Sensor R\&D Collaborations \cite{Muenstermann:2009iq,daVia:2011iq}, have been retained for prototyping with the FE-I4A IC in view of IBL construction. 

The prototype IBL sensor design is based on established prototypes \cite{Aad:2008zz,DallaBetta:2009ud} using the FE-I3 IC, but is modified to match the FE-I4 geometry, to minimize the radial envelope of the IBL and to minimize the material thickness of the IBL layer. To match the FE-I4 geometry, the pixel size is 50$\,\mu$m$\times$250$\,\mu$m. The pixel matrix is enlarged to 336 rows $\times$ 80 columns so that the area of an FE-I4 IC covers nearly six times the active area of an FE-I3 IC. The high fabrication yield of planar sensors has allowed the adoption of Multi Chip Sensor modules (MCS) having 2 FE-I4 ICs. For 3D sensors, yield limitations mandate the use of Single Chip Sensor modules (SCS). The small radial space between the beam pipe and the current b-layer prevents the use of shingled modules to ensure hermiticity. Due to this constraint, a flat arrangement of the modules on the staves is foreseen for the IBL. To guarantee a sufficient hermeticity of the detector layer, the inactive sensor edge must be minimised in the z-direction along the stave.

An additional change concerns the position of bump pads to access the bias-grid ring (DGRID) and the outer guard implantation which takes up all edge leakage currents (DGUARD). These bumps are routed via the read-out chip\footnote{Within the FE-I4A, the two bumps are shorted together while for the FE-I4B, they will be separately accessible.} and should normally be DC-connected to GND to be able to channel leakage currents.  
In the FE-I4 design, these bump pads  are placed within the second and second last column (DGRID) and within the third and third last column (DGUARD). The FE-I4 planar design is compared with that of the existing pixel sensor in figure~\ref{fig:BumpCompApixIBL}. For 3D sensors, no bias grid mechanism exists and DGRID is generally not used. There are, however, guard fences which can be connected to GND via the DGUARD or DGRID.

\begin{figure}[ht!]
 \centering
 \includegraphics[width=0.8\textwidth]{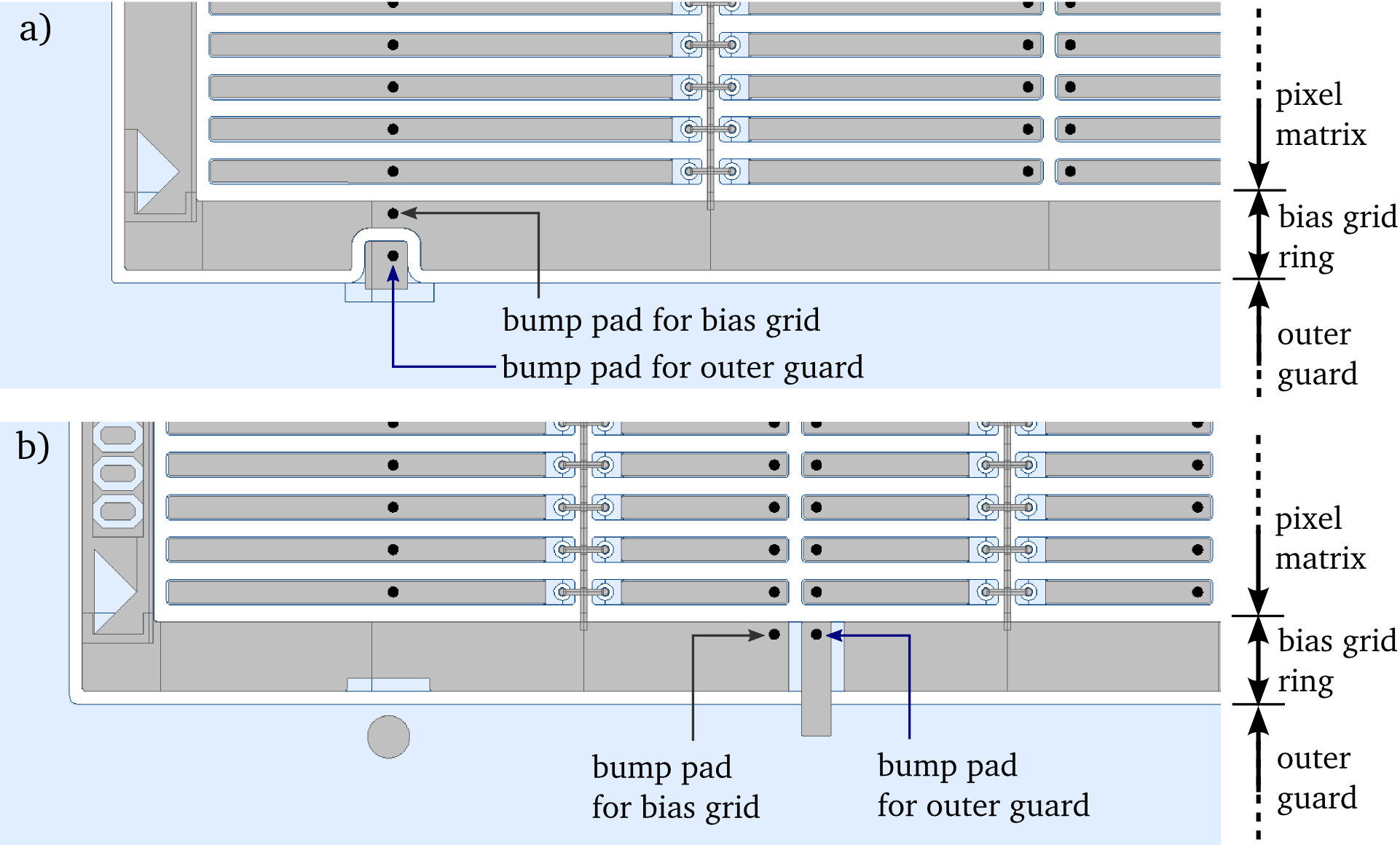}
 \caption{Top view of the corner of the active area of the ATLAS Pixel (a) and the IBL planar design (b). The n$^+$ implantation is blue, the metal grey. The additional bump pads (black circles) which ground the bias grid ring and the outer guard are marked.}
 \label{fig:BumpCompApixIBL}
\end{figure}

Regardless of technology, the following sensor requirements follow from the operational conditions.
\begin{itemize}
\item An inactive edge width $< 450\,\mu$m for 2-chip sensors (MCS) and $< 225\,\mu$m for 1-chip sensors (SCS) is required. This value translates into a geometric efficiency of $97.8\%$ (without taking into account any necessary gaps) and is deemed to be the upper tolerable limit for geometric inefficiencies. 
\item A sensor thickness between 150$\,\mu$m and 250$\,\mu$m is required. The current APS sensor thickness of 250$\,\mu$m is a conservative upper limit. Thinner sensors would reduce the material budget and for planar sensors would also yield more charge at a fixed bias voltage after irradiation \cite{RD50:2011a}.
\item A power dissipation $< 200$mW$/$cm$^2$ at V$\rm{_{b}}= -1000$~V bias voltage is specified. This specification limits the sensor power dissipation to approximately that dissipated by the FE-I4. The specification is used as input for cooling design and thermal runaway calculations.
\item The maximum leakage current is specified to be $< 100$ nA$/$pixel. This is the maximum pixel current allowed by the FE-I4 compensation specification. 
\item The sensor operating temperature should be $<  -15^{\circ}\ \mathrm{C}$ at $< 200$mW$/$cm$^2$. This maximum temperature is specified to engineer the cooling system.
\item The in-time hit efficiency is specified to be $> 97\%$ after a benchmark fluence of $\rm5{\times 10^{15} n_{eq}/cm^2}$ at a maximum $\rm{|V_{b}|}$< 1000 V. The specification does not include geometric inefficiencies. This specification is chosen to limit the degraded performance after irradiation. 
\end{itemize}

The IBL will consist of a combination of n$^+$-in-n MCS modules using planar sensors and n$^+$-in-p SCS modules using 3D sensors. Specific details of the two designs are discussed in the following sub-sections. However, both planar and 3D prototypes discussed in this paper are of the SCS type.

\subsection{The planar sensor design} {\label{section:sensor-design-planar}}

The baseline IBL planar sensor is an electron-collecting n$^+$-in-n silicon sensor design fabricated by CiS\footnote{CiS Forschungsinstitut fur
Mikrosensorik und Photovoltaik GmbH, Konrad-Zuse-Strasse 14, 99099 Erfurt, Germany.}. The 200$\,\mu$m thick lightly n-doped substrate contains a highly n$^+$-doped implantation on the front side and a highly p$^+$-doped implantation on the back side. It is based on the current APS design \cite{Alam:2001217} also fabricated by CiS.

The p$^+$ implantation is made as a single large high voltage pad opposite the pixel matrix, that is surrounded by 13 guard ring implantations with a total width of approximately 350$\,\rm{\mu m}$. The purpose of the guard rings is to provide a controlled potential drop from the high voltage pad to the grounded cutting edge. The guard ring scheme is adopted from the original APS design that used 16 guard rings. It was optimized during previous studies \cite{Goessling:2010410} to find the combination of number of guard rings and dicing street position to ensure the slimmest possible edge while still allowing safe depletion before irradiation\footnote{n$^+$-in-n sensors deplete from the p$^+$-implant and must be operated fully depleted before type inversion to ensure inter-pixel isolation.}.

The n$^+$ implantation of the prototype SCS design is segmented into a matrix of 80 columns and 336 rows of mostly 250$\times$50$\,\rm{\mu m^{2}}$ pixels surrounded by an inactive edge region. The inter-pixel isolation is adapted from the APS sensor and follows the moderated p-spray concept \cite{Wuestenfeld:2001iq}. The outermost columns contain long pixels that are extended to 500$\,\rm{\mu m}$ length and overlap the guard ring structure by 250$\,\rm{\mu m}$. The active area, defined by the 50\% hit efficiency, is required to be 210~$\pm$ 10 $\mu$m from the cutting edge (see figure~\ref{fig:Edge-comp}). Due to the non-vertical inhomogenous electric field, before type inversion \cite{Altenheiner:2011iq}, the region has been shown to remain active for 200$\,\rm{\mu m}$ thick detectors at $\rm{V_{b}}= -60$~V until the edge of the pixel implant.  After type inversion, the hit efficiency further improves at the pixel implant, because the depletion zone grows from the n$^+$ pixel implant.

\begin{figure}[ht!]
 \centering
 \includegraphics[width=0.95\textwidth]{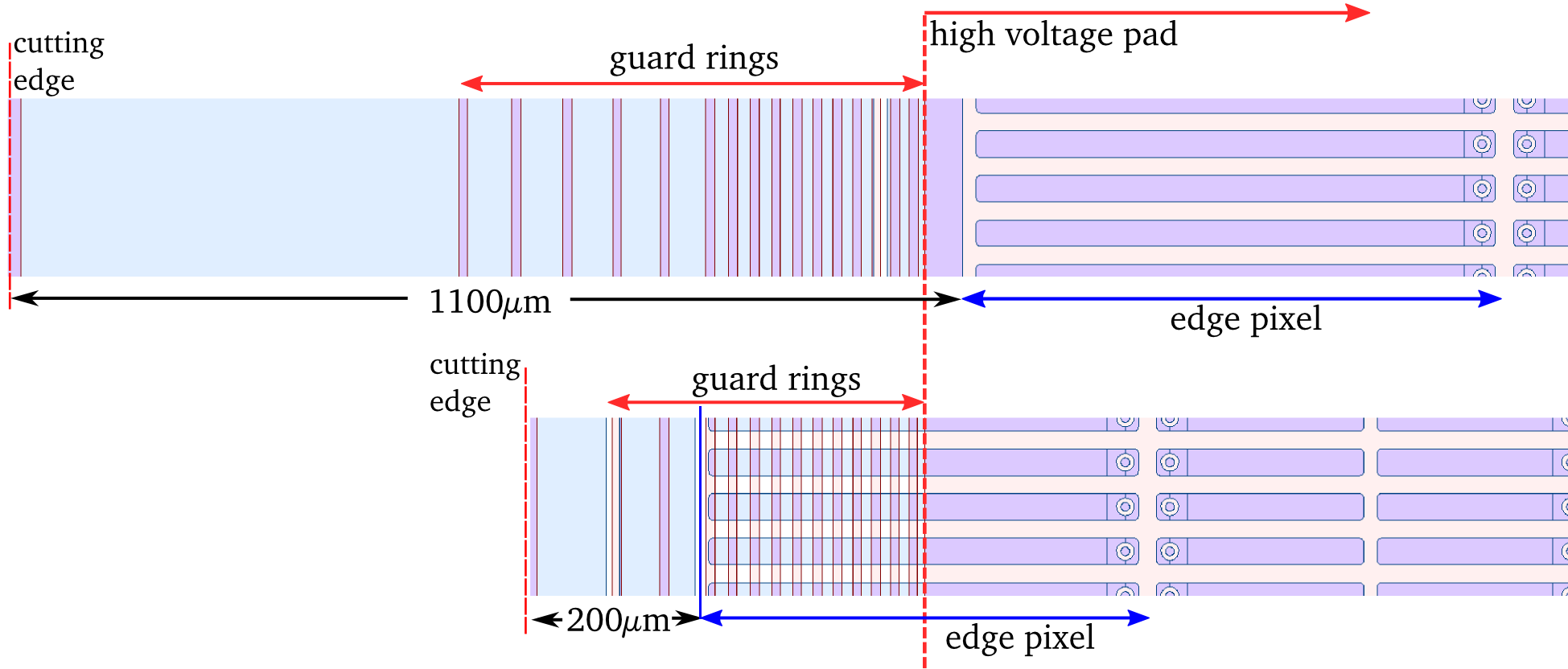}
 \caption{Comparison of the edge region of the current ATLAS Pixel (APS) design (upper) and the IBL planar sensor design (lower).}
 \label{fig:Edge-comp}
\end{figure}

To ease characterization and to avoid a floating potential on pixels having an open bump connection, a punch-through network (bias grid) following the APS design was implemented even though this is known to lead to reduced charge collection efficiency in the bias-dot region after irradiation. The bias dots are always located at the opposite side of a pixel cell with respect to the contact bump (see figure~\ref{fig:BumpCompApixIBL}). The bias grid is connected to an approximately 90$ \,\mu$m wide bias grid ring which surrounds the pixel matrix. Outside the bias ring, a homogeneous n$^+$-implantation (designated as the outer guard, edge implant or DGUARD) extends to the dicing streets and ensures that the sensor surface outside the pixel matrix and the cutting edges share the same potential.

Each pixel, the bias grid and the outer guard are connected to the FE-I4 read-out chip via bump-bonds. As already noted, there are two bumps each for the bias grid (DGRID) and outer guard (DGUARD).

The prototype wafer mask contained two versions of FE-I4 sensors, the slim-edge design described above and a conservative design where the edge pixels were only 250$\,\mu$m long without any overlap between pixel and guard rings. Both designs behaved identically except for the edge efficiency where the conservative design showed the expected 450$\,\mu$m inactive edge. It is the slim-edge design that is described in this paper.

The production used n-doped FZ silicon wafers with a $<$111$>$ crystal orientation and a bulk resistivity of $\rm{2-5~k\Omega~cm}$, thinned to thicknesses of 250, 225, 200, 175 and 150$\,\mu$m. All wafers were diffusion oxygenated for 24 hours at $1150^{\circ}$C after thinning, as for the current APS production \cite{Hugging:2001iq}. The remaining production steps are as for the APS sensor: thermal oxide deposition, n$^+$-implantation,  tempering, p$^+$-implantation, tempering, nitride deposition, p-spray  implantation, tempering, nitride openings, oxide openings, aluminium deposition and patterning, and passivation deposition.

The production of 5 different thicknesses aimed at obtaining experience of the production yield without the use of support wafers; thin sensors are preferred because they can be operated at lower bias voltage and because of the reduced detector material. After irradiation, they also tend to give more collected charge for the same bias voltage. The production yield was stable down to the 175$\,\mu$m batch. However, the bump-bonding vendor required at least 200$\,\mu$m thick 4" wafers to be able to apply Under-Bump Metallisation (UBM)\footnote{UBM is a post-processing galvanic application of a sandwich of metals to make the aluminium bump-bonding pads solderable. The choice of UBM depends on the bump-bonding process (Pb/Sn, Ag/Sn, Indium reflow, Indium stud bumps.).} without the need for an additional support wafer. A  200$\,\mu$m sensor thickness was therefore proposed for the IBL production. No yield difference was measured between the slim-edge and conservative designs and so the slim-edge design was selected as the baseline geometry.

The floor plan of the IBL prototype production includes one MCS and two SCS sensors for each of the slim-edge and conservative designs in the central part of the wafer. In addition, several FE-I3 compatible sensors, diodes and dedicated test structures are included in the periphery of the wafer.

\subsection{The 3D Sensor Design} {\label{section:sensor-design-3D}}

The 3D sensor fabrication uses a combination of two well established industrial technologies:
MEMS (Micro-Electro-Mechanical Systems) and VLSI. The micromachining used in MEMS is
applied to etch deep and very narrow apertures within the silicon wafer using the so-called Bosch
process \cite{Ayon:0964-1726-10-6-302} followed by a high temperature thermal diffusion process to drive dopants in to form the n$^+$ and p$^+$ electrodes. Two etching options have been considered for the prototype 3D sensors: Full3D with active edges and double-side 3D with slim fences. For the first option, etching is performed from the front side, with the use of a support wafer and at the same time implementing active edge electrodes, but this requires extra steps to attach and remove the support wafer. In the second option, etching is made from both sides (n+ columns from the front side, p$^+$ columns from the back side) without the presence of a support wafer.

For the prototype 3D sensors reported here, the double-side option with slim fence was chosen since all the technological steps were reliable and well established. It should be noted that all of the remaining processing steps after the electrode etching and doping are identical to those of a planar silicon sensor. In particular, both the 3D and planar sensors have the same handling and hybridization requirements.

The 3D silicon sensors use 4" FZ p-type high resistivity wafers having specifications normally used for fabrication of high resistivity p-type silicon sensors. The wafers were supplied by TOPSIL\footnote{Topsil Semiconductor Materials A/S, Linderupvej 4, DK-3600 Frederikssund, Denmark.} to two manufacturers in the 3D Pixel Collaboration: CNM-CSIC\footnote{Centro Nacional de Microelectronica (CNM-IMB-CSIC), Campus Universidad Autonoma de Barcelona, 08193 Bellaterra (Barcelona), Spain. See http:// www.imbcnm. csic.es.} and FBK\footnote{Fondazione Bruno Kessler (FBK), Via Sommarive 18, 38123 Povo di Trento, Italy. See http://www.fbk.eu.}. Figure~\ref{CV_figure_fabrication3D} shows details of the 3D layout for the CNM (a) and and FBK (b) sensors. 

The main difference between the two sensor versions regards the column depth: in CNM sensors, columns do not traverse the substrate but stop at a short distance from the surface of the opposite side, whereas FBK sensors have traversing columns. Another difference concerns the isolation implantation between the n$^+$ columns at the surface: p-stops are implanted on the front side of CNM sensors while FBK sensors use p-spray implantations on both sides. The slim edge guard ring design in CNM sensors is made using the combination  of a n$^+$ 3D guard ring that is grounded, and fences that are at the bias voltage from the ohmic side. In FBK sensors, the slim edge fence consists of several rows of ohmic columns that effectively stop the lateral depletion region from reaching the cut line, thus significantly increasing the shielding of the active area from edge effects \cite{Povoli:2012a}.

\begin{figure}[ht!] 
\centering
\subfigure[a][]{\includegraphics[width=0.49\linewidth] {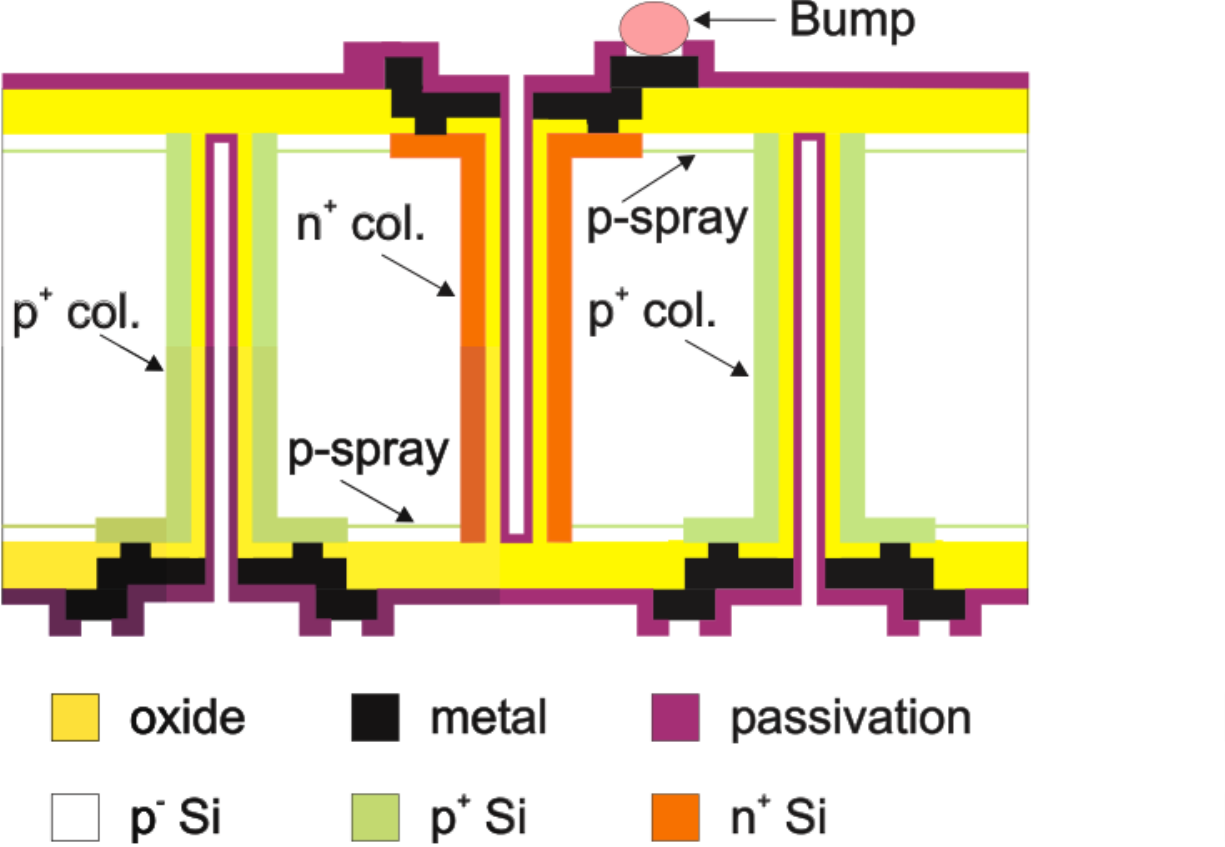}}
\subfigure[b][]{\includegraphics[width=0.49\linewidth] {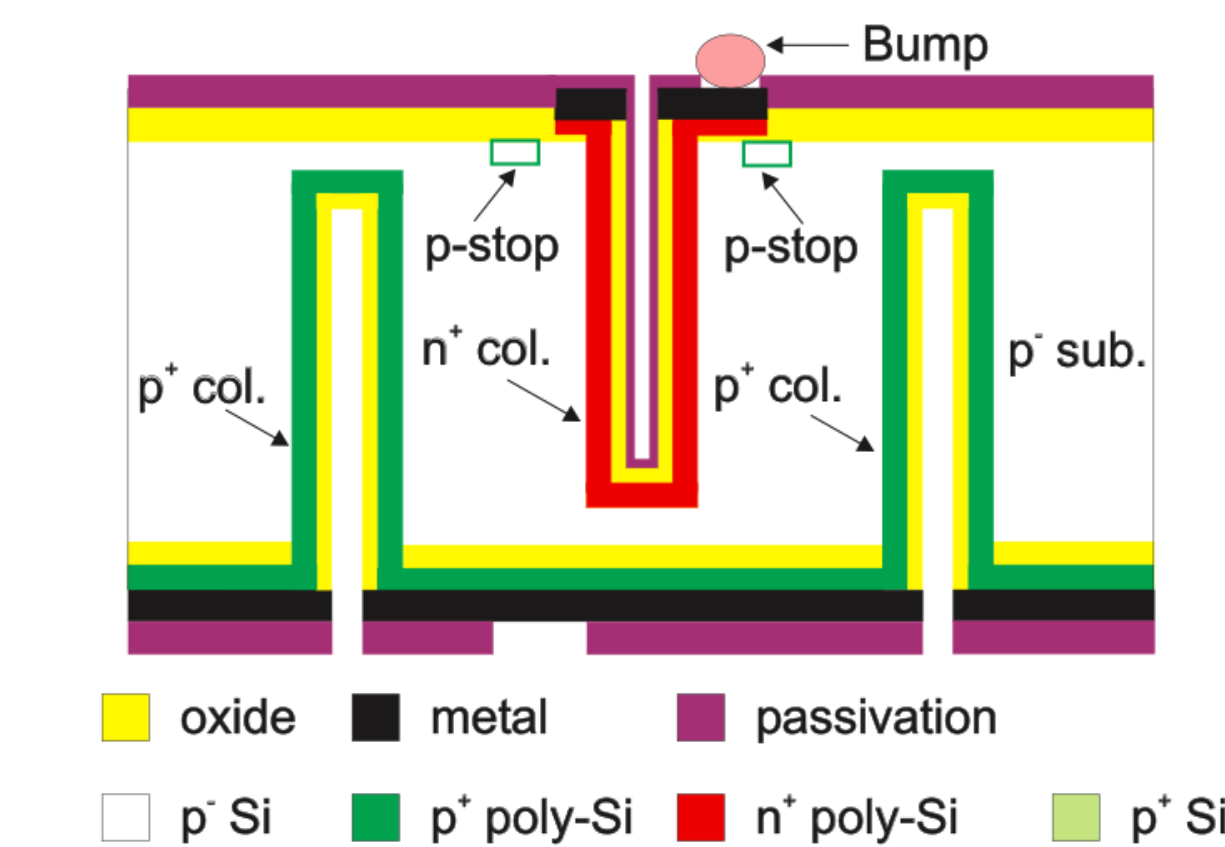}}
  \caption{3D etched columns from the pixel sensor design of the FBK (a) and CNM (b) fabrication facilities.}
  \label{CV_figure_fabrication3D}
\end{figure}

The core of the prototype wafer layout is common for both CNM and FBK sensors, and contains 8 SCS sensors adapted for the FE-I4A IC, 9 single chip sensors compatible with the currently installed ATLAS FE-I3 IC, and 3 pixel sensors compatible with the CMS-LHC experiment front-end readout IC. At the wafer periphery, test structures that are foundry specific are added to monitor the process parameters and to perform electrical tests. 
  
\begin{table*}[ht!]
\footnotesize
\begin{center}
\caption{3D sensor specifications.}
\medskip
\begin{tabular}{|l|l|}
\hline
\hline
         \textbf{Item} &\textbf{Sensor Specification} \\
\hline
\hline
 Module type & single  \\
 Number of n$^+$ columns per 250 $\mu$m pixel & 2 (so-called 2E layout) \\
 Sensor thickness & 230 $\pm$ 20 $\mu$m \\
 n$^+$-p$^+$ columns overlap & $>$ 200 $\mu$m \\
 Sensor active area & 18860 $\mu$m $\times$ 20560 $\mu$m \\
                              & (including scribe line) \\
 Dead region in Z & $<$ 200 $\mu$m guard fence $\pm$ 25 $\mu$m cut residual \\
 Wafer bow after processing & $<$ 60 $\mu$m \\
 Front-back alignment & $<$ 5 $\mu$m \\
\hline      
\hline
\end{tabular}
\label{specs-3D}
\end{center}
\end{table*} 

\subsection{Measurements of fabricated sensors} {\label{section:sensor-measurements}}

The main measurements used for sensor production quality assurance (QA) are $\rm{I-V_{b}}$ curves that are made at all stages of production: on the wafer (for all sensors), after dicing and when fully assembled into modules. 

For planar sensors, the $\rm{I-V_{b}}$ curve is a measure of the sensor leakage current via the bias grid by placing the sensor with the n-side onto a metal chuck and applying high voltage to the p-side HV pads. The metal chuck connects the n-implanted cutting edge and a punch-through is formed between the outer guard and bias grid ring. $\rm{C-V_{b}}$ curves measuring the depletion voltage V$\rm{_{d}}$ by sensing the plateau in capacitance that is reached after the bulk is fully depleted, were also made for some devices. To ensure safe depletion, a breakdown voltage V$\rm{_{bk}}$ > V$\rm{_{d}}$ + 30 V is required for accepted devices.

For both 3D manufacturers, $\rm{I-V_{b}}$ measurements are made on each sensor at the wafer scale using probe stations, using a removable temporary metal (FBK) or measuring the guard ring current (CNM). Working 3D sensors were required to satisfy the following electrical specifications at room temperature  ($20-24^{\circ}\ \mathrm{C}$):
\begin{itemize}
\item $\rm{|V_{d}|} \leq$~20~V and $\rm{|V_{b}|~\geq~|V_{d}|}$ +10~V (by construction V$\rm{_{b}}$ and V$\rm{_{d}}$ are much lower than for planar sensors);
\item Leakage current I(V$\rm{_{b}}$) $<$ 2 $\mu$A per sensor and I($\rm{|V_{b}|}$)/I($\rm{|V_{b}|}$ - 5 V) $<$ 2;
\item Guard-ring current before bump-bonding  I$\rm{_{guard}}$(V$\rm{_{b}}$) $<$200 nA per sensor;
\item Breakdown voltage $\rm{|V_{bk}|} >$ 25 V.
\end{itemize}
After processing, the accepted 3D FE-I4A SCS-type sensors also satisfied the geometrical specifications of table~\ref{specs-3D}.

For the planar IBL qualification production, successive measurements have been made on all batches after production, after UBM application and after dicing. Figure~\ref{fig:PPS_IV-curves} shows $\rm{I-V_{b}}$ measurements of 200$\,\mu$m thick sensors from 4 wafers after the UBM and dicing steps. Only a few sensors exhibit early breakdown; the depletion voltage was measured to be $\rm{|V_{d}|} < $50 V and hence a requirement $\rm{|V_{bk}|}$ > 80 V was applied. The production yield including UBM and dicing post-processing was approximately 90\%. Also shown are the $\rm{I-V_{b}}$ measurements for those sensors from other wafers subsequently used to fabricate the PPS modules described in Section \ref{section:IBL-module}.

\begin{figure}[ht!]
 \centering
 \includegraphics[width=0.9\textwidth]{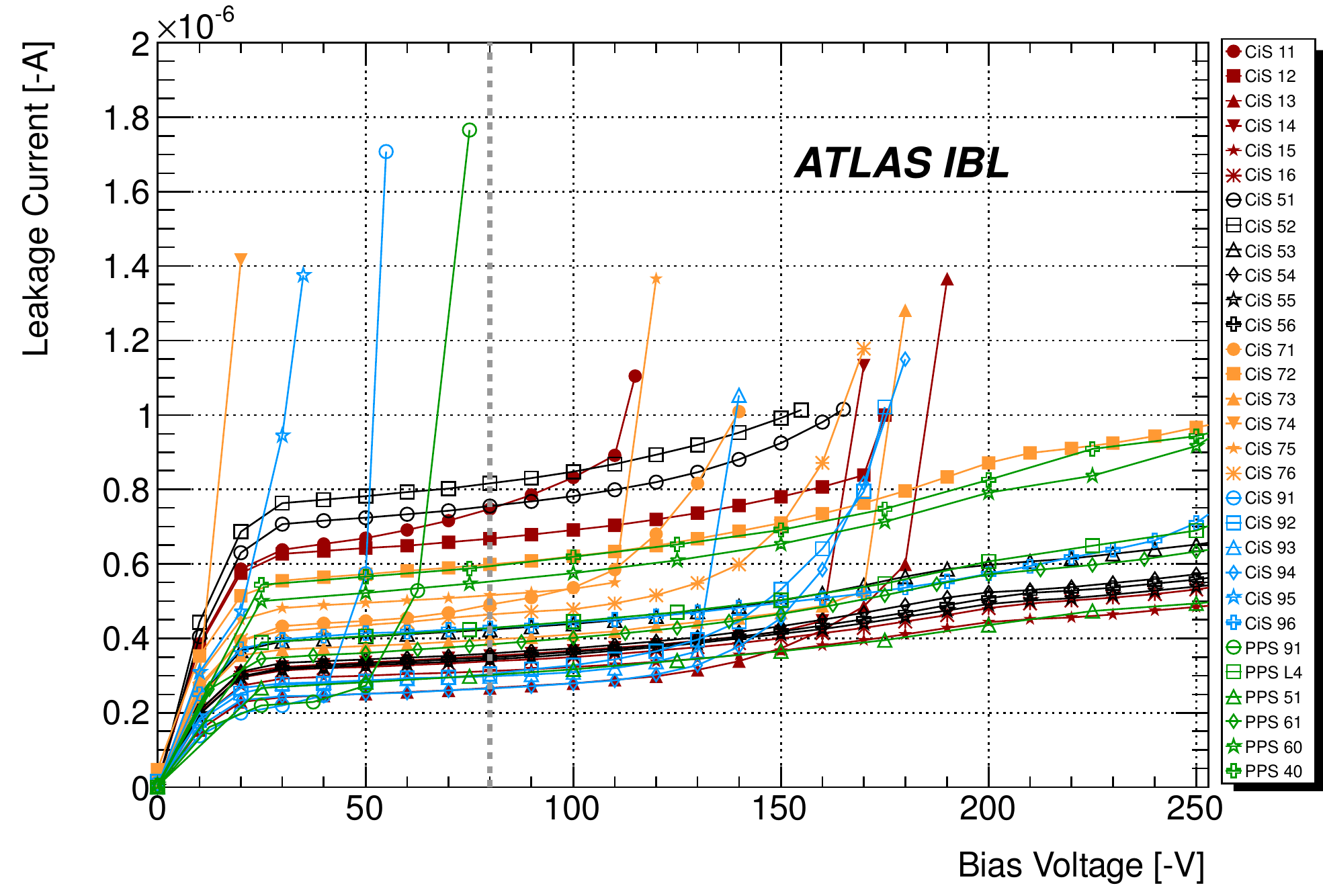}
 \caption{$\rm{I-V_{b}}$ curves of single-IC planar sensors from four 200$\,\mu$m thickness wafers, after the UBM and after slim-edge dicing steps. Only a few sensors exhibit breakdown at less than the $\rm{|V_{bk}|}$ > 80 V requirement shown in the figure. Also shown are the $\rm{I-V_{b}}$ measurements for sensors from different wafers subsequently used to fabricate the PPS modules described in the text. }
 \label{fig:PPS_IV-curves}
\end{figure}

\begin {figure}[ht!]
\centering
\subfigure[a][]{\includegraphics[width=0.49\linewidth] {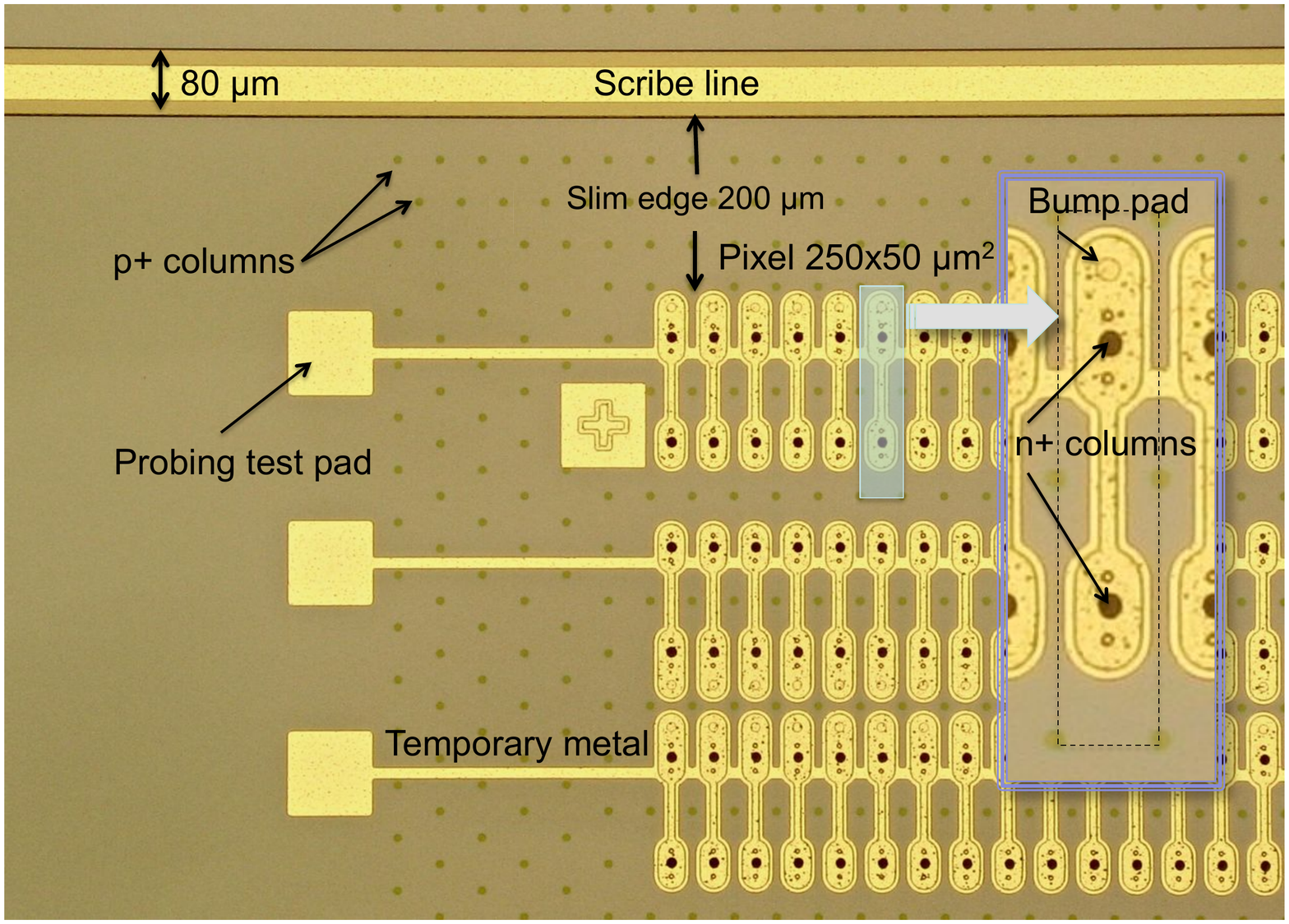}}
\subfigure[b][]{\includegraphics[width=0.49\linewidth] {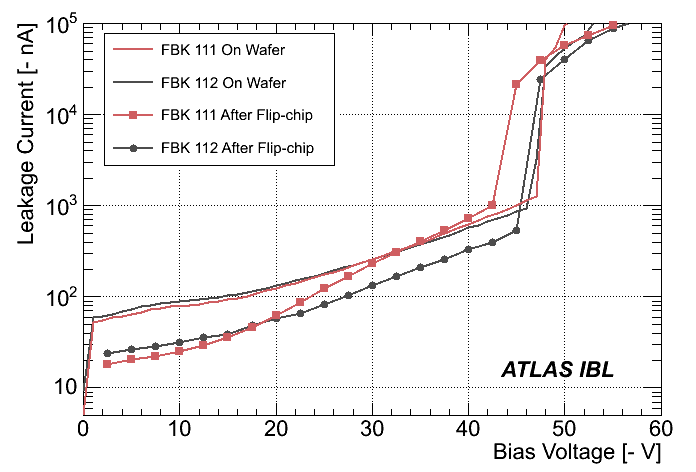}}
\caption{(a) FBK temporary metal used for sensor selection on wafer. Details are visible for two columns. On the left, the metal line termination on probing pads are outside the active region to avoid surface damage. (b)  $\rm{I-V_{b}}$ measurements of the FBK 111 and FBK 112 modules, before and after bump-bonding but before irradiation. The on-wafer curve corresponds to the sum of the 80 columns $\rm{I-V_{b}}$ characteristics of 336 pixels joined together by an aluminium strip. Each measurement is recorded twice to check reproducibility.}
\label{CV_figure_Meas13D}
\end{figure}

 The temporary metal selection method used by FBK allows a measurement of $\rm{I-V_{b}}$  at the column level in each sensor. A temporary metal line is deposited after the completion of the process, as in figure~\ref{CV_figure_Meas13D} (a). Probing pads, on the left hand side of the figure and placed outside the active region to avoid surface damage, are used to measure the $\rm{I-V_{b}}$ at column level. This operation is made automatically on pixels corresponding to the 80 FE-I4A columns by using a dedicated probe card. Each $\rm{I-V_{b}}$  measurement therefore tests the performance of 336 pixels, allowing a fine definition of potential defects. After this operation is completed and $\rm{I-V_{b}}$  curves obtained, the temporary metal is removed. An example of the measurement is shown in figure~\ref{CV_figure_Meas13D} (b) where $\rm{I-V_{b}}$ measurements from all 80 columns are added together to provide an $\rm{I-V_{b}}$ measurement of the full sensor for the FBK 111 and FBK 112 modules. $\rm{I-V_{b}}$ measurements after bump-bonding show only a small increase of the sensor leakage current, and confirm the measurement method.

The guard ring $\rm{I-V_{b}}$ measurement used by CNM is sensitive to defects on the sensor edges that have been shown to be indicative of the bulk sensor behaviour. A photograph of the guard ring probing pad is visible in figure~\ref{CV_figure_Meas23D} (a). The guard ring current is not a measurement of the full sensor current but should provide a reproducible selection test based on V$\rm{_{bk}}$. Tests performed before and after bump-bonding support this measurement method, as can be seen in figure~\ref{CV_figure_Meas23D} (b) where $\rm{I-V_{b}}$ measurements have been recorded. The difference in leakage current is due to the reduced volume probed through the guard ring pad. The improvement in V$\rm{_{bk}}$ has not been systematically studied, but is thought to result from the release of thermal stresses during the bump-bonding process, or from non-controlled effects such as humidity.  

Further $\rm{I-V_{b}}$ measurements of FBK and CNM sensors after bump-bonding are discussed in Section \ref{section:IBL-module-prepost-rad}.

\begin {figure}[ht!]
\centering
\subfigure[a][]{\includegraphics[width=0.49\linewidth] {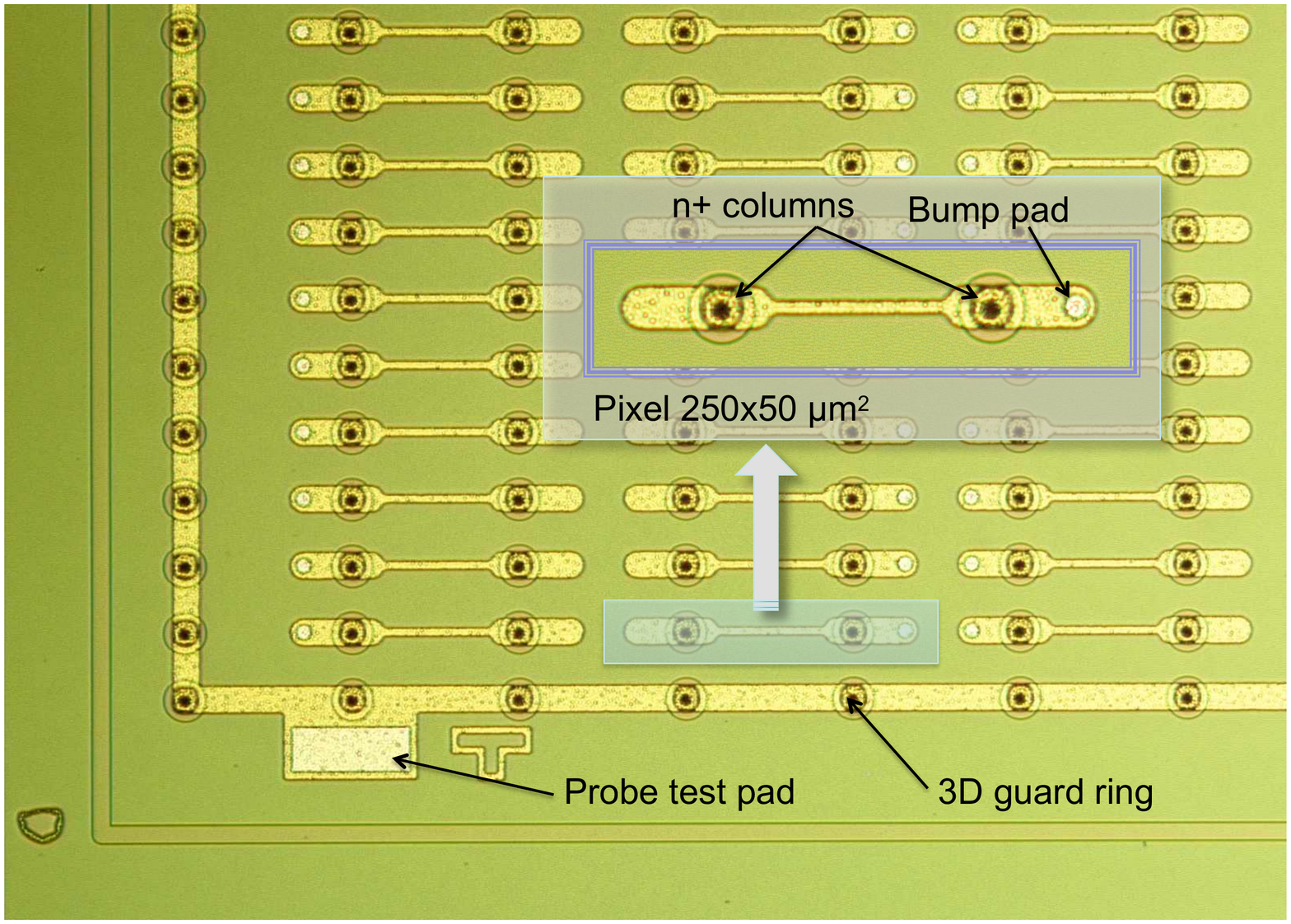}}
\subfigure[b][]{\includegraphics[width=0.49\linewidth] {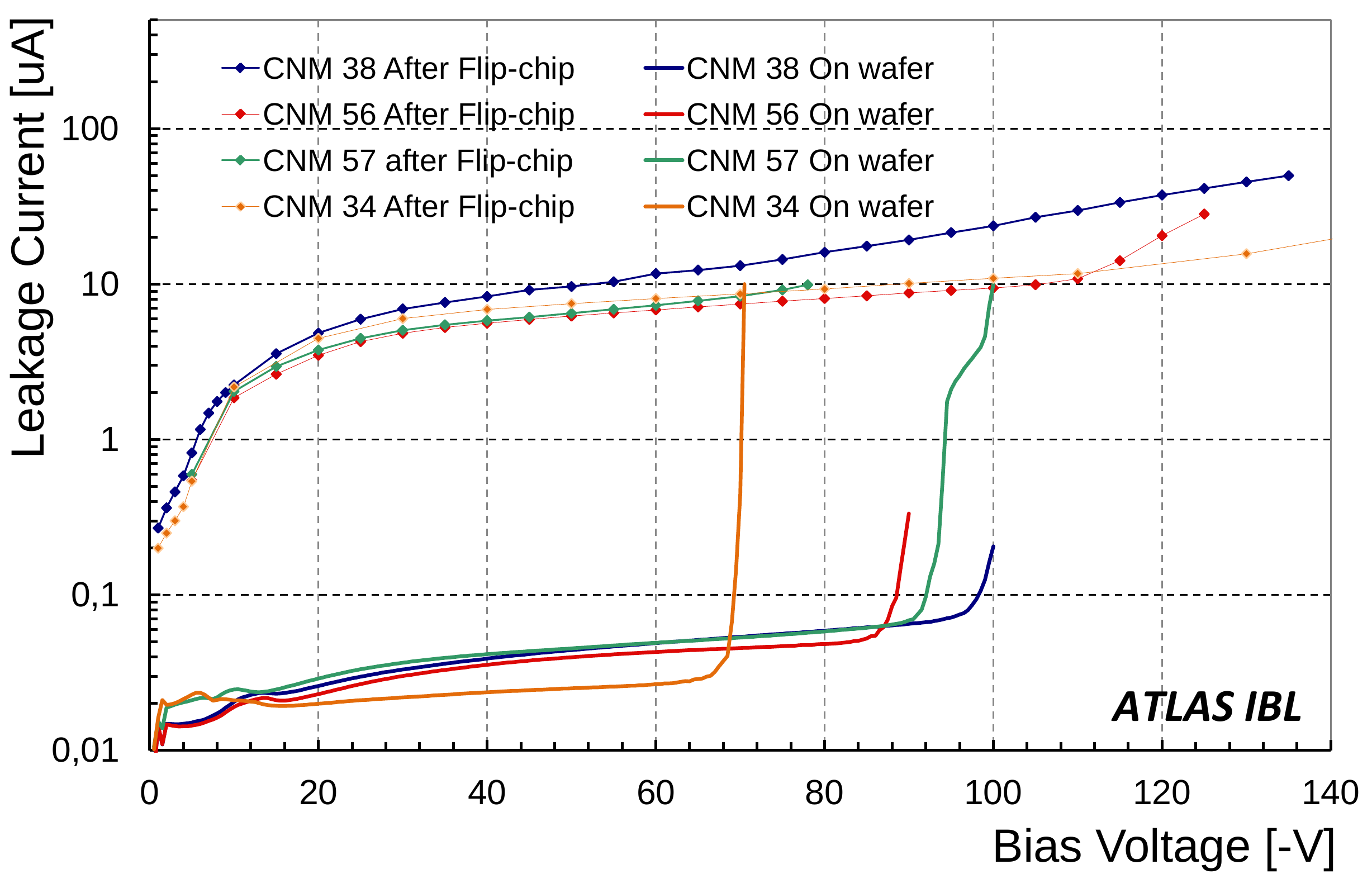}}
\caption{(a) Corner picture of one of the CNM 3D sensors showing the guard ring surrounding the pixel matrix active area and guard ring probing pad. (b) Guard ring leakage current measured as a function of the bias voltage for 3 CNM 3D sensors before bump-bonding. The full sensor current is also shown after bump-bonding to the FE-I4 readout electronics.}
\label{CV_figure_Meas23D}
\end{figure}

\section{The IBL Module } {\label{section:IBL-module}}
   
\subsection{Assembly of the IBL module} {\label{section:IBL-module-assembly}}

As noted in Section \ref{section:sensor-design-performance}, prototype FE-I4A compatible sensors have been fabricated by CiS (planar sensors) and by FBK and CNM (3D slim edge sensors). The planar sensors are of two types: MCS sensors as foreseen for the final IBL planar modules, and smaller SCS sensors as foreseen for the final 3D modules. All of the prototype modules described in this paper, either planar or 3D silicon, are of the SCS type, with a single FE-I4A IC bump-bonded to the sensor. 

An IBL prototype module consists of a sensor integrated to a FE-I4A IC via flip-chip bump-bonding, connecting each pixel on the sensor side to its dedicated FE-I4A pixel pre-amplifier input. Bump-bonding requires a low connection defect rate (nominally $<$10$^{-4}$), with a bump density of order 8000 per cm$^{2}$ and a bump pitch in the r$\phi$ direction of 50$\,\rm{\mu m}$. The principal bump-bonding provider for this prototyping phase has been IZM\footnote{Fraunhofer IZM-Berlin, Gustav-Meyer-Allee 25, 13355 Berlin.}. IZM also provided a large fraction of the bump-bonding for FE-I3 based modules of the current ATLAS pixel detector. 

A complication is that the FE-I4A IC is thinned to reduce the IBL material budget. The procedure for thin IC bump-bonding required specific R\&D development. After wafer level thinning, a glass handling wafer is temporarily glued to the FE-I4A wafer backside. Once Ag-Sn solder bumps are formed on the FE front side, the FE are diced, and flip-chipping to the sensors is carried out. The carrier chip is then detached from the assembly by laser exposure. With this method, FE-I4A ICs have been thinned to 150$\,\rm{\mu m}$ and 100$\,\rm{\mu m}$, and successfully flip-chipped. Working with thin ICs also brings constraints to all subsequent steps in the module assembly, for example the module manipulation and the wire-bonding steps needed to power the module. As noted in Section \ref{section:sensor-design-performance}, the sensors are also thin (nominally 200$\,\rm{\mu m}$ unless otherwise stated for planar sensors and 230$\,\rm{\mu m}$ for 3D sensors).

Contrary to FE-I3 based modules, no additional steering IC for FE-I4A based modules is required: the FE-I4A needs only 2 LVDS inputs (40 MHz clock and 40 Mb/s command) and streams out data on one LVDS transmitter pair at 160 Mb/s.

\subsection{Pre- and post-irradiation module performance} {\label{section:IBL-module-prepost-rad}}

The focus of this section is to provide test characterizations of the constructed SCS modules, before and after irradiation to the expected integrated fluence, and where possible to compare the data with that obtained using the individual sensors. As noted previously, $\rm{I-V_{b}}$  measurements are made, as well as measurements of the ENC. In addition ToT measurements of the collected charge using a $^{90}$Sr source allow comparisons of the charge collection before and after irradiation, for a given threshold setting\footnote{Since no reliable ToT charge calibration exists, uncalibrated ToT spectra, as well as the peak (most probable value or MPV) are presented in this paper, in units of the 25 ns bunch crossing clock (BC).}. Unless explicitly stated, all un-irradiated pixel modules are characterized at room temperature ($20-24^{\circ}\ \mathrm{C}$), while irradiated modules are characterized at $-15^{\circ}\ \mathrm{C}$. 

Data from irradiated modules result from exposure at KIT \footnote{Karlsruhe Institute of Technology, Karlsruhe, Germany.} with a 25 $\,\mathrm{MeV}$ proton beam or with neutrons at the TRIGA reactor \footnote{TRIGA reactor, Jozef Stefan Institute, Ljubljana, Slovenia.}. The particle fluences are scaled to 1 MeV equivalent neutrons per square centimeter ($\mathrm{n_{eq}/cm^2}$) with hardness factors of 1.85 and 0.88 for 25 MeV protons and reactor neutrons, respectively. The uncertainty in the irradiation fluences is smaller than $10 \%$. Because of the low beam energy, modules irradiated at KIT received a TID dose of up to 750~Mrad, three times the rated radiation hardness of the FE-I4A IC. These modules were unpowered and maintained at a temperature of $-25^{\circ}\ \mathrm{C}$ during irradiation. At the TRIGA reactor, the modules were irradiated unpowered at the ambient temperature of $45-50^{\circ}\ \mathrm{C}$. After irradiation, all irradiated modules were stored at < $0^{\circ}\ \mathrm{C}$, but were annealed at $60^{\circ}\ \mathrm{C}$ for two hours before the first measurements, effectively removing any annealing effects of the previous irradiation history. 

Planar modules (PPS) and 3D modules of each type (CNM and FBK) were retained for comparison with irradiated samples at both the test bench and test beam levels. Unfortunately, for logistic reasons, there are no consistent measurements of $\rm{I-V_{b}}$ at each of the sensor, pre-radiation and post-irradiation stages for individual modules. The modules used, together with the level of irradiation, are shown in table~\ref{table:duts}. 

\begin{table*}
\centering
\footnotesize
\caption{Sensors characterized following irradiation at the KIT 25 MeV proton beam or the TRIGA reactor neutron source (see text). The quoted fluences are normalised to the equivalent damage of 1 MeV neutrons. Also listed are un-irradiated modules used for comparison. The PPS L1 module is characterized before and after irradiation. Modules that were used in the test beam are noted.
}
\medskip
\begin{tabular}{|l|l|l|c|l|}
\hline\hline
Sample ID & Type                                   & Irradiation 	 & Dose   &\\
	        &                                           & Facility & $10^{15} \ \mathrm{n_{eq}/cm^2}$  &\\
\hline \hline
PPS 15    & PPS Slim Edge $250 \ \mathrm{\mu m}$   &                      & un-irradiated   &   \\
PPS 31    & PPS Slim Edge $250 \ \mathrm{\mu m}$   &                      & un-irradiated   & Test beam     \\
PPS 40    & PPS Slim Edge $200 \ \mathrm{\mu m}$   &                      & un-irradiated   & Test beam        \\
PPS 51    & PPS Slim Edge $200 \ \mathrm{\mu m}$   & KIT (p)              & 6     		   &  \\
PPS 60    & PPS Slim Edge $200 \ \mathrm{\mu m}$   & KIT (p)              & 6 		   & Test beam   \\
PPS 61    & PPS Slim Edge $200 \ \mathrm{\mu m}$   & KIT (p)              & 6  		   & Test beam  \\
PPS 91    & PPS Slim Edge $200 \ \mathrm{\mu m}$   &                      & un-irradiated    &   \\
PPS L1    & PPS Slim Edge $250 \ \mathrm{\mu m}$   &                      & un-irradiated    &   \\
              &                                                                   & TRIGA (n)         & 3.7 		   &   \\
PPS L2    & PPS Slim Edge $250 \ \mathrm{\mu m}$   & TRIGA (n)         & 3.7 		   & Test beam  \\
PPS L4    & PPS Slim Edge $200 \ \mathrm{\mu m}$   & TRIGA (n)         & 5 			   & Test beam  \\
              &                 						  &			  &			   &     \\
CNM 34  & CNM IBL design                         		  & KIT (p)              & 5 			   & Test beam   \\
CNM 36    & CNM IBL design                         		  & KIT (p)              & 6 			   &     \\
CNM 55    & CNM IBL design                         		  & n/a                   & un-irradiated   & Test beam    \\ 
CNM 81    & CNM IBL design                         & TRIGA (n)            & 5 & Test beam   \\
CNM 82    & CNM IBL design                         & TRIGA (n)            & 5 & Test beam   \\
CNM 97    & CNM IBL design                         & KIT (p)                 & 5 & Test beam   \\
CNM 100  & CNM IBL design                         & TRIGA (n)            & 2 & Test beam   \\
CNM 101  & CNM IBL design                         &                            & un-irradiated &   \\ 
                 &                 			      &				&		       &     \\
FBK 13     & FBK IBL design                         &                      & un-irradiated       & Test beam                      \\
FBK 87     & FBK IBL design                         & KIT (p)                  & 5 & Test beam   \\
FBK 90     & FBK IBL design                         & KIT (p)                  & 2 & Test beam   \\
FBK 104   & FBK IBL design                         &     		   & un-irradiated       &		   \\
FBK 111   & FBK IBL design                         &                      & un-irradiated       &                      \\
FBK 112   & FBK IBL design                         &                      & un-irradiated       &                      \\
\hline\hline
\end{tabular}
\label{table:duts}
\end{table*}

The performance of un-irradiated planar sensors is well understood, and therefore details of only two assemblies (PPS L1 and PPS 91) are shown as a reference for comparison to the irradiated behaviour. In particular, the PPS L1 module has been measured before and after irradiation. Changes in V$\rm{_{bk}}$ have not been systematically studied, but are thought to result from the release of thermal stresses during the bump-bonding process, or from environmental effects. Figure~\ref{fig:PPS91_1} illustrates the expected diode-like $\rm{I-V_{b}}$ dependence of PPS L1 with a plateau extending much beyond the working point of $\rm{V_{b}}= -80$~V followed by a (in this case) rather slow breakdown. The beam-spot of a collimated $^{90}$Sr  source is clearly visible within the hitmap of all events having a single-pixel cluster, for the PPS 91 module. 

\begin {figure}[ht!]
\centering
\subfigure[a][]{\includegraphics[width=0.46\linewidth] {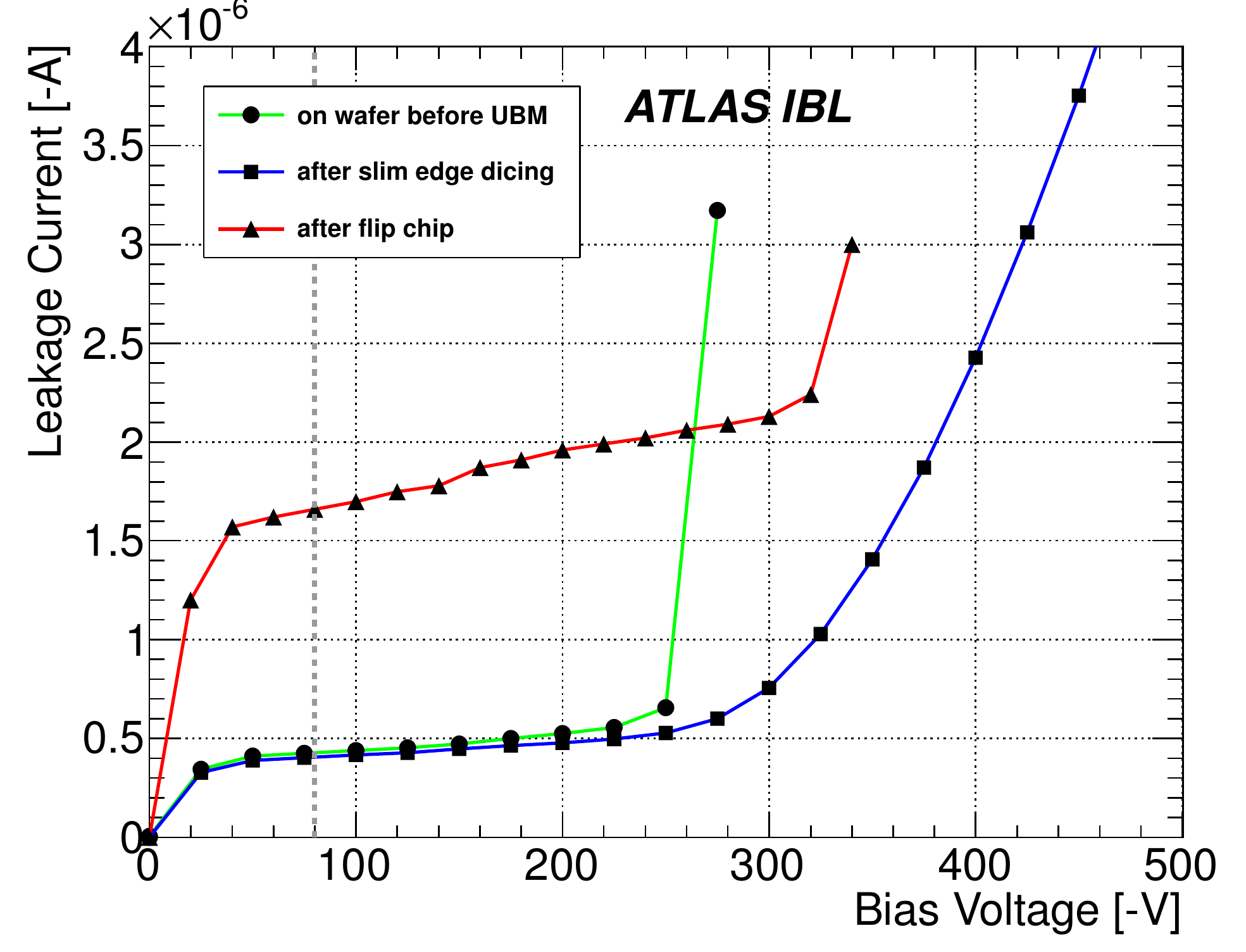}}
\subfigure[b][]{\includegraphics[width=0.53\linewidth] {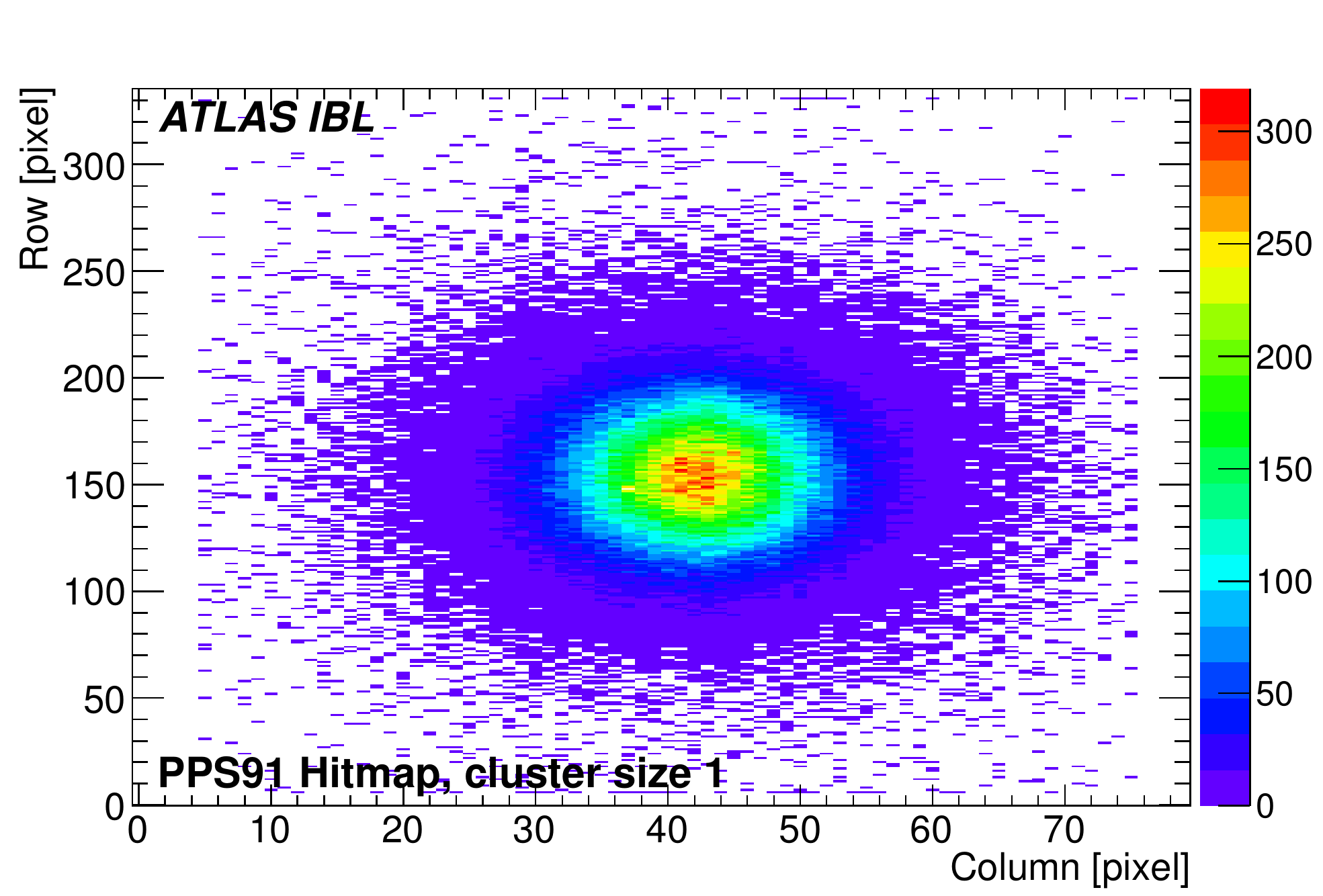}}
 \caption{(a) $\rm{I-V_{b}}$ measurement of the PPS L1 sensor and module successively before the UBM process, after the slim-edge dicing of the sensor, and after bump-bonding. The increased leakage current after bump-bonding can be attributed to the increased temperature, due to the FE-I4A IC. (b) Hit map of a strongly collimated $^{90}$Sr  source on the PPS 91 module. Only hits with a cluster size of 1 pixel were selected for the hit map to avoid hits with stronger electron scattering that would make the beam spot less clear.}
\label{fig:PPS91_1}
\end{figure}

Post-irradiation, the planar module performance remains satisfactory, with stable operation at a module temperature of  $-15^{\circ}$C. Following irradiation, no module breakdowns have been observed. $\rm{I-V_{b}}$ measurements are shown in figure~\ref{fig:Irr-IV} for the PPS~51 and PPS~61 planar modules (see table~\ref{table:duts}). Both show a dominantly ohmic behaviour as normally expected after heavy irradiation. In each case, the temperature is controlled using a Pt-1000 temperature sensor on the module. Taking into account the active area of approximately $3.44 $ cm$^2$, the power dissipation (leakage current) satisfied the specified value of 200 mW/cm$^2$ (200 $\mu$A/cm$^2$ at $\rm{V_{b}}= -1000$~V). Test-beam data for PPS61 are also shown in Section \ref{section:modules-test-beam}. 
\begin{figure}[!hbt]
\centering
 \includegraphics[width=.80\linewidth]{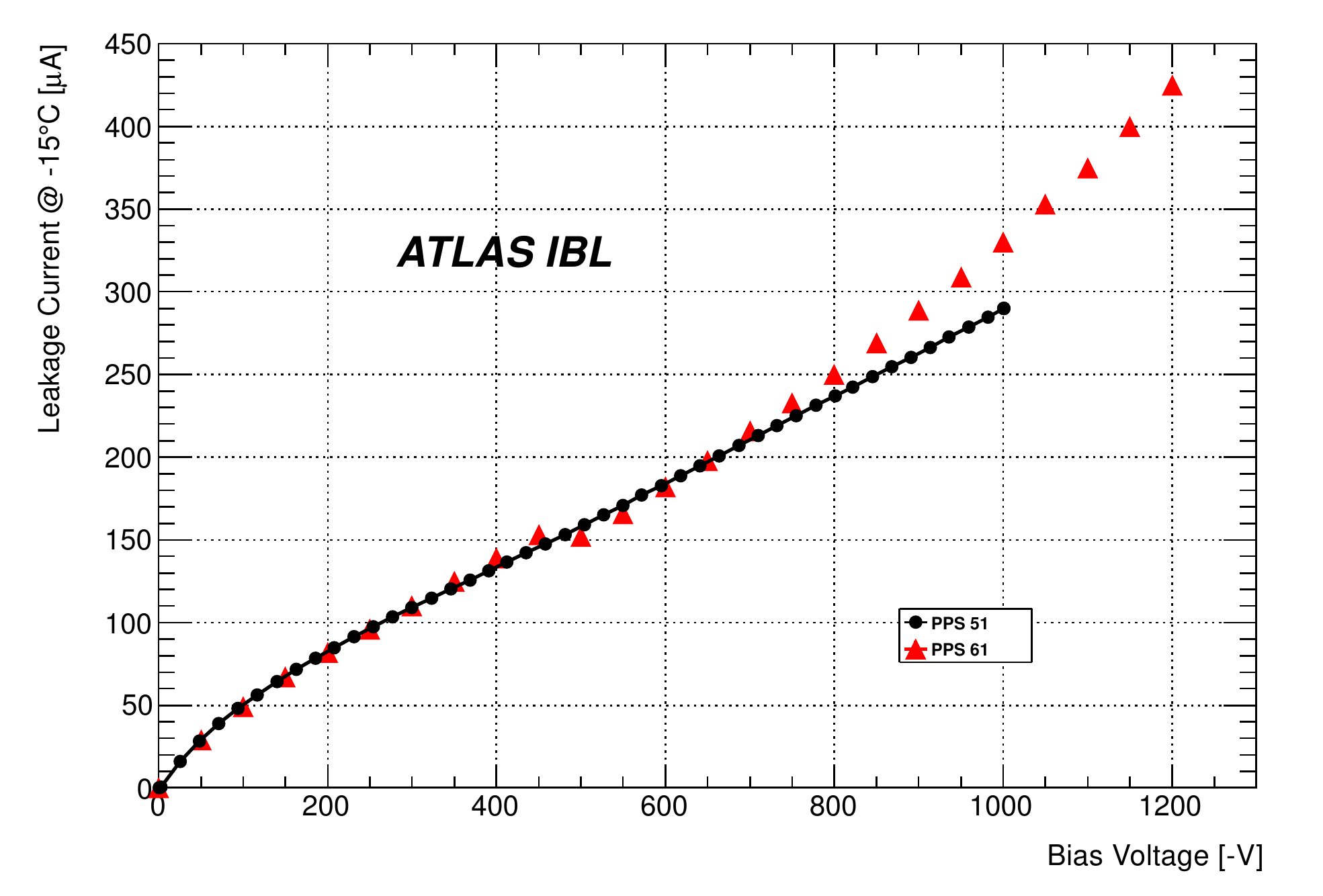} 
  \caption{$\rm{I-V_{b}}$ measurement after irradiation of the PPS 51 and PPS 61 modules measured at a sensor temperature of $-15^{\circ}$C.}
  \label{fig:Irr-IV}
\end{figure}

Stable threshold tunings as low as 1000 e$^{-}$ are possible with only a modest noise increase, even following heavy irradiation. While neutron-irradiated modules were in all cases well-behaved, proton-irradiated modules suffered from a significant fraction of digitally unresponsive pixels. This effect was also measured in test beam data. It was later found to result from the non-optimized value for an internal bias controlling the feedback of the second stage amplifier in the pixel. The leakage current on this high impedance node after irradiation was sufficient to exceed the second stage operating range in some pixels. This can be fully compensated by adjusting the bias (all biases are adjustable via the chip configuration), but this adjustment was not made for the results presented. Instead, pixels that were not responding were excluded from consideration. It was nevertheless later confirmed that proper bias adjustment fully recovers the excluded pixels.

Using the $^{90}$Sr  source, figure~\ref{fig:PPS91_PPSL4_TOT} shows the MPV of the ToT charge distribution measured as a function of V$\rm{_{b}}$  for the un-irradiated PPS 91 module, and for the irradiated PPS L4 module. The tuning of PPS 91  (PPS L4) was made at V$\rm{_{b}}$= $-80$ V (V$\rm{_{b}}$= $-500$ V), aiming for a mean 1600 e$^{-}$ threshold and a ToT measurement over 5 bunch crossings for a 10000 e$^{-}$ signal. While PPS 91 is expected to have full charge collection (the MPV dependence is small), that of PPS L4 suggests an increase of collected charge towards higher V$\rm{_{b}}$ as expected after irradiation from the increased operating voltage, and at the highest voltages from charge multiplication. The lack of a reliable ToT calibration for these modules precludes any measurement of the relative charge collection efficiency.

\begin{figure}[!hbt]
\centering
 \includegraphics[width=.80\linewidth]{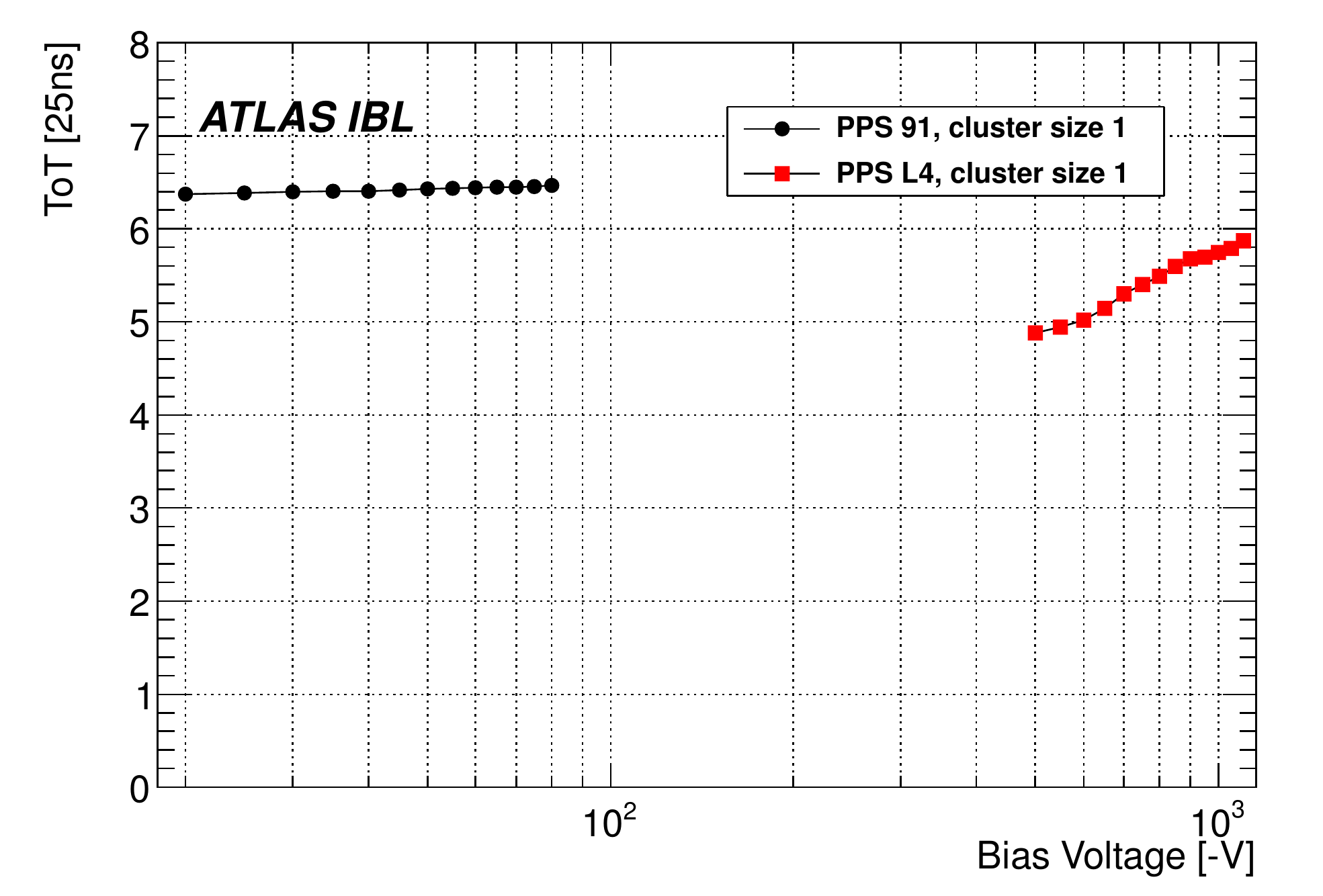} 
\caption{The MPV of the ToT spectrum for single-hit clusters measured as a function of the bias voltage using a $^{90}$Sr source, for the un-irradiated PPS 91 and irradiated PPS L4 modules. The tuning was made at V$\rm{_{b}}$= $-80$ V (respectively V$\rm{_{b}}$= $-500$ V) for a mean 1600 e$^{-}$ target threshold and a ToT measurement of 5 bunch crossings for a 10000 e$^{-}$ signal.}
\label{fig:PPS91_PPSL4_TOT}
\end{figure}

As discussed in Section {\ref{section:bare-test-results}}  the ENC depends on the operating point for the FE-I4A IC. Figure~\ref{fig:PPS91_PPSL4_ENC} shows the measured ENC as a function of V$\rm{_{b}}$ for the un-irradiated PPS 15 and PPS 91 modules, and for the irradiated PPS L4 module. The data are shown for several threshold settings, and IC working points. The $120-150$~ENC of PPS 91 is typical of un-irradiated modules. The ENC of PPS L4 is measured at several threshold values: typical noise values for irradiated PPS modules at a threshold of 1600 e$^{-}$ are in the range $150-200$~ENC.  

\begin{figure}[!hbt]
\centering
 \includegraphics[width=.80\linewidth]{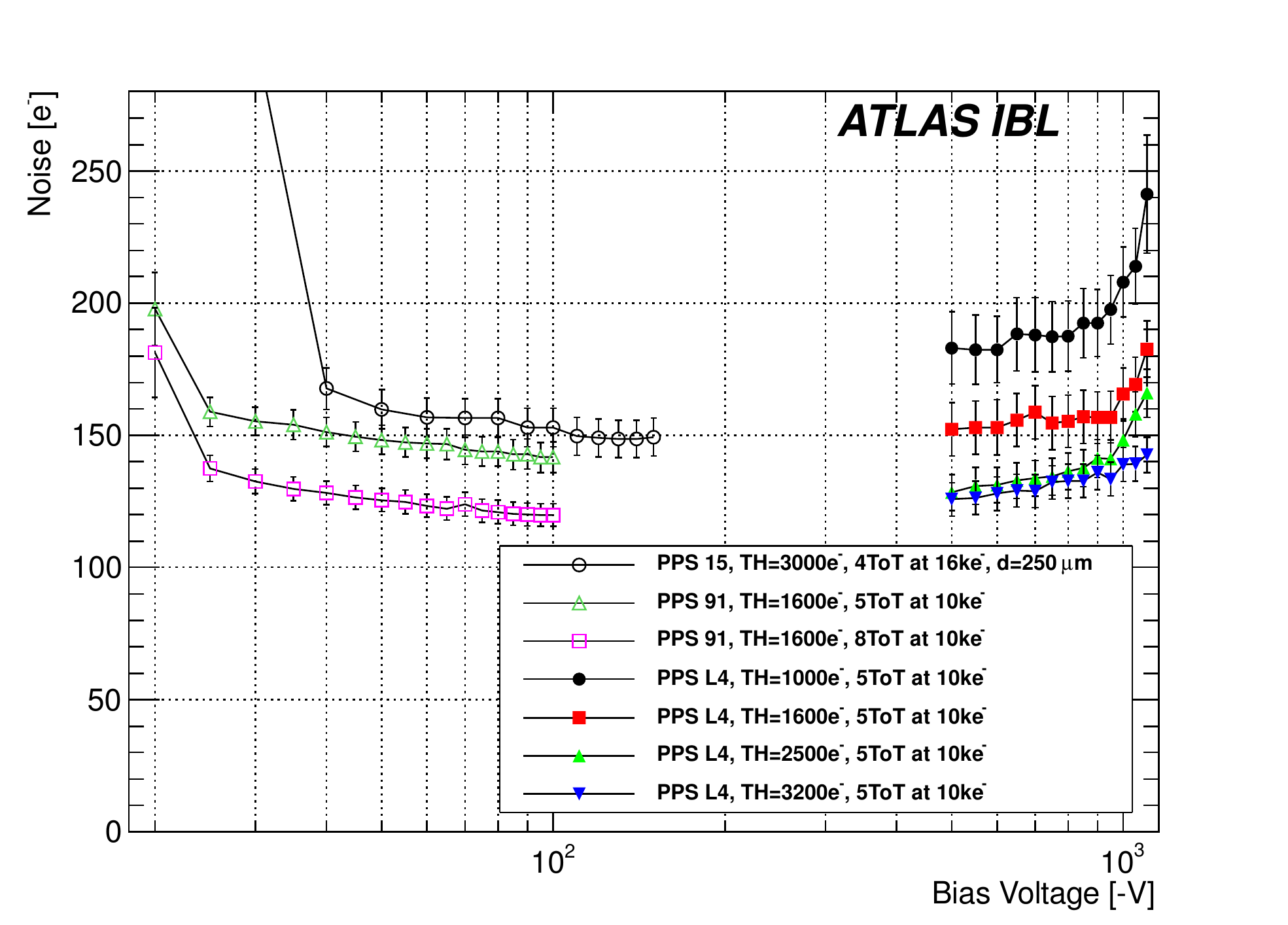} 
\caption{The ENC as a function of the bias voltage for the un-irradiated PPS 15 and PPS 91 modules, and for the irradiated PPS L4 module. The noise of the PPS 15 and PPS 91 modules are measured for different ToT tunings, while the noise of the PPS L4 module is measured at several threshold values. The PPS 15 sensor has a thickness of 250 $\rm{\mu m}$. }
\label{fig:PPS91_PPSL4_ENC}
\end{figure}

As an example of proton-irradiated assemblies, Figures~\ref{fig:SCC60-ThN} (a) and (b) show the threshold and ENC distribution  for each channel of the PPS 60 module, with a mean threshold tuned to 1000 e$^{-}$ and a ToT measurement of 5 bunch crossings for a 10000 e$^{-}$ signal. Figures~\ref{fig:SCC60-ThN} (c) and (d) show the projections of each measurement. The measured noise for these tuning parameters is 142 ENC. 

\begin {figure}[ht!]
\centering
\subfigure[a][]{\includegraphics[width=0.48\textwidth] {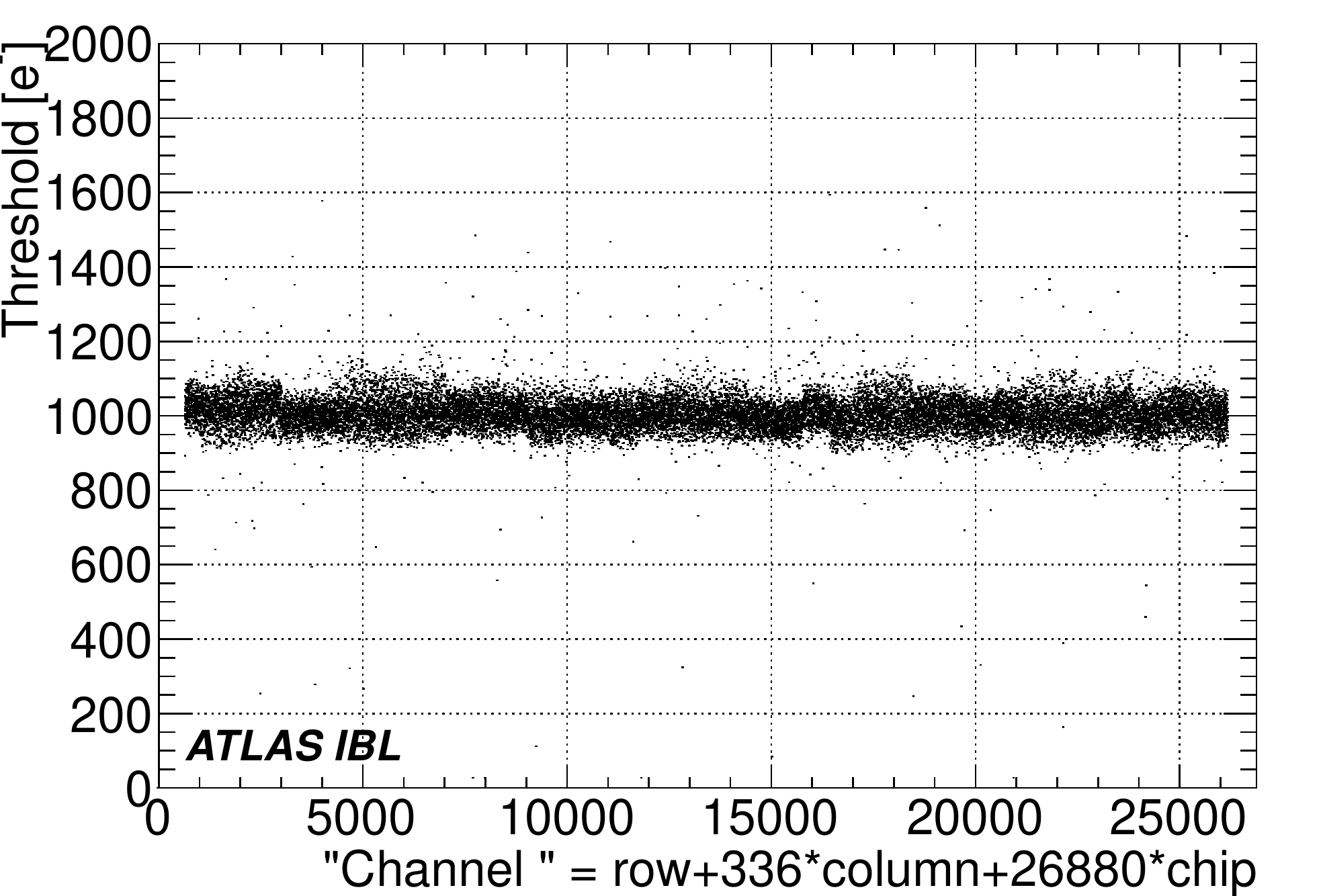}}
\subfigure[b][]{\includegraphics[width=0.48\textwidth] {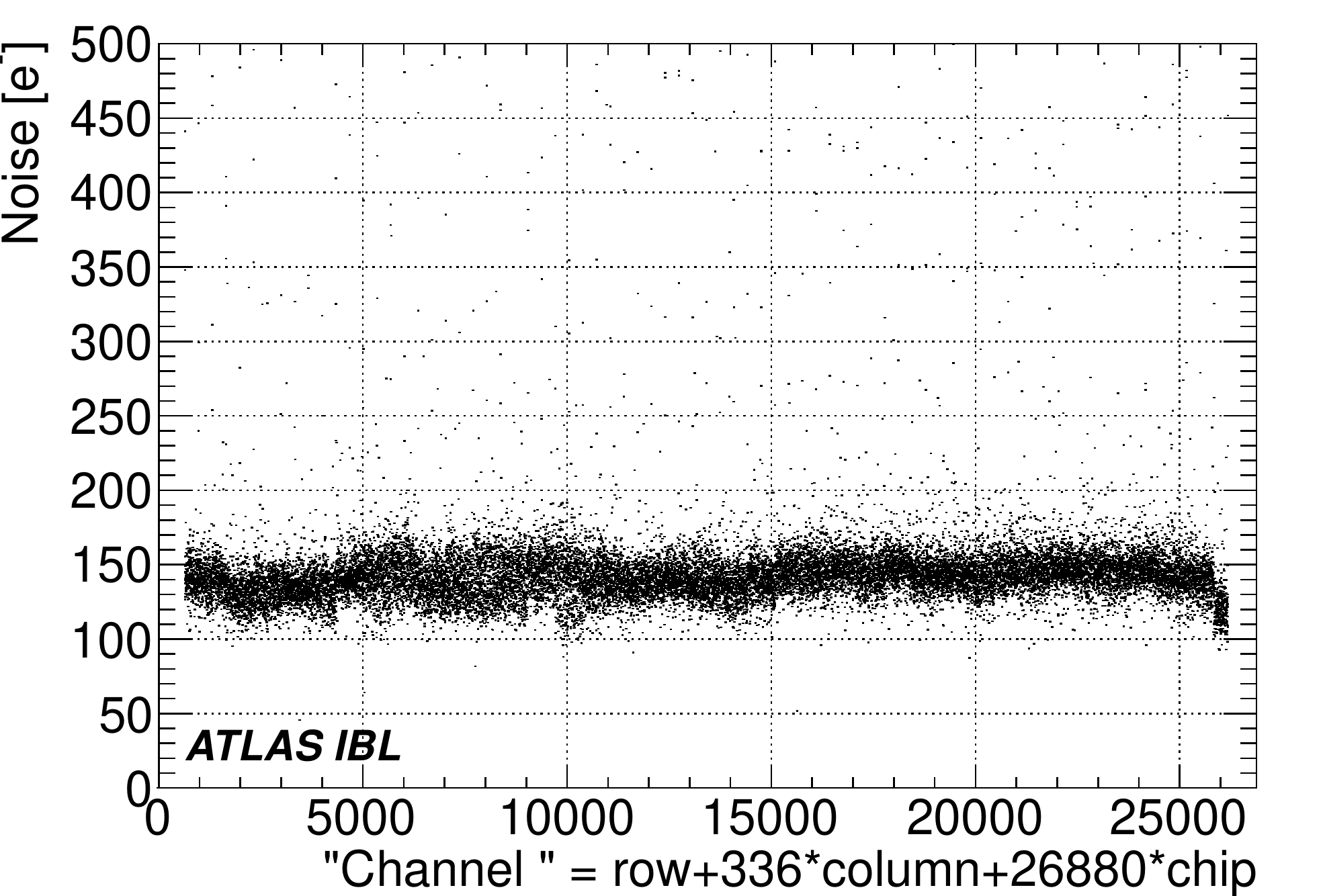}}
\subfigure[c][]{\includegraphics[width=0.48\textwidth] {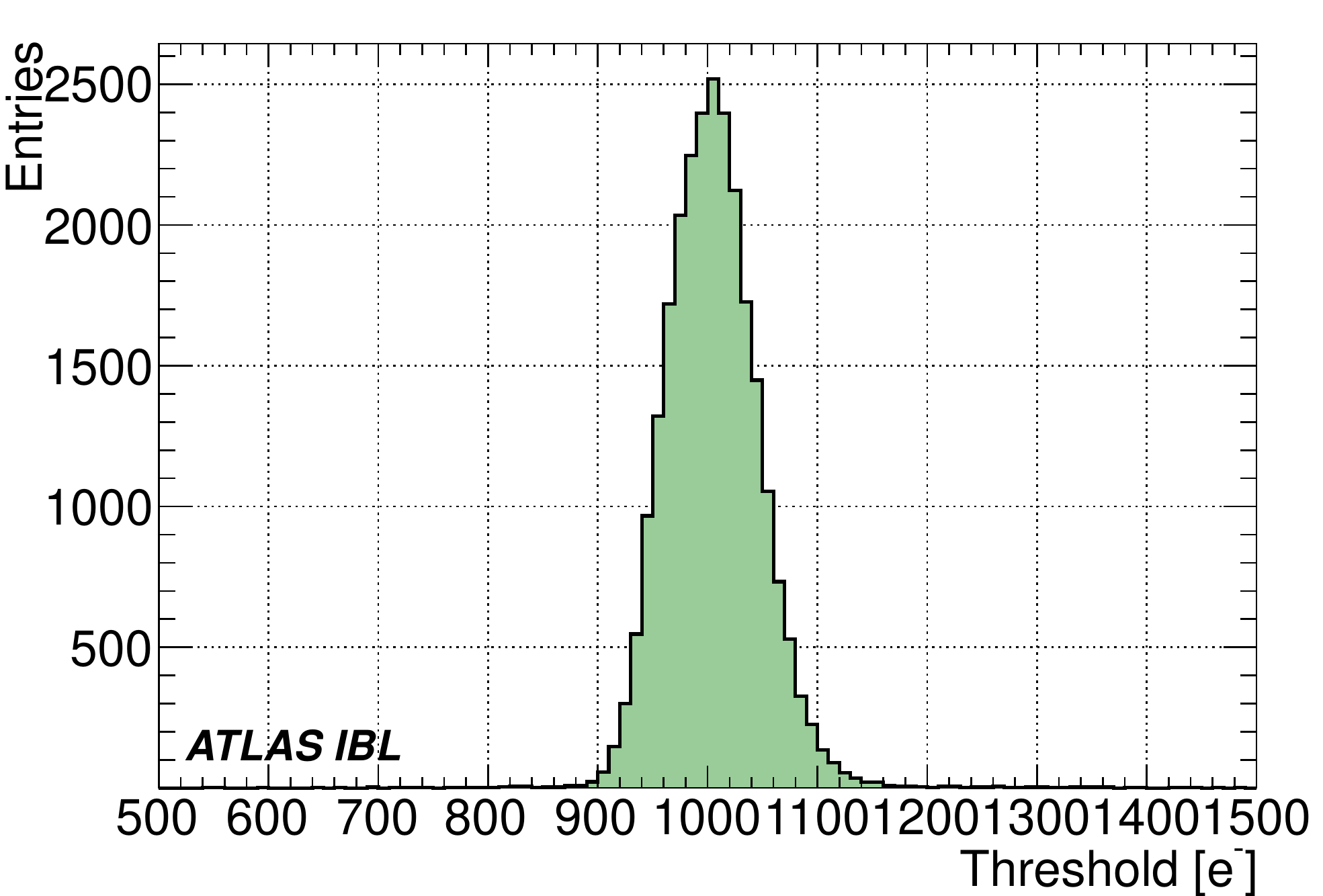}}
\subfigure[d][]{\includegraphics[width=0.48\textwidth] {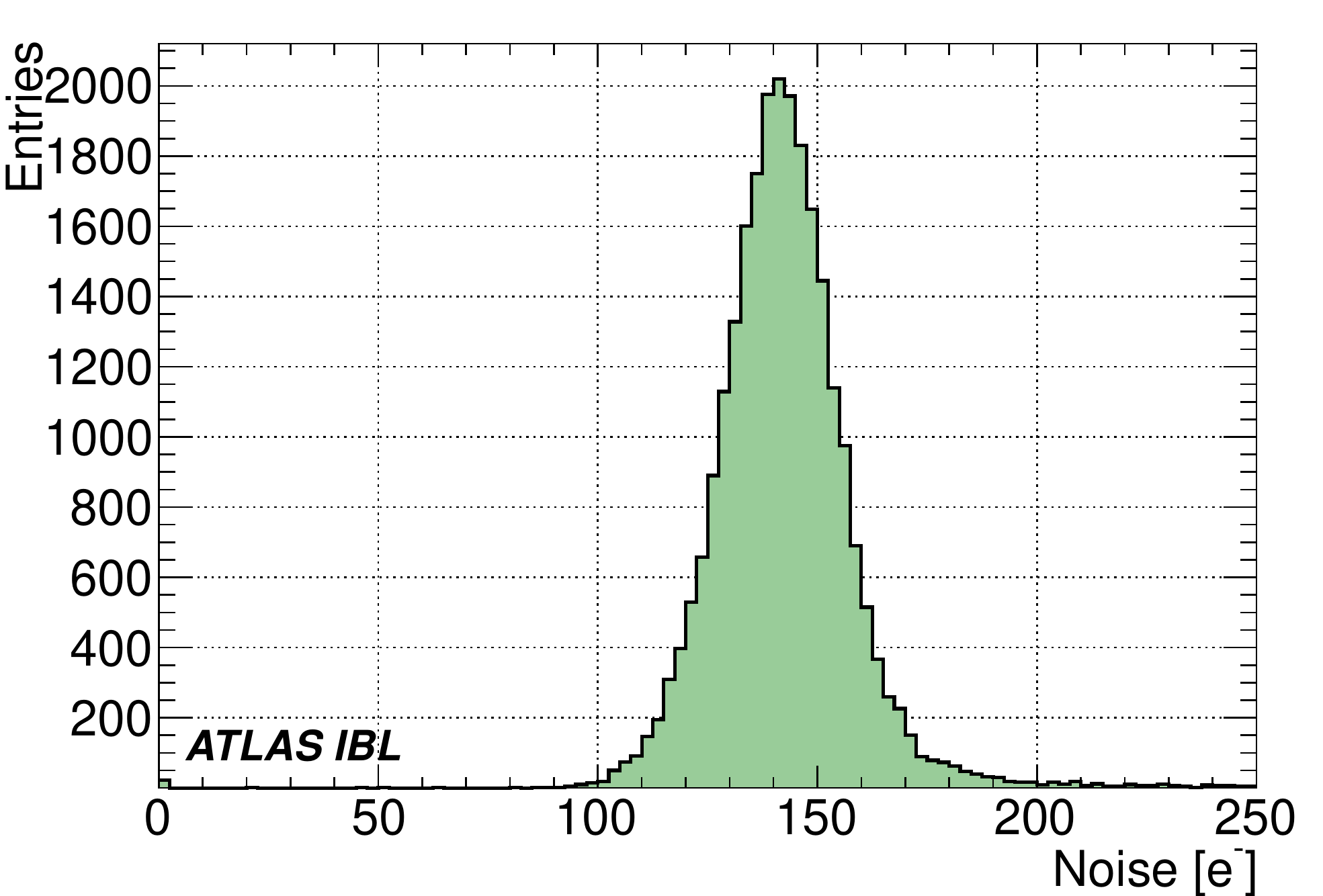}}
\caption{(a) Channel-by-channel threshold measurement for the PPS 60 module, with (c ) the histogram of all channels. (b) The measured noise distribution of each channel, for the threshold values quoted,  and (d) the histogram of the measured noise for all channels.}
\label{fig:SCC60-ThN}
\end{figure}

The qualification of 3D modules was made on 32 bump-bonded sensors from the CNM and FBK prototype batches. $\rm{I-V_{b}}$ measurements are shown for the full qualification set in figure~\ref{CV_figure_IV}, for modules using FBK (a) and CNM (b) sensors. Clearly visible are the break-down points which for most assemblies are $\rm{|V_{bk}|}>30$~V, reaching, in the case of CNM assemblies values greater than $\rm{|V_{bk}|}>100$~V. The lower values of V$\rm{_{bk}}$ for FBK sensors, as compared to CNM sensors,  results from the p-spray concentration that is optimized to yield much large V$\rm{_{bk}}$ values following irradiation. 

\begin {figure}[ht!]
\centering
\subfigure[a][]{\includegraphics[width=0.49\linewidth] {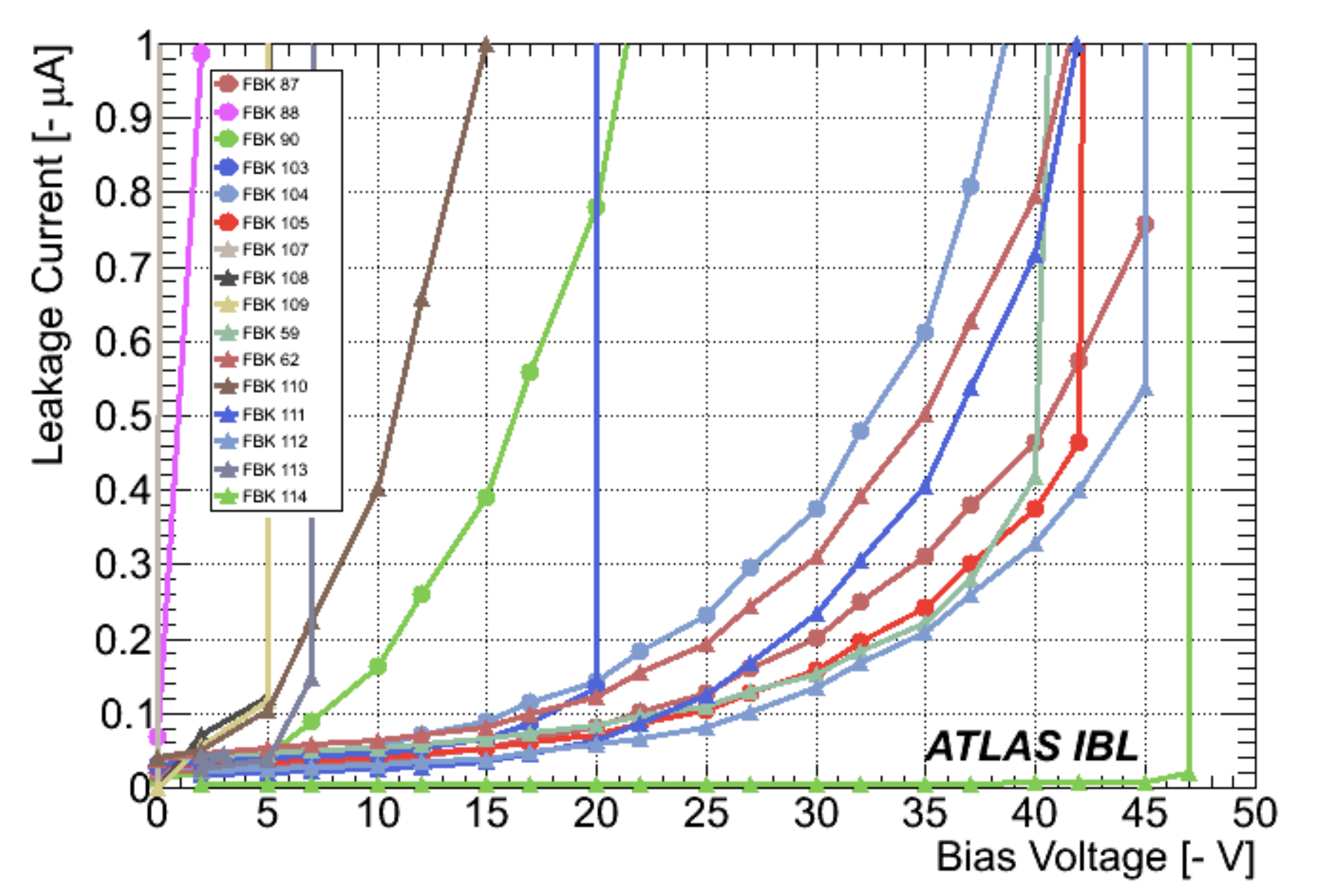}}
\subfigure[b][]{\includegraphics[width=0.49\linewidth] {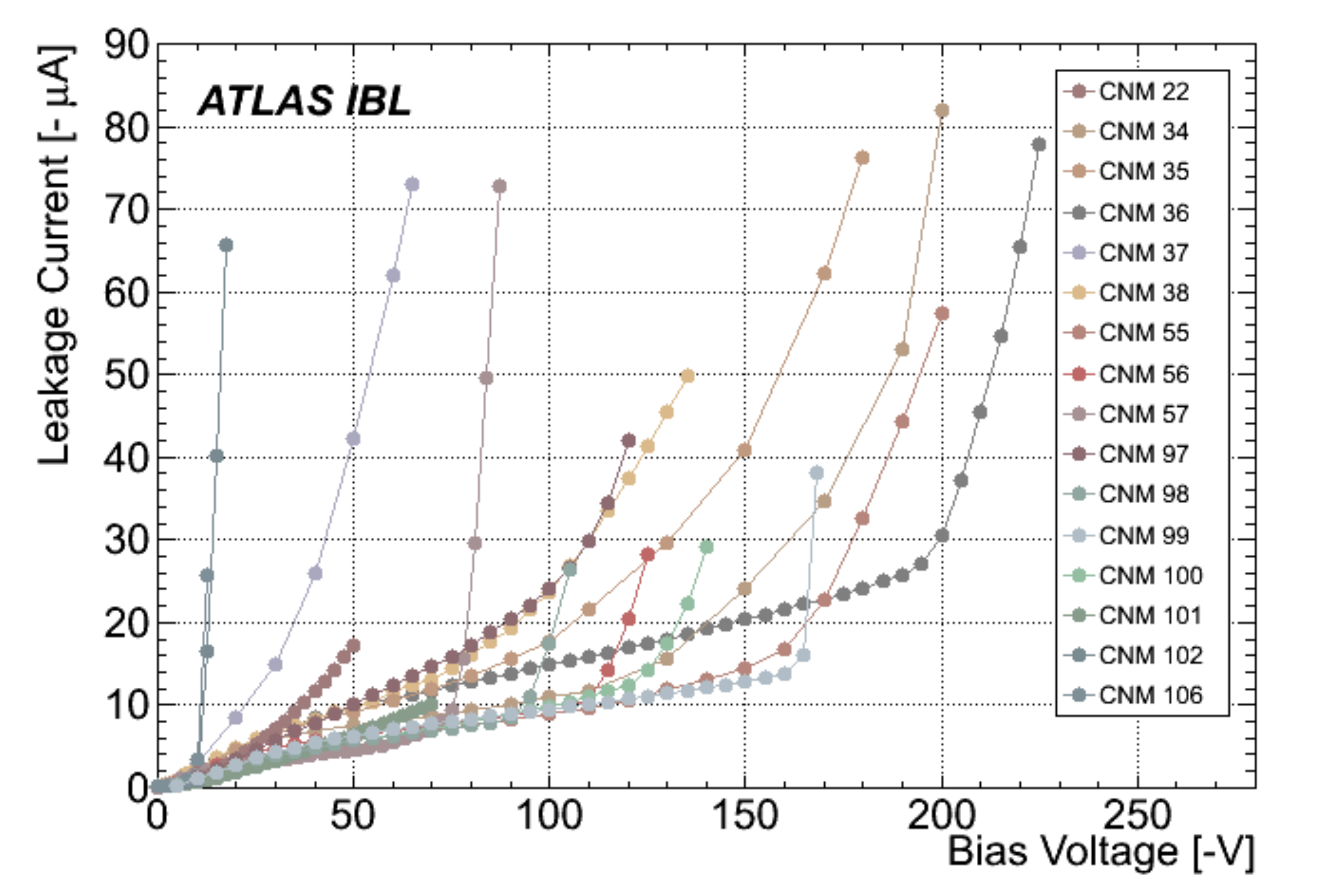}}
\caption{Measurement of the leakage current as a function of the bias voltage for modules using (a) FBK sensors and (b) CNM sensors. In each case the measurements are for a module temperature of $20^{\circ}$C.}
\label{CV_figure_IV}
\end{figure}

$\rm{I-V_{b}}$ measurements as a function of bias voltage following irradiation (5$\times10\mathrm{^{15}n_{eq}/cm^{2}}$) are shown in figure~\ref{CV_figure_IVtemp3D}  for FBK 87  (a) and CNM 36 (b) modules, measured at different sensor temperatures. These leakage currents grow with the bias voltage but there is no indication of breakdown or thermal runaway at the temperatures considered. 

\begin {figure}[ht!]
\centering
\subfigure[a][]{\includegraphics[width=0.49\textwidth] {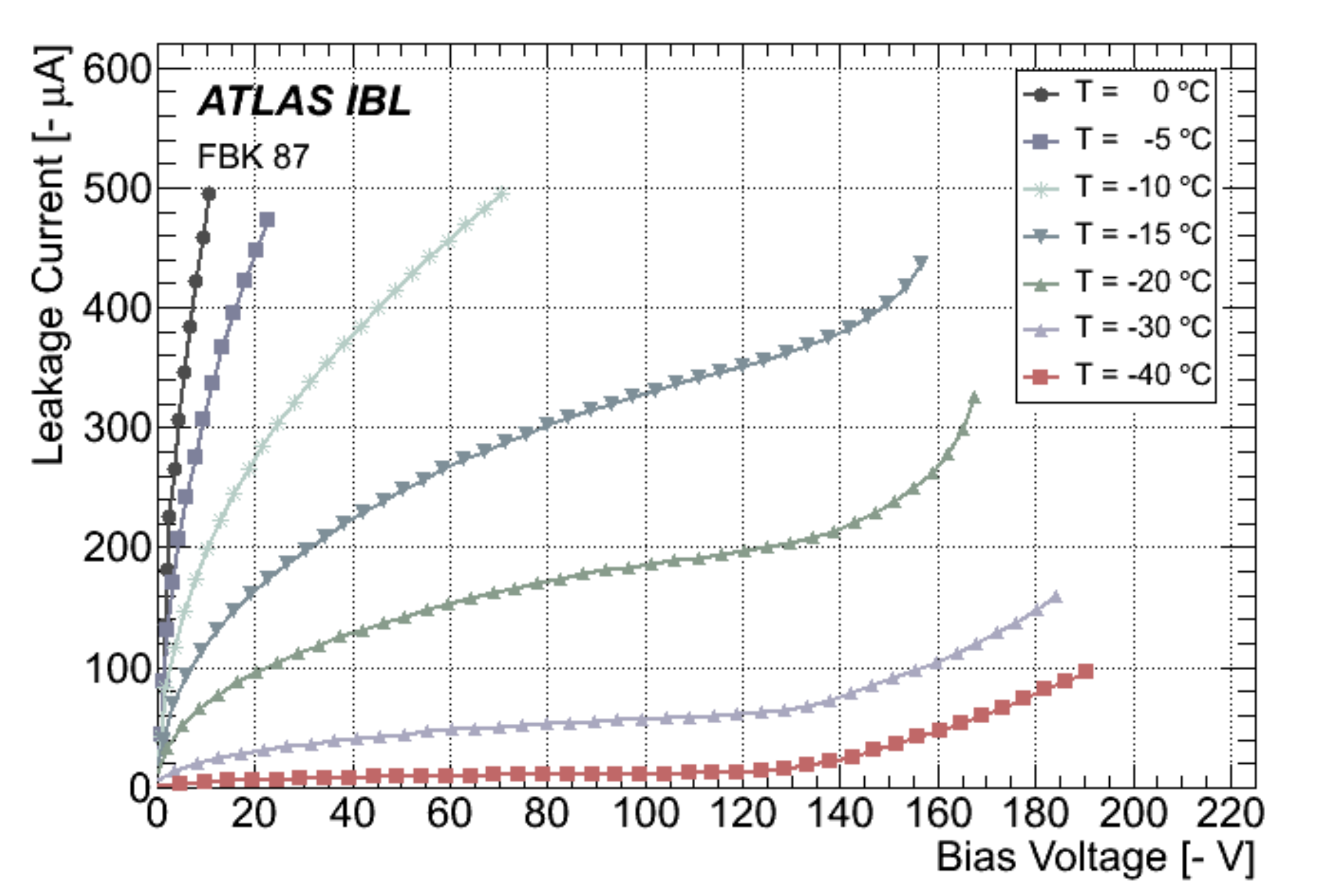}}
\subfigure[b][]{\includegraphics[width=0.49\textwidth] {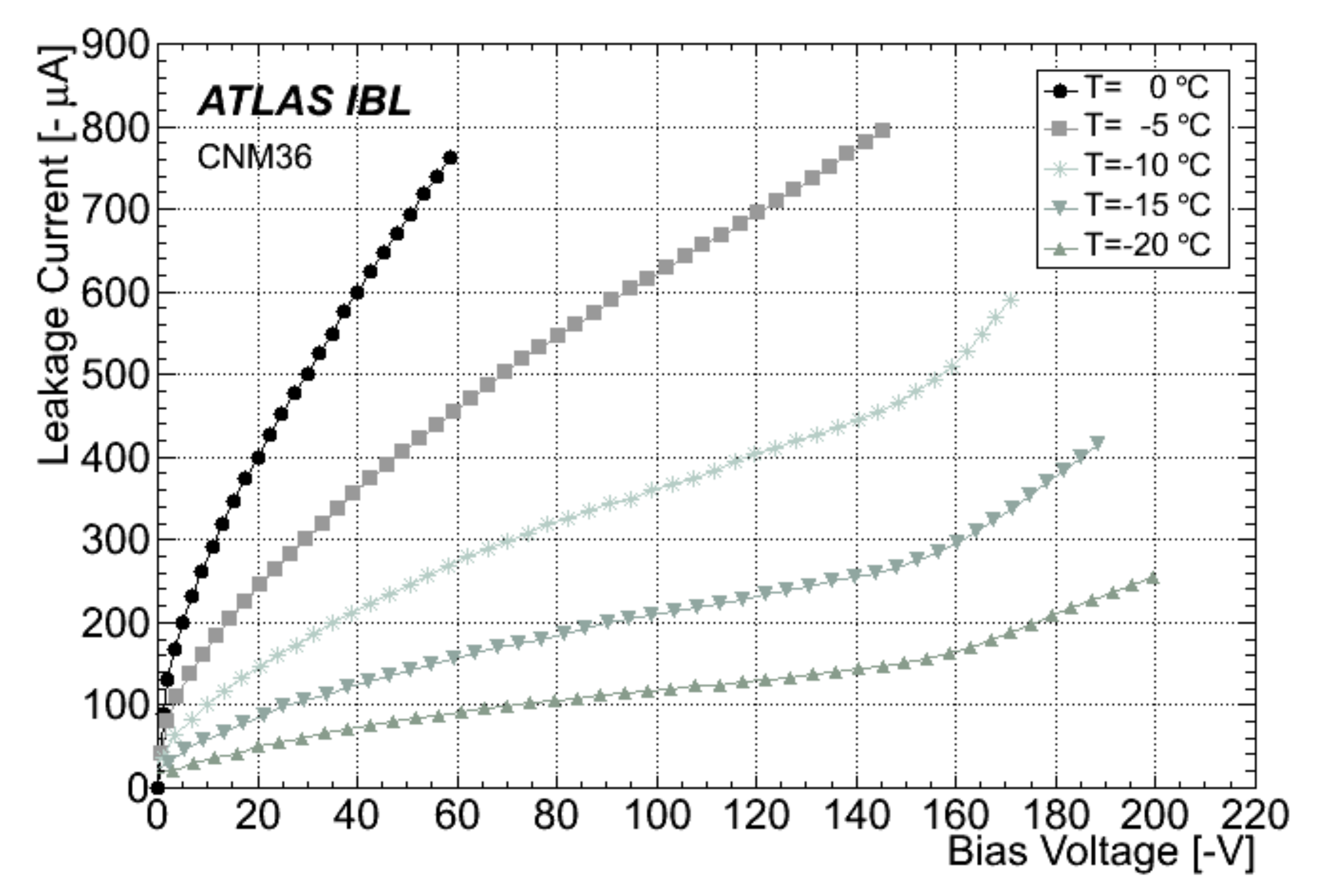}}
\caption{$\rm{I-V_{b}}$ measurements shown at different module temperatures,  after irradiation to a fluence of 5$\times10\mathrm{^{15}n_{eq}/cm^{2}}$, for (a) the FBK 87 module and (b) the CNM 36 module.}
\label{CV_figure_IVtemp3D}
\end{figure}

As with the planar modules, $^{90}$Sr radioactive sources have been used with an external electron trigger to test the measurement reproducibility of the 3D bump-bonded assemblies. Figure~\ref{CV_figure_TOT3D} shows the measured ToT signal as a function of the bias voltage using selected FBK and CNM modules, before and after irradiation. From such measurements it can be seen that full depletion and consequently maximum signal before irradiation is reached at about 10~V for the FBK module, and at about 20~V for the CNM module. Because of the lack of calibration, no quantitative comparisons of the charge collection efficiency are possible. Variations of the MPV with bias voltage for a given detector are dependent on the charge collection efficiency, but the data are less well controlled than test beam measurements. Nevertheless, operation at $\rm{V_{b}}\gtrsim 160$~V indicates an acceptable charge collection efficiency following irradiation.

Figure~\ref{CV_figure_noise3D} shows the measured ENC as a function of bias voltage for the FBK 104 and CNM 101 modules (not irradiated) and the CNM 34 and CNM 97 modules, irradiated using protons to a level of 5$\times$10$^{15}$ n$_{eq}$/cm$^{2}$. For bias voltages of $\rm{V_{b}}\gtrsim 80$~V, the ENC remains stable. The measured ENC before irradiation does not depend significantly on the choice of FBK or CNM modules and test beam data show that to be the case after irradiation. The ENC of 3D modules and of planar modules is not significantly different.  These data were collected using a threshold of 3200 e$^{-}$ and a ToT measurement of either 5 or 8 bunch crossings for a 20000 e$^{-}$ signal.  

\begin{figure}[ht!] 
\centering
\includegraphics[width=0.70\textwidth] {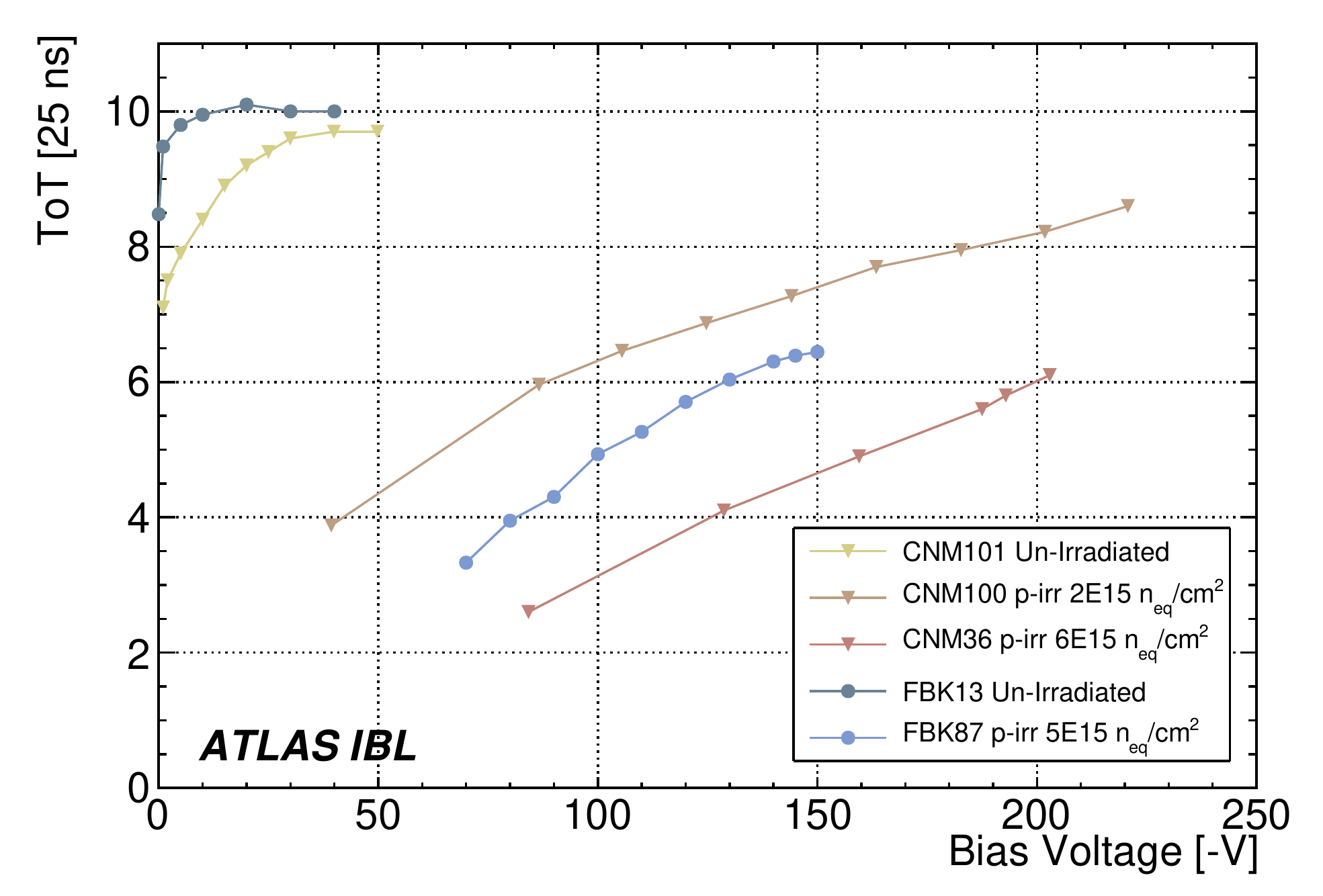} 
  \caption{ The MPV of the ToT spectrum measured for selected FBK and CNM 3D modules after bump-bonding, using a $^{90}$Sr source. The data are shown in units of the 25 ns bunch crossing clock as a function of V$\rm{_{b}}$. Irradiated modules at several fluences are compared with non-irradiated samples. }
  \label{CV_figure_TOT3D}
\end{figure} 

\begin{figure}[!htb]
\centering
\includegraphics[width=0.70\textwidth]{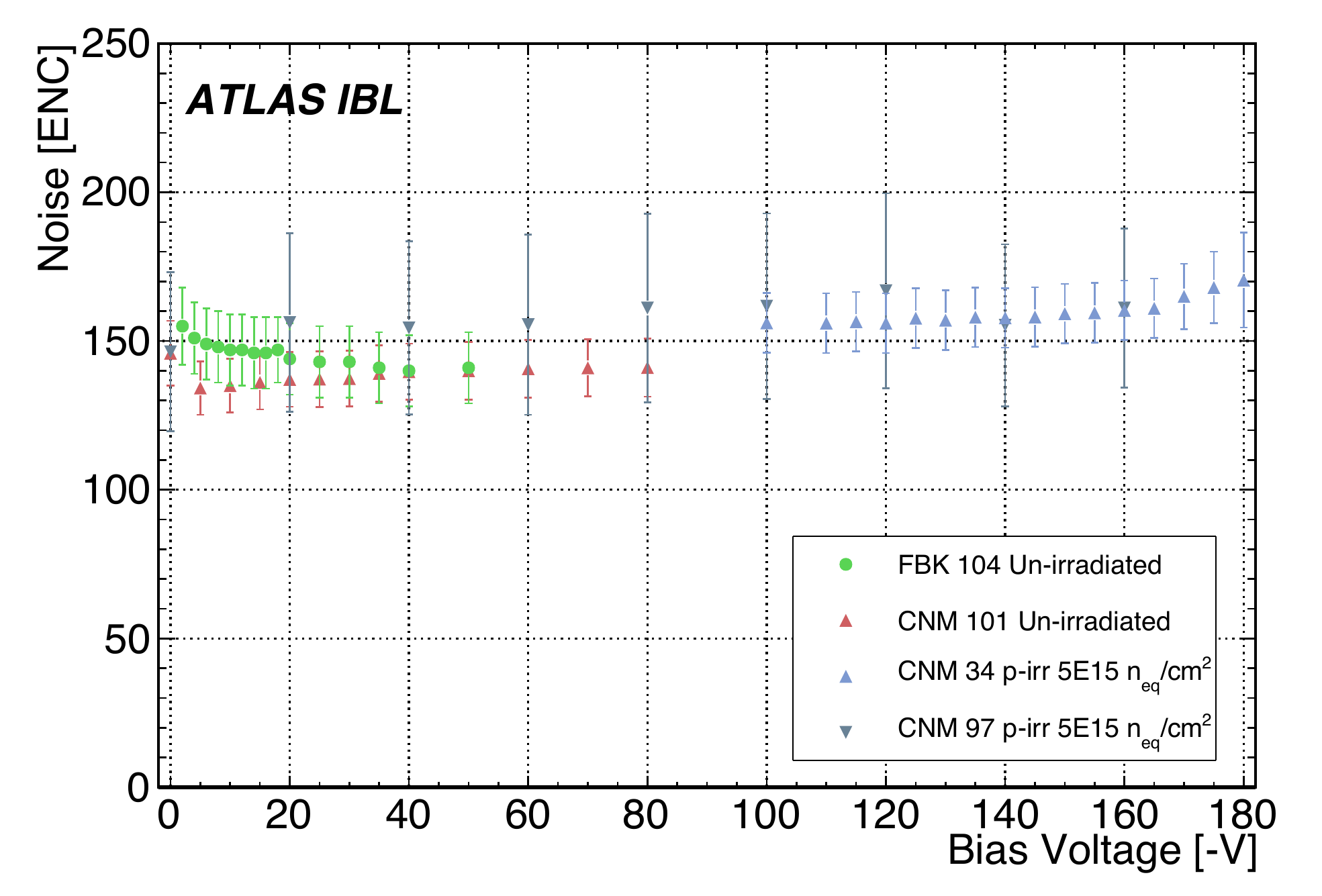}
\caption{The ENC of FBK and CNM modules, measured as a function of the bias voltage before and after irradiation. The measurements use a threshold setting of 3200 e$^{-}$ and a ToT measurement of either 5 or 8 bunch crossings for a 20000 e$^{-}$ signal. The operating temperature of the modules is $-15^{\circ}\ \mathrm{C}$.}
\label{CV_figure_noise3D}
\end{figure}
      
\section{Test Beam Measurements of the IBL Module} {\label{section:IBL-module-test-beam}}

Detailed test beam studies of both non-irradiated and irradiated modules are essential to understand the module performance. 
These studies allow a comparison with bench-top ENC and ToT data, and as well allow both an understanding and an optimization of the module operating parameters. Several IBL prototype modules have been characterized using 4~GeV positrons at a DESY beam line in April 2011, and 120~GeV pions at the CERN SPS H6 and H8 beam lines in respectively June and September 2011. More extensive test beam campaigns are underway. 

\subsection{Test beam setup} {\label{section:test-beam-setup}}

Three modules, together with a non-irradiated FE-I4A reference module, were normally included in the test beam at any given time. For part of the run period, measurements were possible with the modules mounted within the bore of the 1.6 T superconducting Morpurgo dipole magnet \cite{Morpurgo:2009aa}, simulating the ATLAS solenoid field direction. However, because of limited beam time, only a few measurements were made with the magnet powered. Most of the measurements described use perpendicular incident tracks, but some measurements were also made using non-perpendicular incidence tracks to replicate features of the IBL geometry. 

The modules under test were mounted normally with respect to the incoming beam using mechanical holders so that the long pixel direction (corresponding to the z direction in the IBL) was horizontal. Small, well-defined tilt angles around that horizontal axis, referred to as tilts in the $\phi$ direction (see figure~\ref{IBL_layout_A19}), were achieved by mounting the modules on wedges machined to the desired angle. Rotations around the vertical axis, corresponding to different pseudo-rapidity values ($\rm{\eta}$), were made using specially designed spacers allowing rotations equivalent to tracks in the pseudo-rapidity range $0.88\leq\eta\leq 4.74$ (figure~\ref{fig:High_Eta_Photo}).

\begin {figure}[ht!]
\centering
\subfigure[a][]{\includegraphics[width=0.70\textwidth] {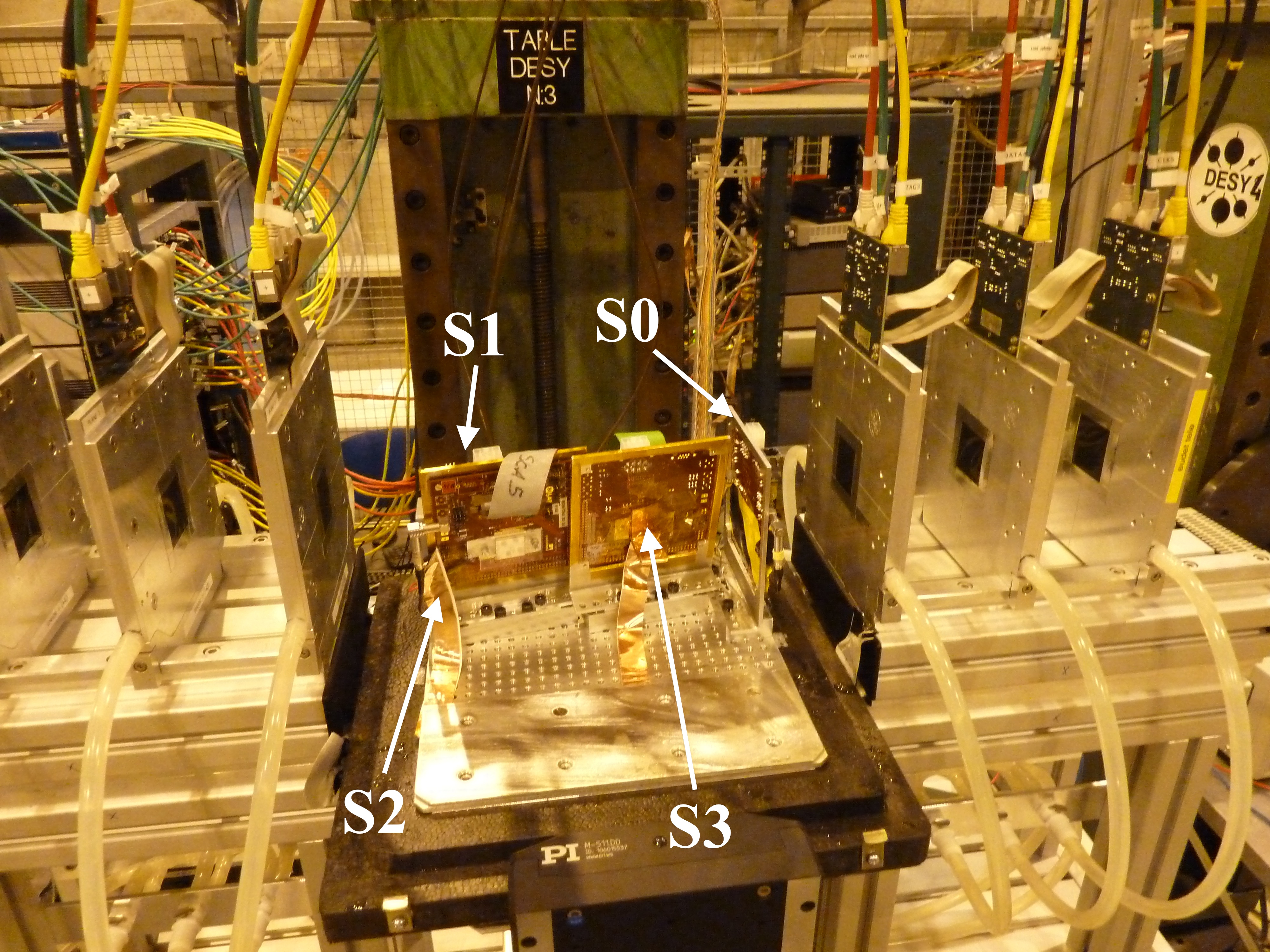}}
\subfigure[b][]{\includegraphics[width=0.70\textwidth] {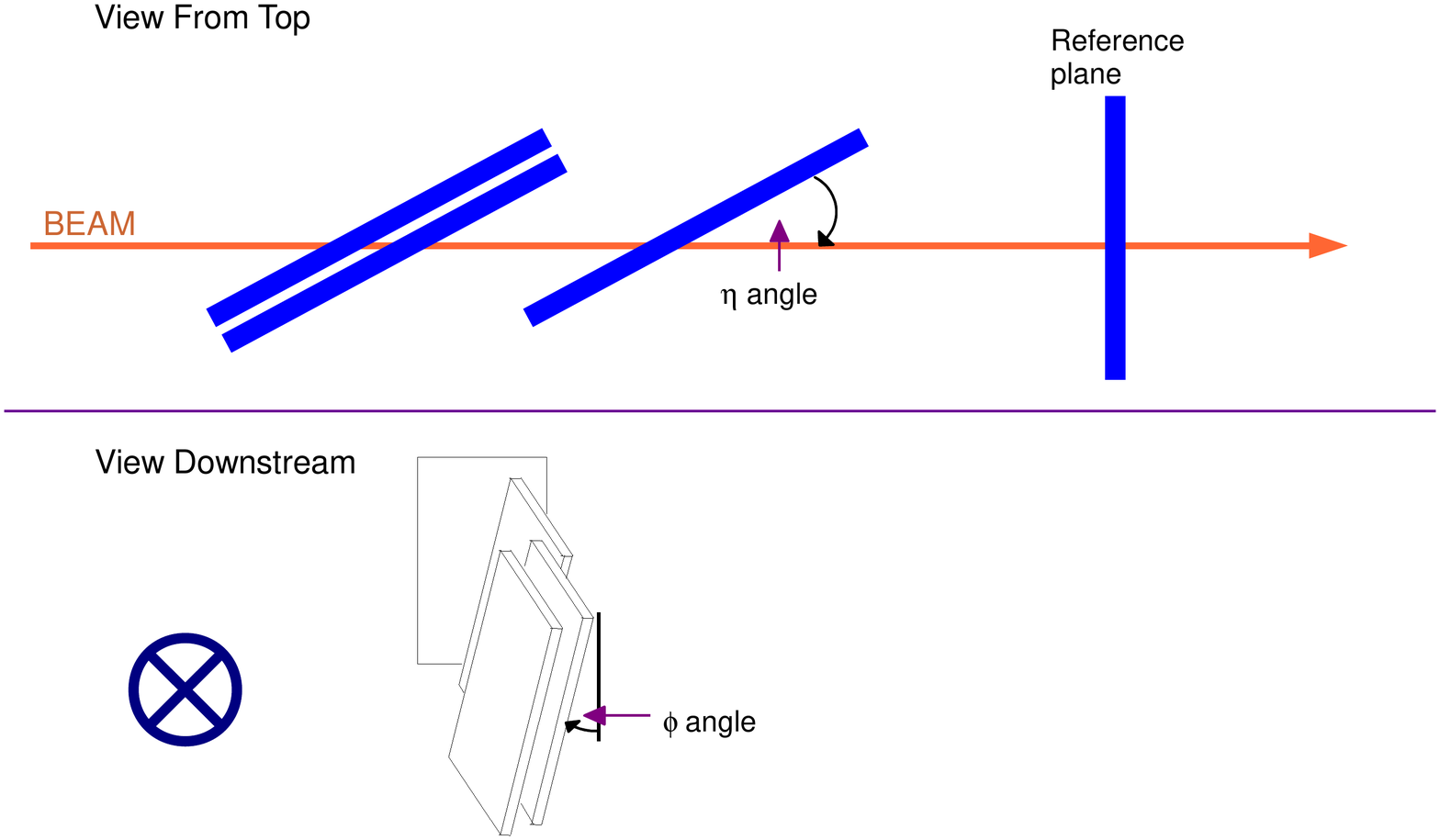}}
	\caption{(a) Photograph of the high-$\rm{\eta}$ setup. The two arms of the EUDET/AIDA test beam spectrometer each with three Mimosa26 sensors are shown on each side of the photograph. Between the two arms, jigs support four test samples, shown as S0 - S3.  (b) Schematic of the high-$\rm{\eta}$ set-up.} 
	\label{fig:High_Eta_Photo}
\end{figure}

The three test modules were placed in an insulated thermally controlled box together with the reference plane. For some measurements, the modules were also mounted with the long pixel direction vertical, such that the magnetic field of the dipole pointed in the same direction as in the IBL (see figure~\ref{fig:High_Eta_Photo}). To test the modules under IBL operating conditions, they were cooled to a sensor temperature of approximately $-15^{\circ}\ \mathrm{C}$. 

Beam particle trajectories were reconstructed using the high resolution EUDET telescope \cite{Bulgheroni:2010zz,EUDET2:2011zz}, consisting of six planes instrumented with Mimosa26 active pixel sensors with a pitch of 18.4~$\mu$m. Each plane consists of 1152~$\times$~576 pixels covering an active area of 21.2~$\times$~10.6~mm$^2$. A coincidence of four scintillators was employed for triggering resulting in an effective sensitive area of 2~$\times$~1~cm$^2$. The track position uncertainty when interpolated to the test modules was estimated to be $\sim$~2~$\mu$m. 

The Trigger Logic Unit \cite{Cussans:2007iq} issues a trigger signal for the Mimosa26 and FEI4 readout systems upon a coincidence of the four scintillators once both readout systems are free to accept a trigger. The Mimosa26 VME-based readout employs a continuous rolling shutter and is at the same time trigger-based. This means that the system is configured to drop frames if there is no explicit trigger received to send the frame. The FEI4 readout based on USBPIX system is configured to be sensitive to only first 400 ns upon trigger arrival, while Mimosa26 frame integration time is 115.2~$\mu$s. Tracks passing through the telescope planes during the sensitive time of the FEI4 modules (in-time tracks) were selected by requiring at least one hit in another module, known as the reference plane.

The telescope planes are read by a custom-made VME system controlled by a single board PC for each telescope arm (see figure~\ref{fig:High_Eta_Photo}). Each of the PCs sends a separate ethernet data stream to a run control PC. The test modules are read using the USBpix system (see Section \ref{section:USBpix-test-system}) connected  to the EUDET telescope trigger interface \cite{Cussans:2007iq}.

\subsection{IBL modules measured in the test beam}{\label{section:modules-test-beam}}

Several modules of the PPS slim edge design but with different sensor thickness were studied in the test beam. In addition modules using both FBK and CNM 3D sensors were studied. The modules used, as well as their operating conditions, are indicated in table~\ref{tab:efficiency}. 

\begin{table*}[ht!]
\centering
\caption{Tracking efficiency for CERN beam test samples. The magnetic field was 1.6 T with the field aligned as for the ATLAS solenoid. The efficiency measurement for the FBK 87 module was made at the DESY beam in April 2012, all other measurements were made at the CERN test beams.}
\medskip
\label{tab:eff}
\footnotesize
\begin{tabular}{|l|c|c|c|c|c|c|c|c|}
\hline\hline
Sample & Bias  		& Magnetic 					& Tilt Angle		& 	& Dose ($\rm{10^{15}}$)  				& Hit  & 
ToT/ Signal & Threshold \\
 ID                   & (V)  & Field \cite{Morpurgo:2009aa} 	& $\rm{\phi}$ ($\rm{^o}$)  & 			& $\rm{n_{eq}/cm^2}$ 	&Efficiency ($\%$) 		& (ke) Tuning   	       	& $\rm{e^{-}}$ \\
\hline\hline
PPS 40    &  -150  & on    & 0          & un-irr	& -	& 99.9                  & 5.8/10  & 2700	 \\
              &  -150  & off    & 0          & un-irr	& -	& 99.9                  & 5.8/10  & 2700          \\
PPS 60    &  -940  & off   & 15         & p-irr  	& 5  	& 97.7                  & 8/10  & 1600       \\
PPS 61    & -1000 & off   & 15         & p-irr  	& 5  	& 96.8                  & 5/20  & 1400       \\
       	     &  -800  & off   & 15         & p-irr	& 5  	& 93.9                  & 5/20  & 1600       \\
              &  -600  & off   & 15         & p-irr	& 5  	& 86.4                  & 5/20  & 1600       \\
PPS L2    & -1000  & off  & 15         & n-irr	& 4  	& 98.3                  & 5/20  & 1600       \\
              &  -800  & off   & 15         & n-irr	& 4  	& 98.3                  & 5/20  & 1600       \\
              &  -600  & off   & 15         & n-irr	& 4  	& 97.2                  & 5/20  & 1600       \\
              &  -400  & off   & 15         & n-irr	& 4  	& 95.3                  & 5/20  & 1600       \\
PPS L4    & -1000  & off  & 0          & n-irr	& 5  	& 97.9                  & 8/10  & 1600       \\
              &  -600  & off   & 0          & n-irr	& 5  	& 97.2                  & 8/10  & 1600       \\

CNM 34 &  -140  & on    & 0          & p-irr	& 5  	& 95.4                  & 10/20  & 3200       \\
             &  -140  & off    & 0          & p-irr	& 5  	& 95.7                  & 10/20  & 3200       \\
	    &  -160  & off   & 0           & p-irr	& 5  	& 94.5                  & 10/20  & 3200       \\
	    &  -180  & off   & 0		& p-irr	& 5  	& 98.4                  & 10/20  & 1500       \\
	    &  -160  & off   & 0		& p-irr	& 5	& 98.1                  & 10/20  & 1500       \\
	    &  -140  & off   & 0		& p-irr	& 5	& 97.6                  & 10/20  & 1500       \\
	    &  -120  & off   & 0		& p-irr	& 5	& 97.2                  & 10/20  & 1500       \\
	    &  -100  & off   & 0		& p-irr	& 5	& 96.6                  & 10/20  & 1500       \\
	    &  -160  & off   & 15         & p-irr	& 5	& 99.0                  & 10/20  & 1500       \\
CNM 55 	&   -20   & on    & 0     & un-irr	& -  	& 99.5                  & 10/20  & 1600       \\
            	&   -20   & off    & 0      & un-irr	& -  	& 99.6                  & 10/20  & 1600       \\
CNM 81   &  -160  & off   & 0       & n-irr	& 5	& 97.5                  & 10/20  & 1500       \\
CNM 97   &  -140  & off   & 15     & p-irr	& 5	& 96.6                  & 10/20  & 1800       \\
FBK 13    &   -20    & off   &  0     & un-ir	& -	& 98.8                  & 10/20  & 1500       \\
FBK 87    &  -140   & off   & 15    & p-irr	& 5	& 95.6                  & 10/20  & 2000       \\
              &  -160  & off   & 15     & p-irr	& 5	& 98.2                  & 10/20  & 1500       \\
FBK 90    &  -60   & off   & 15    & p-irr	 	& 2	& 99.2                  & 8/10  & 3200      \\
\hline\hline
\end{tabular}
\label{tab:efficiency}
\end{table*}

\subsubsection{Charge collection}{\label{section:charge-collection}}

The FE-I4A front-end IC provides a  4-bit digital ToT measurement in bins of the $25 \ \mathrm{ns}$ LHC bunch crossing rate. The raw ToT distributions of four representative samples for a $15^{\circ}$ incident azimuthal beam angle are shown on figure~\ref{fig:chargecollection}. In the absence of a ToT-to-charge calibration for each module at its operating point, the distributions cannot be directly compared. Also, different operating conditions prevent a quantitative comparison with equivalent source measurements, but the data are compatible. For some measurements in the June test beam, including those shown, irradiated CNM and FBK modules were operated at $\rm{V_{b}} = -140$~ V, below the requirement for full charge collection.

\begin{figure}
\begin{center}
\includegraphics[scale=0.350]{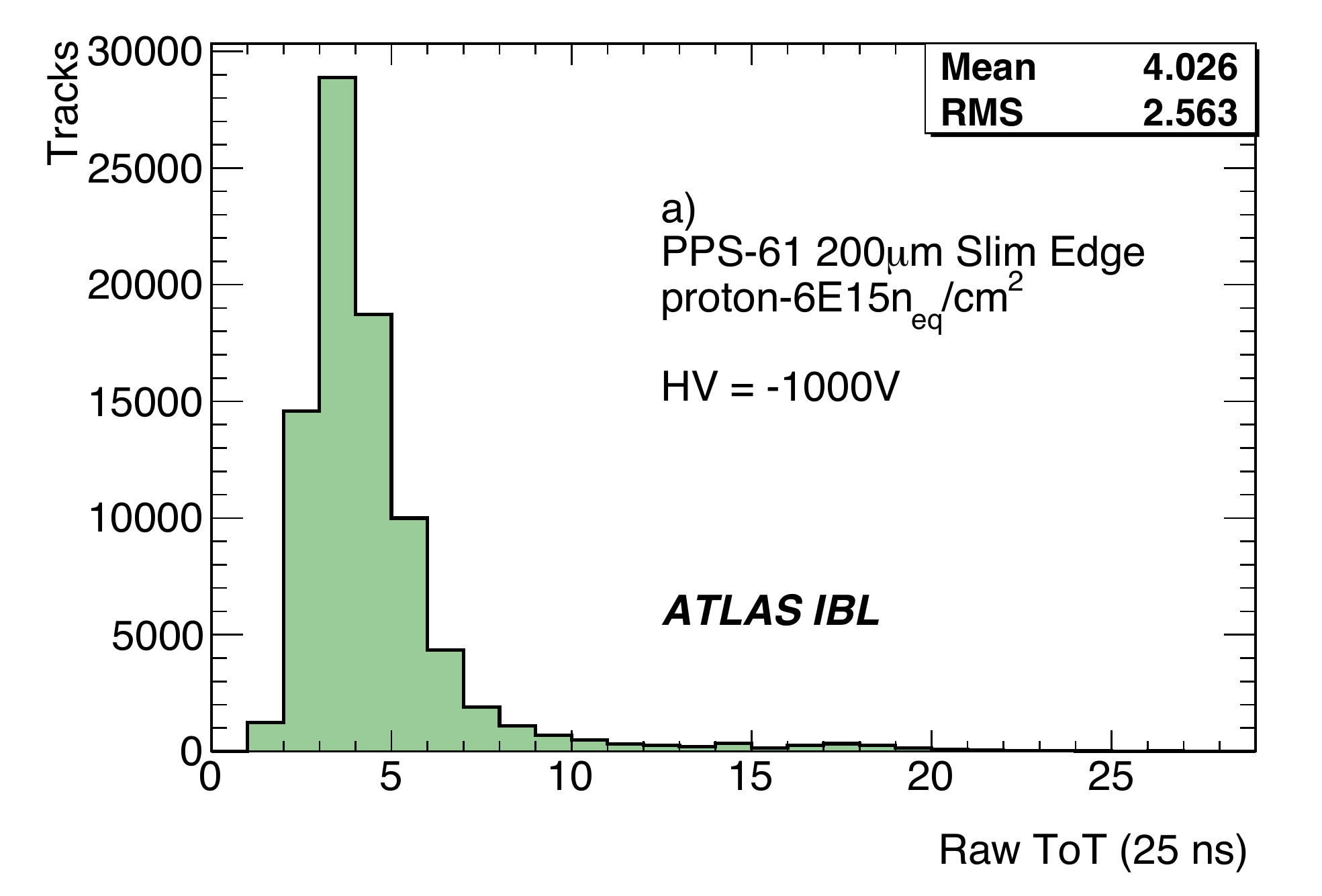}
\includegraphics[scale=0.350]{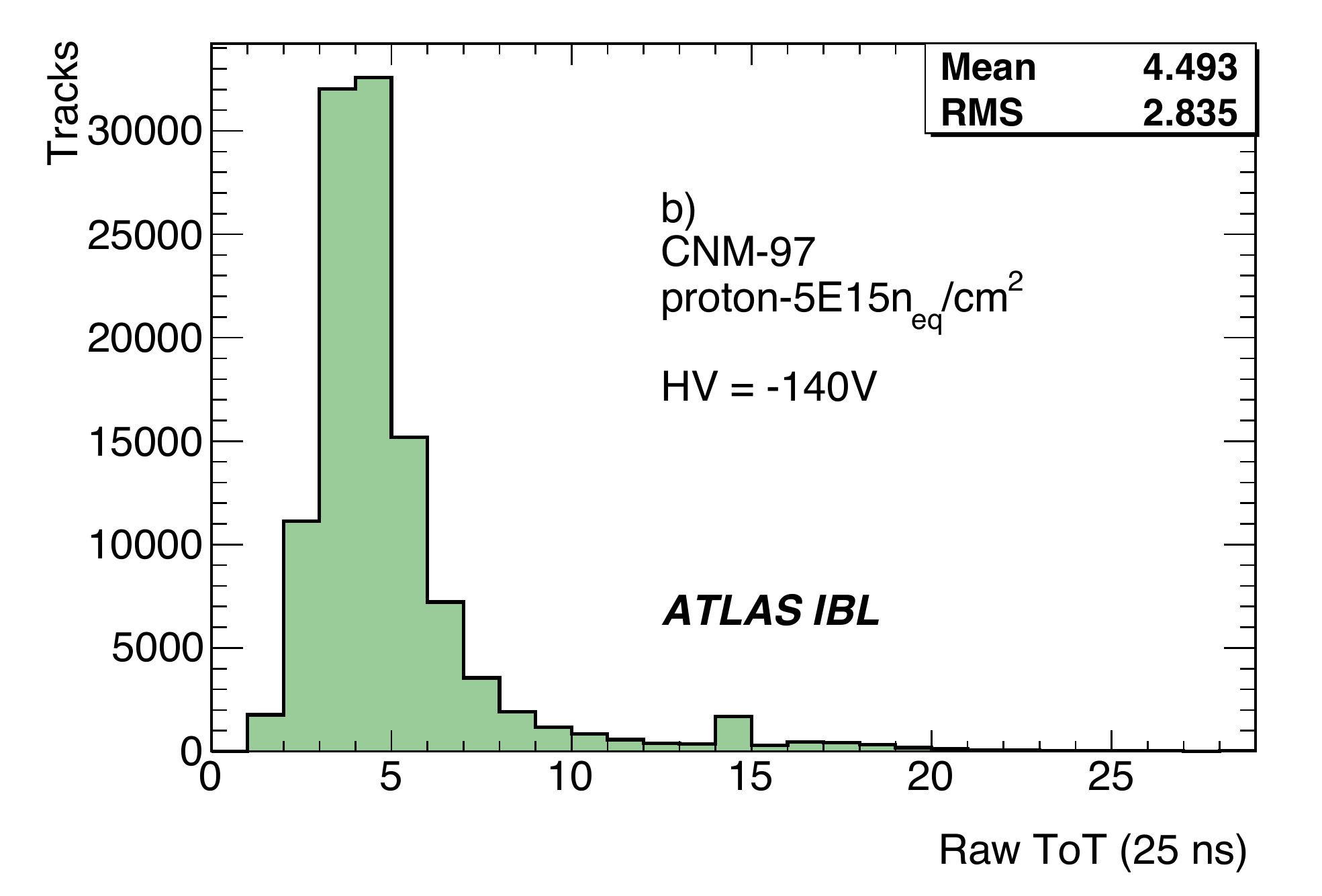}
\includegraphics[scale=0.350]{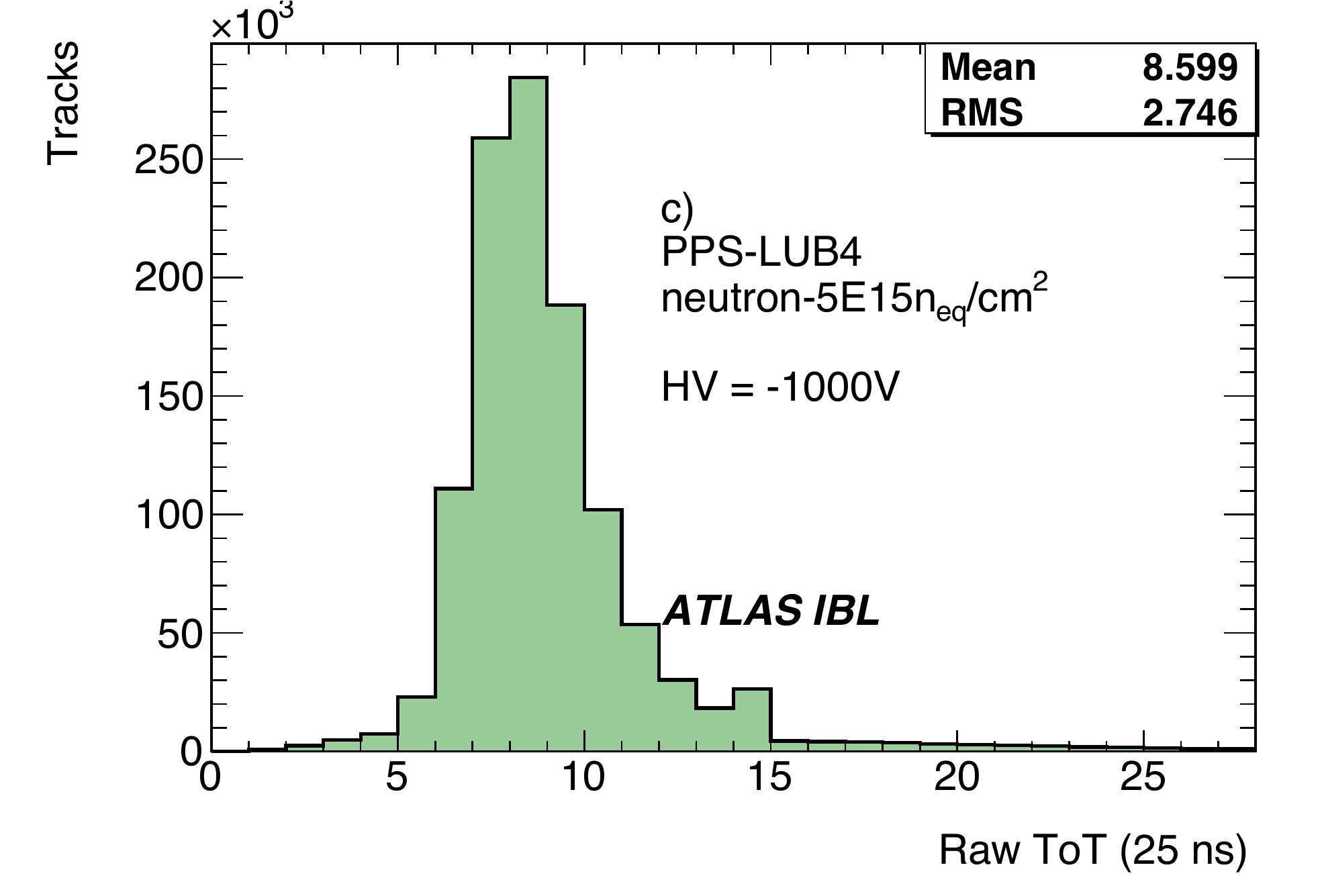}
\includegraphics[scale=0.350]{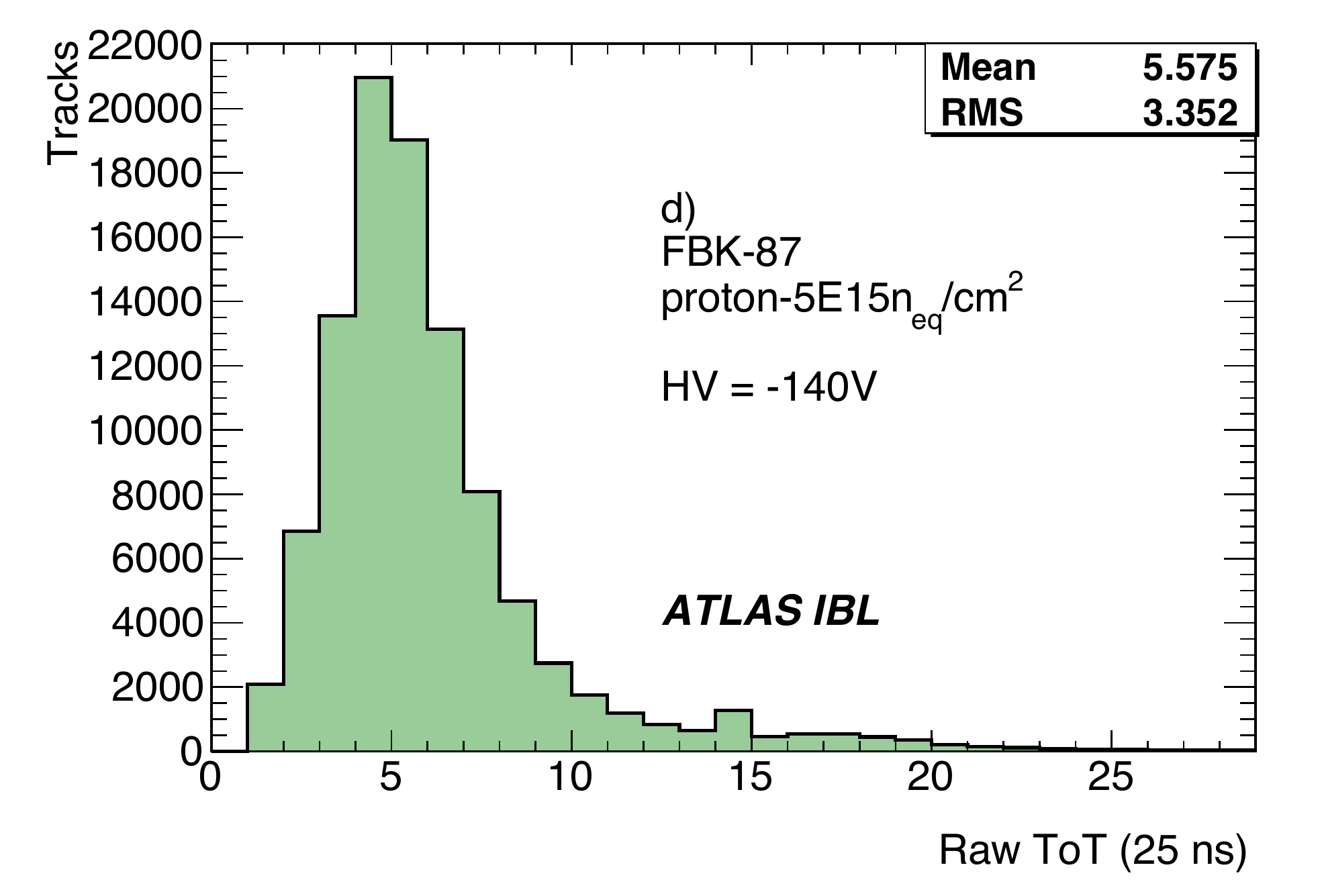}
\end{center}
\caption{\label{fig:chargecollection}
The raw ToT spectra for (a) PPS 61 (threshold setting 1400 e$^{-}$), (b) CNM 97 (threshold setting 2950 e$^{-}$), (c) PPS L4 (threshold setting 1600 e$^{-}$) and (d) FBK 87 (threshold setting 2450 e$^{-}$) modules, with no magnetic field and with a $15^{\circ}$ incident beam angle. 
The distributions are shown in units of the 25 ns bunch crossing clock.}
\end{figure}

\subsubsection{Hit efficiency}{\label{section:hit-efficiency}}

The hit efficiency is a key performance parameter for pixel modules, in particular because the sensors are required to survive the large radiation doses expected at the IBL.

The overall hit efficiency is measured using tracks reconstructed with the telescope and interpolated to the test modules to search for a matching hit. The number of tracks with a matching hit is divided by the total number of tracks passing through the sensor. To remove fake tracks that would bias the efficiency measurement of a particular test module, a matching hit in at least one other test module is required, as described above. 

As discussed in Section \ref{section:IBL-module-prepost-rad}, the large TID dose received by the FE-I4A IC during proton irradiation led to some noisy or dead pixel cells. To assess the intrinsic effect of radiation dose on the sensor behavior, tracks pointing to very noisy or dead pixels and surrounding pixels were excluded from the efficiency measurement of proton irradiated modules. These losses are fully recoverable following chip reconfiguration, but this was not done in the test beam studies so far.  

The tracking efficiency measurements for all samples and all operating conditions are summarized in 
table~\ref{tab:efficiency}. As expected the measured efficiency for the un-irradiated PPS sample is nearly $100 \ \%$. The un-irradiated 3D sensor has a slightly reduced efficiency due to the tracks that pass through the empty electrodes where the charge is not collected. Full efficiency is recovered for tilted tracks \cite{Grenier:2011ba}. The efficiencies measured with the magnetic field on and off are comparable for the 3D samples confirming that the effect of magnetic field is negligible on 3D sensors \cite{Grenier:2011ba}. 

As shown in  table~\ref{tab:efficiency}, the irradiated modules show very good behavior. At the expected operating conditions of bias voltage ($\rm{|V_{b}|}$ > 800~V for planar sensors and $\rm{|V_{b}|}$ > 160V for the 3D sensors) and pixel threshold ( < 2000 e$^{-}$), the module efficiency exceeds $97.5 \ \%$. For all the irradiated planar sensors, the hit efficiency increases with increased bias voltage because of the increased charge collection. 

\subsubsection{Cell efficiency}{\label{section:cell-efficiency}}

\begin{figure}
\begin{center}
\includegraphics[scale=0.50]{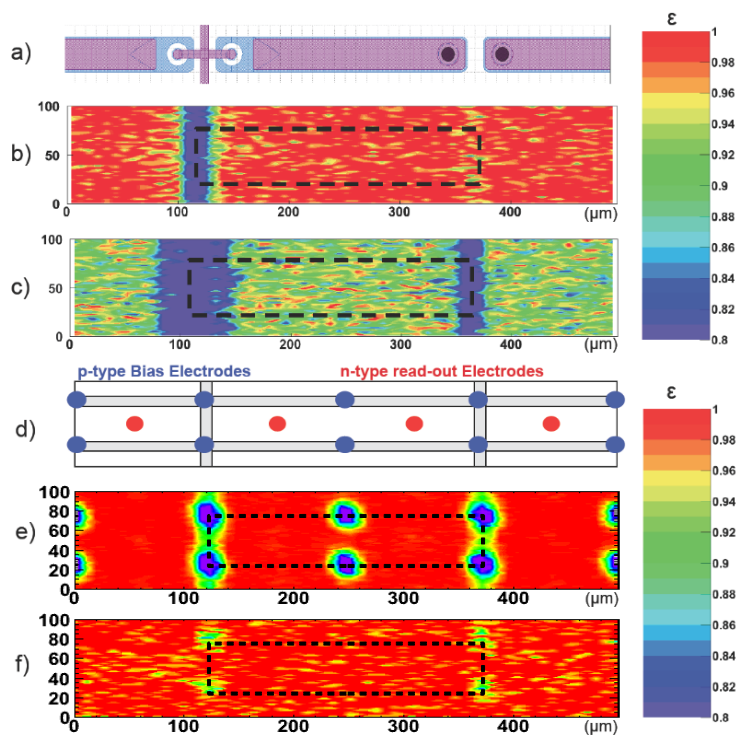}
\end{center}
\caption{\label{fig:2deffmap}
Cell efficiency maps: a) lithography sketch for PPS, b) and c) 2D efficiency maps for PPS 61 at -1000 V 
(mean efficiency 96.9\%) and at -600 V (mean efficiency 86.4\%) using $15^{\circ}$ inclined tracks, d) lithography sketch for 3D, e) the 2D efficiency map for the CNM 81 module using normal incident tracks (mean efficiency 97.5\%) and f)  the 2D efficiency map for the CNM 34 module using $15^{\circ}$ inclined tracks (mean efficiency 99\%). Both the CNM 34 and CNM81 are operated at a bias voltage of $160$~V. All dimensions are in microns. See the text for explanations.}
\end{figure}

The in-depth behavior of the sensor and the relative loss of efficiency after irradiation are better 
assessed by looking at the efficiency distribution within pixel cells. To improve the statistics and assuming that they behave similarly, all cells have been added together. 

Figure~\ref{fig:2deffmap} (a) shows the lithography sketch for the PPS 61 module. The bias grid and dots, and the solder bumps can be seen on the left and right of the sketch. Two cells are actually plotted: a central cell (dashed line) surrounded by two half cells in both the vertical and horizontal directions. Also shown are the 2D efficiency maps for the module at V$\rm{_{b}}$ =$-1000$~V (b) and V$\rm{_{b}}$ =$-600$~V (c) for $\rm{\phi}=15^{\circ}$ track inclination.    Efficiency loss occurs at the edge of the cell. At V$\rm{_{b}}$ =$-1000$~V the effect is mainly visible on the bias side of 
the cell. When the bias voltage is decreased, the effect also develops on the solder bump side. The loss is 
primarily due to charge sharing between the cells. When charge sharing occurs, less charge is collected by 
the readout cell, reducing the probability to exceed the electronics threshold. The effect 
is more pronounced for highly irradiated samples and for lower bias voltage. In addition, some charge is 
lost and trapped in the bias grid and dots, further decreasing the collected charge and therefore the 
efficiency.

The n-type readout electrodes (red) and p-type bias electrodes (blue) are shown on the 3D lithography sketch (d) for the CNM 34 and CNM 81 modules. As for (a), 2 cells are plotted. The 2D efficiency maps are shown for the for the neutron-irradiated CNM 81 sample at normal incidence (e), and for the proton-irradiated CNM 34 sample using $15^{\circ}$ track inclination (f).  Both the CNM 81 and CNM 34 modules are operated at $-160$~V effective bias voltage, allowing good charge collection efficiency. The loss of efficiency for tracks passing through the electrodes is clearly visible in (e) since the electrodes are empty and do not produce charge. More efficiency loss occurs near the bias electrodes because of charge sharing and because the electric field is less than near the readout electrodes. The CNM 34 data are for inclined tracks and the efficiency loss is less important as the tracks pass through some of the wafer bulk and not entirely through the electrodes. The effect of charge sharing is in this case more evident on the shorter side because of the track inclination. Simulations of the charge collection reproduce these experimental data.

\subsubsection{Edge efficiency}{\label{section:edge-efficiency}}

The size of the inactive area of the sensors can be estimated by measuring the hit efficiency of the edge pixels (see figure~\ref{fig:Edge-comp}). Two dimensional efficiency maps and their one-dimensional projections onto the long pixel direction are built for edge pixels. Figure~\ref{fig:edge} shows the photo-lithography sketch of the edge pixels for both PPS (a) and CNM 3D (b) sensors, and the corresponding efficiency projections. In figure~\ref{fig:edge} (c,d), the efficiency projections have been fit with s-curve functions. 

\begin {figure}[ht!]
\centering
\subfigure[a][]{\includegraphics[width=0.70\textwidth] {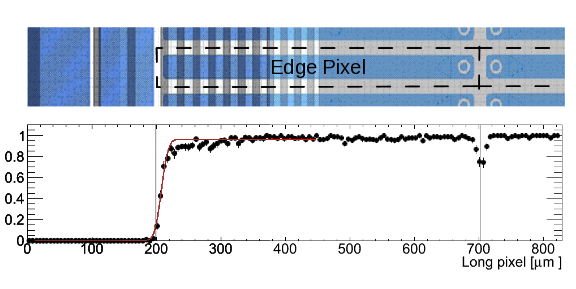}}
\subfigure[b][]{\includegraphics[width=0.70\textwidth] {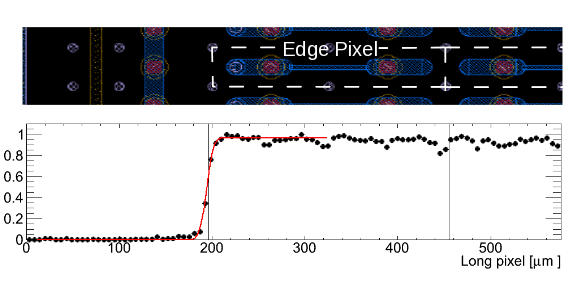}}
	\caption{Edge efficiency measurements following irradiation. (a) A PPS edge pixel photo-lithography drawing with the efficiency projection of the PPS L2 module at V$\rm{_{b}}$= $-1000$ V. (b) A 3D CNM edge pixel photo-lithography drawing with the 1-dimensional efficiency projection of the CNM 34 module at V$\rm{_{b}}$ =$-140$. The CNM 34 module was not fully biased in this measurement.} 
	\label{fig:edge}
\end{figure}

For the PPS modules, the inactive length is measured from the fixed dicing street. For PPS L2 at a bias voltage of V$\rm{_{b}}$= $-1000$ V, the inactive region is estimated to be approximately 200~$\mathrm{\mu m}$ at $50 \%$ efficiency as shown in figure~\ref{fig:edge} (a).  

The 3D edge design has a $200 \ \mu \mathrm{m}$ guard ring extending over the edge pixels. The active area extends to about $20 \ \mu \mathrm{m}$ over the edge pixel making the inactive area also approximately 200~$\mathrm{\mu m}$ at $50 \%$ efficiency. The reduced efficiency in some columns results from a lack of full depletion at V$\rm{_{b}}$~=$-140$~V.  

Overall, following irradiation both PPS and 3D sensors have only small inactive regions, as expected from their geometry.

\subsubsection{Charge sharing}{\label{section:charge-sharing}}

The charge sharing between cells is another important parameter of pixel detectors. Large charge sharing between neighbouring pixels leads to better tracking resolution as the hit position is better determined. However if charge sharing occurs, less charge is available to the hit cells, reducing the probability to pass the electronic threshold and therefore the hit efficiency. This can be a major concern for highly irradiated samples 
as the total available charge is reduced. 

Charge sharing between cells is directly related to the size of the reconstructed clusters. Cluster size 
distributions for both a PPS and a 3D sensor, measured with $15^{\circ}$ beam incident angle, are shown on figure~\ref{fig:clusize}. In the absence of magnetic field, PPS and 3D modules show similar behavior (quantitative comparisons are not possible because of the different operating conditions). Nevertheless, the hit efficiency results of table~\ref{tab:efficiency} indicate that charge sharing is not a concern for the bias voltages noted there.

\begin {figure}[ht!]
\centering
\subfigure[a][]{\includegraphics[width=0.49\textwidth] {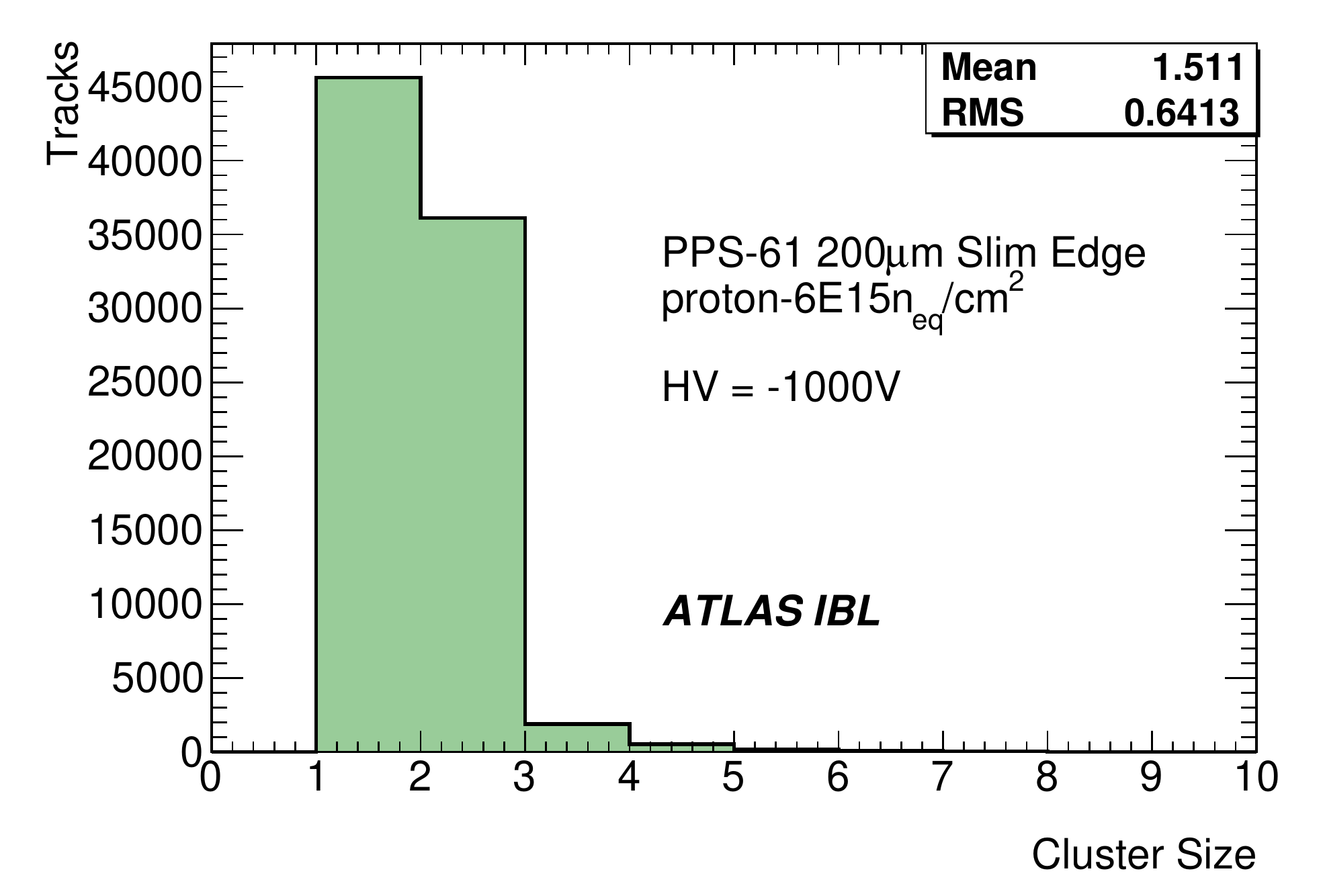}}
\subfigure[b][]{\includegraphics[width=0.49\textwidth] {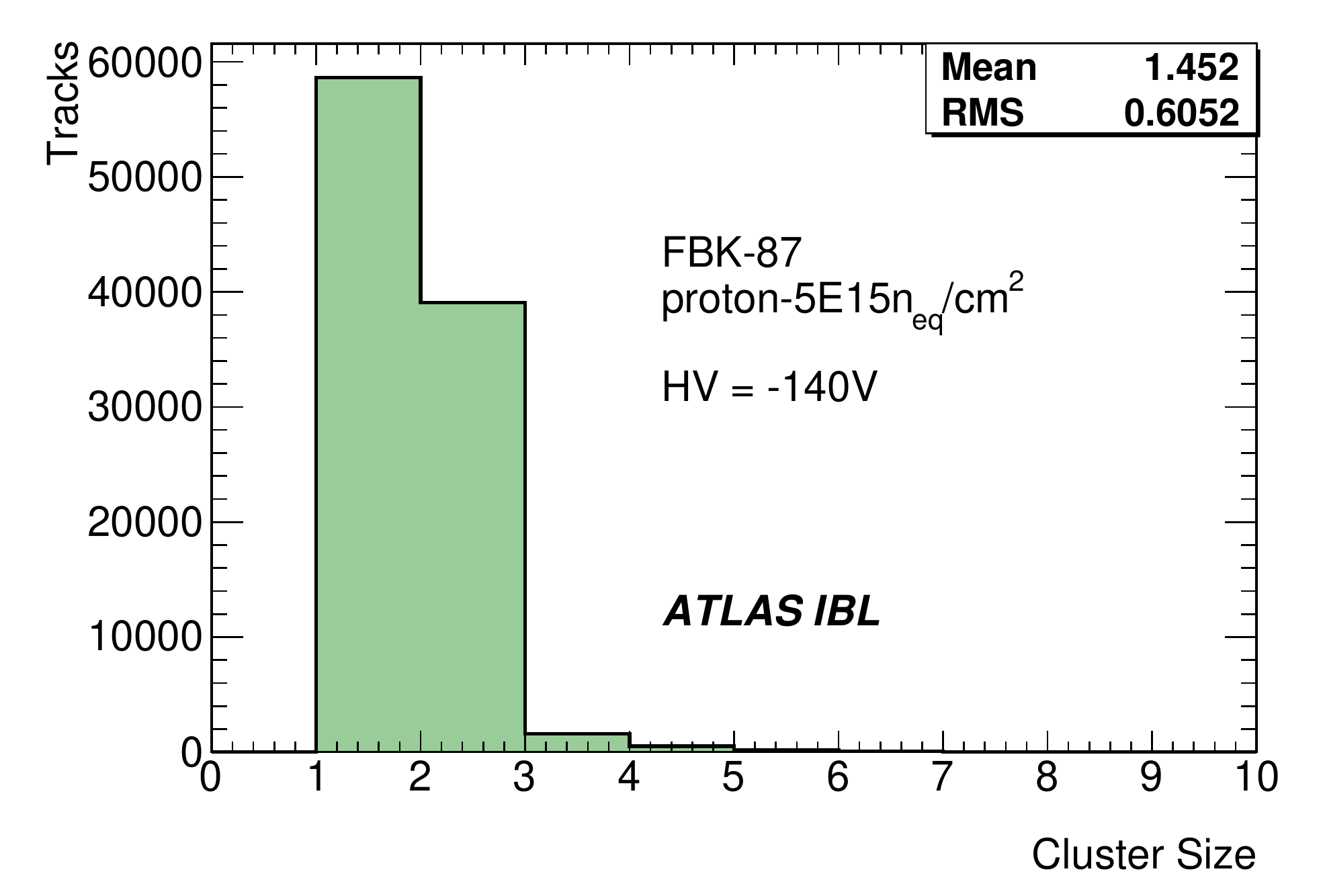}}
\caption{Comparison of the cluster size distributions for (a) the PPS 61 module and (b) the FBK 87 module, with no magnetic field and at $15^{\circ}$ beam incident angle. The threshold settings for the PPS 61 and FBK 87 modules are respectively 1400 e$^{-}$ and 2450 e$^{-}$.}
\label{fig:clusize}
\end{figure}

\subsubsection{Spatial resolution}{\label{section:spatial-resolution}}

Spatial resolution is another key parameter of pixel detectors. For multi-hit clusters the analog information 
of the deposited charge can be used to improve the determination of the track position in the sensor.
The spatial resolution of PPS and 3D samples is measured from residual distributions of all-hit clusters in the short pixel 
direction, where the track position is interpolated from the telescope and the cluster position is calculated using simple charge weighting between the cells. Distributions for the representative PPS L4 and CNM 81 modules, using a $15^{\circ}$ 
beam incident angle, are shown in figure~\ref{fig:figure28-resolution}. The measured resolution of approximately 15~$\rm{\mu}$m is similar for both sensor technologies. This measurement will be significantly improved for both module types when more sophisticated cluster algorithms are implemented. 

\begin {figure}[ht!]
\centering
\subfigure[a][]{\includegraphics[width=0.49\textwidth] {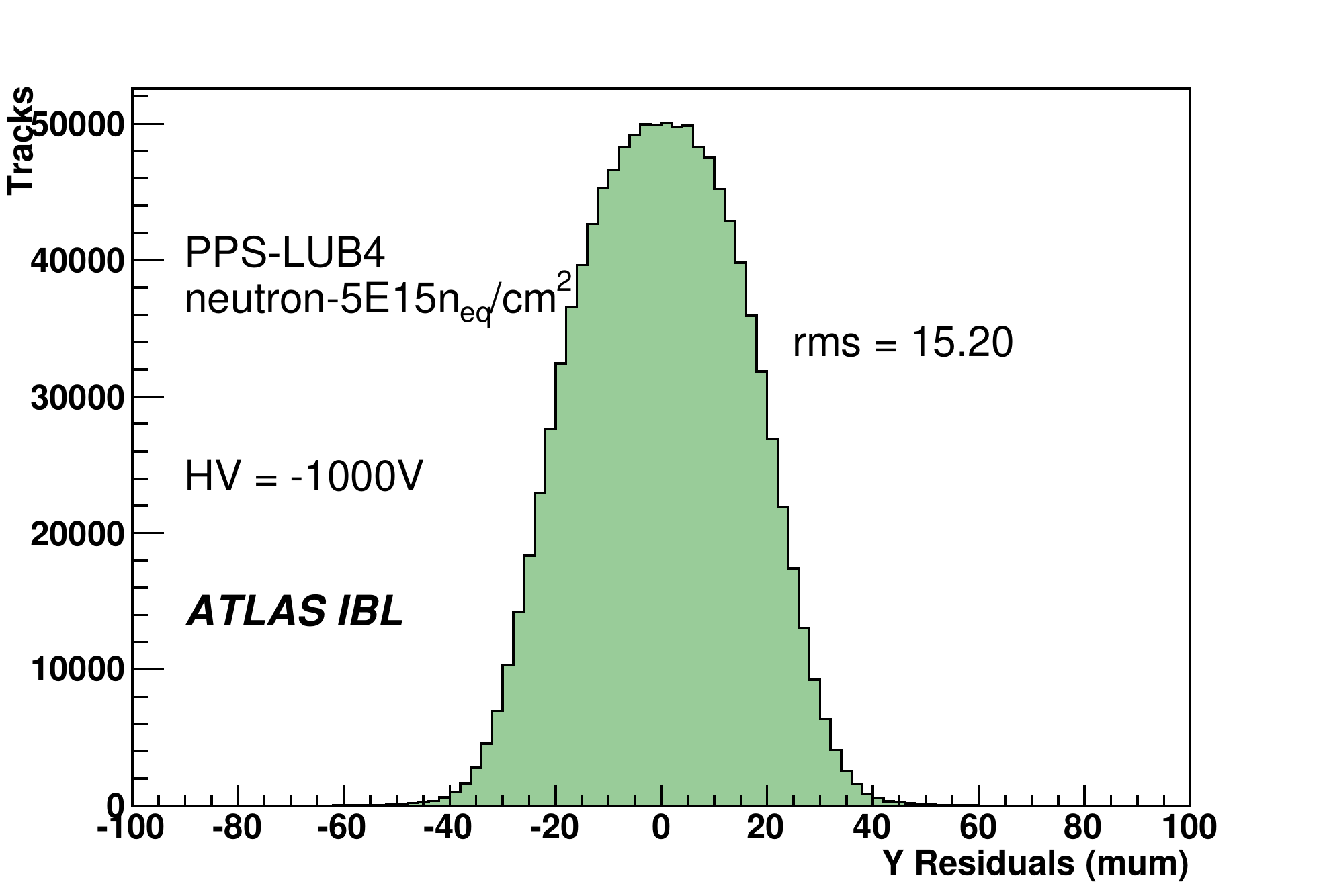}}
\subfigure[b][]{\includegraphics[width=0.49\textwidth] {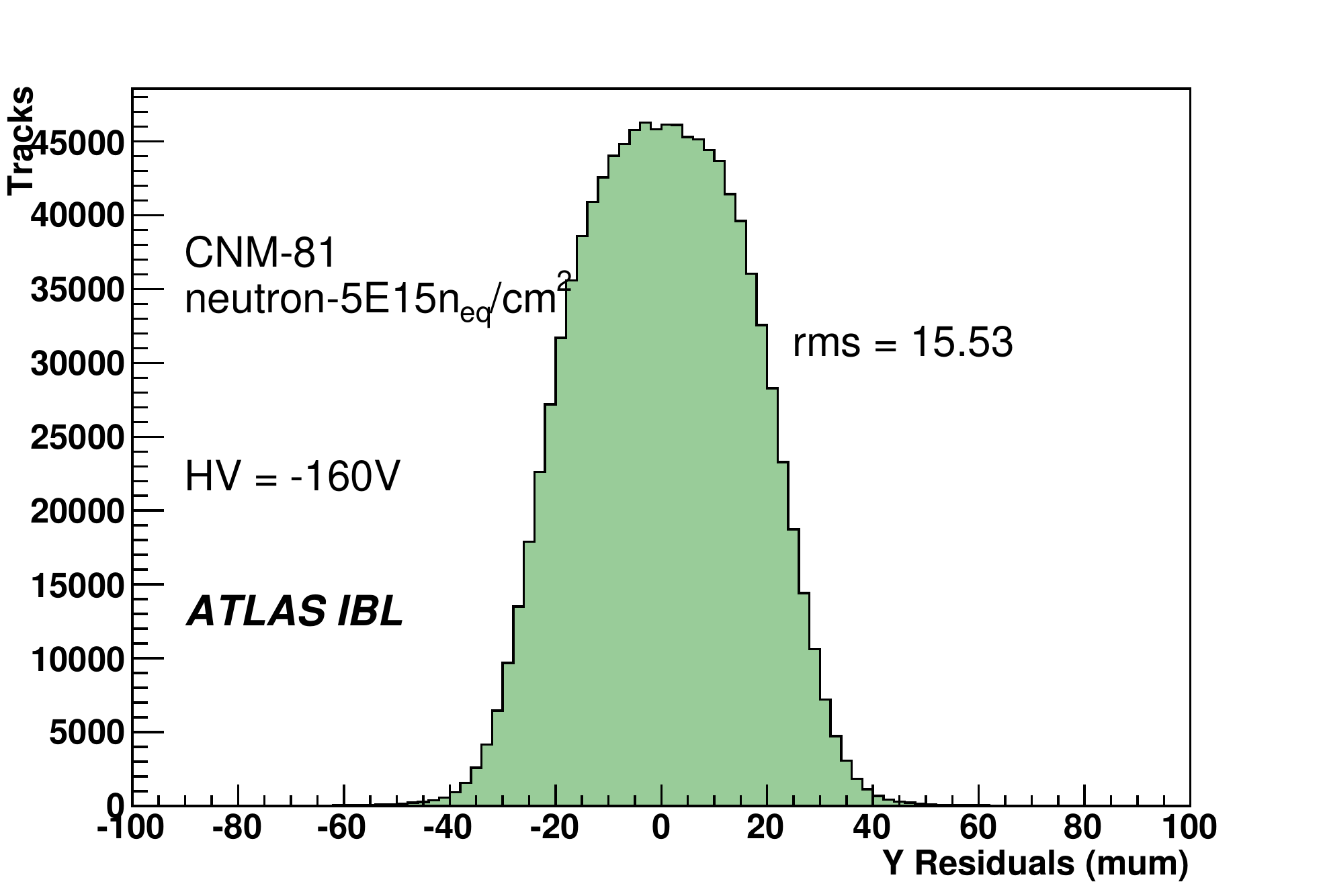}}
\caption{(a) Residual distribution of the PPS L4 module for all-hit clusters in the short pixel direction, for $15^{\circ}$ beam incident angle and for a threshold setting of 1600 e$^{-}$.  (b) Residual distribution for the CNM~81 module at the same incident angle and a threshold setting of 1500 e$^{-}$. The measured rms resolution is similar for both the planar and 3D module types.}
\label{fig:figure28-resolution}
\end{figure}

\subsection{Measurements at small incidence angle}{\label{section:shallow-incidence}}

In the IBL, the largest-$\eta$ modules are positioned at $\eta = 2.9$ which corresponds to an incident track angle $\theta = 84^{\circ}$ with respect to normal incidence. The behavior of planar and 3D modules has been measured for $80^{\circ}$ incident beam angle. Given the sensor thickness and the pixel length, tracks at that angle can traverse up to seven cells in the z-direction. The cluster size has a direct implication on the measured track precision and can be affected by two factors: 
\begin{itemize}
\item Threshold effects: the charge produced in the cluster edge cells may not be sufficient to pass the electronic threshold setting and to register a hit; 
\item Charge collection: following a large radiation dose, the sensors may be required to operate at a bias voltage for which the charge collection is not optimal, therefore reducing the collected charge and hence not registering a hit in one or more cells. Depending on the depletion depth, cells traversed close to the sensor surface may not be recorded. 
\end{itemize}

Figure~\ref{fig:fig29_higheta_CNM} compares the cluster size distribution for two CNM modules, mounted back-to-back: the un-irradiated CNM~55 module and the irradiated CNM~34 module.  As expected, due to a combination of the threshold effect and the reduced charge collection after irradiation, the cluster size distribution is broader and with lower mean value for the irradiated module. This will affect and possibly deteriorate the track hit accuracy. Because of the electrode orientation, this effect is much less important for modules using 3D modules than for planar modules. 

\begin {figure}[ht!]
\centering
\subfigure[a][]{\includegraphics[width=0.49\textwidth] {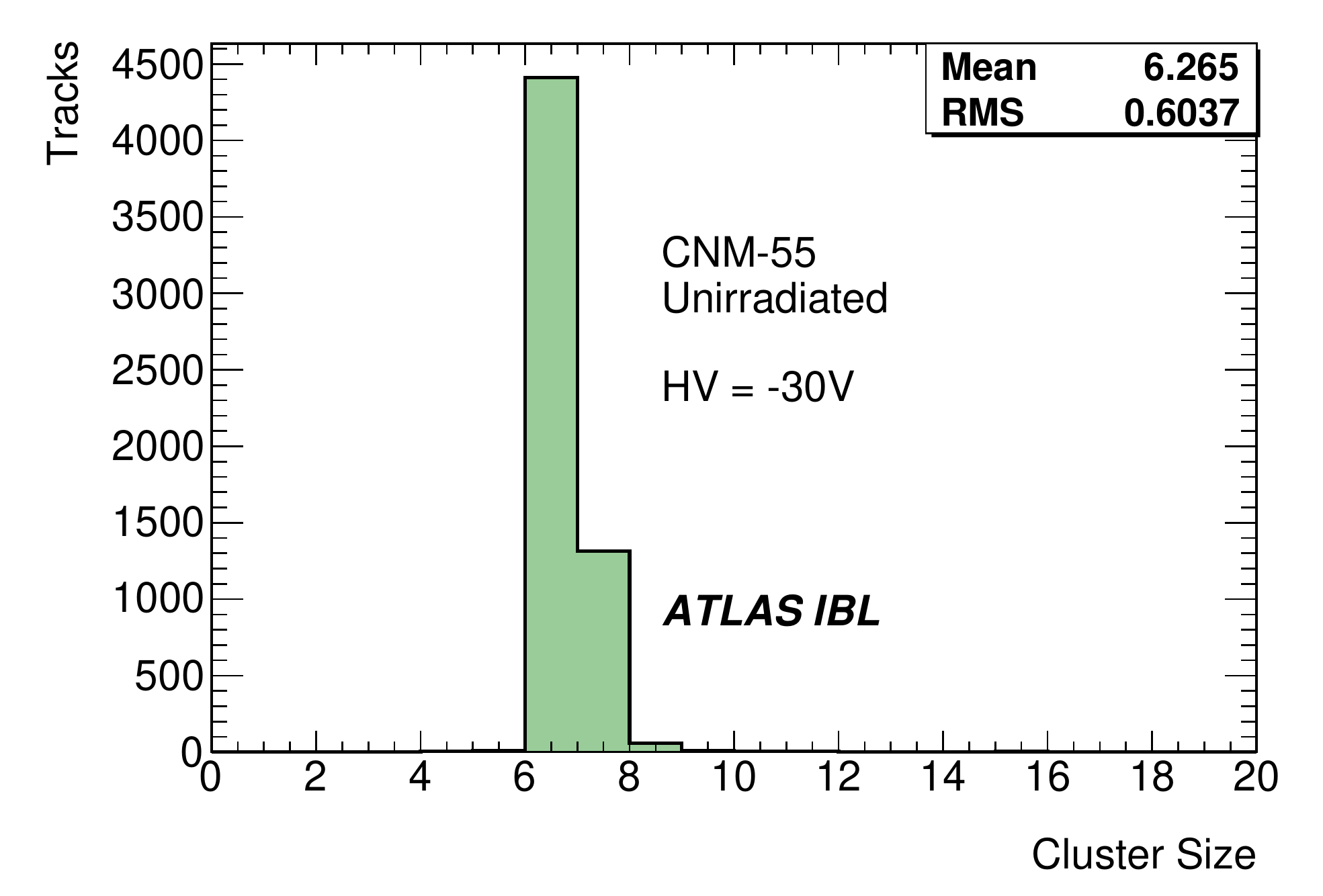}}
\subfigure[b][]{\includegraphics[width=0.49\textwidth] {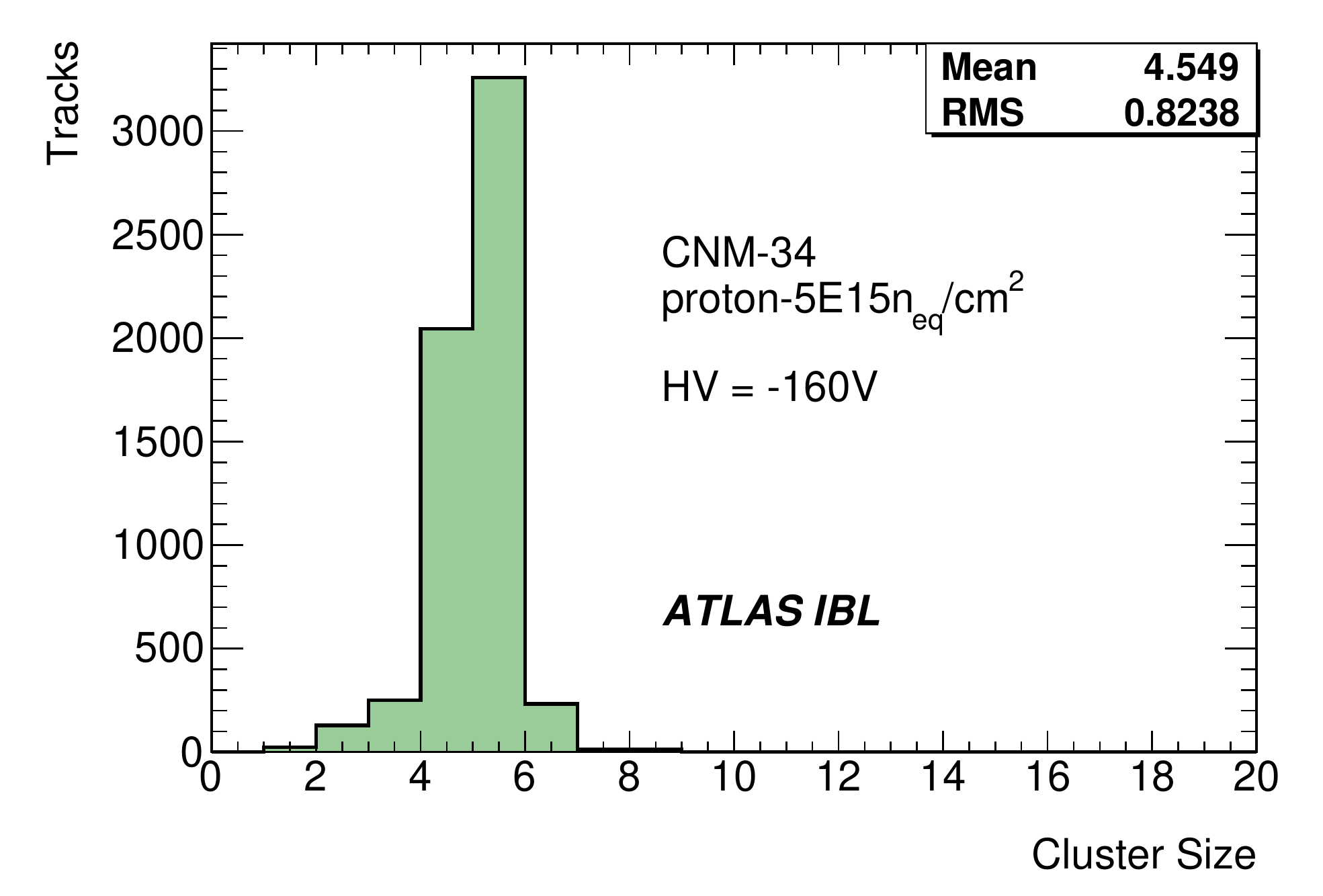}}
\caption{(a) Cluster size distribution for the unirradiated CNM 34 module for tracks at $80^{\circ}$ incident track angle with respect to normal incidence. The bias voltage was V$\rm{_{b}}$ = -30~V and the nominal threshold setting was 1500 e$^{-}$. (b) Similar data for the CNM 55 module, following proton irradiation to a fluence of  $5\times10^{15} \ \mathrm{n_{eq}/cm^2}$.  The data are collected with V$\rm{_{b}}$= -160 V and a threshold setting of 1600 e$^{-}$.}
\label{fig:fig29_higheta_CNM}
\end{figure}

After irradiation, type inversion occurs in planar sensors and the devices may be operated at under-depleted bias voltages. Depending 
on the depletion depth, the collected charge in cells located close to the sensor surface may not exceed the electronic threshold. As a consequence, recorded clusters may have a reduced number of cells and a subsequent degradation of the track resolution. 
The cluster size distributions of the (irradiated) PPS L4 module, operated at V$\rm{_{b}}= -400$~V  and V$\rm{_{b}}= -1000$~V are shown on figure~\ref{fig:fig30_higheta_PPS}. Because of the reduced charge collection, the number of cells is significantly reduced at V$\rm{_{b}}= -400$~V.

\begin {figure}[ht!]
\centering
\subfigure[a][]{\includegraphics[width=0.49\textwidth] {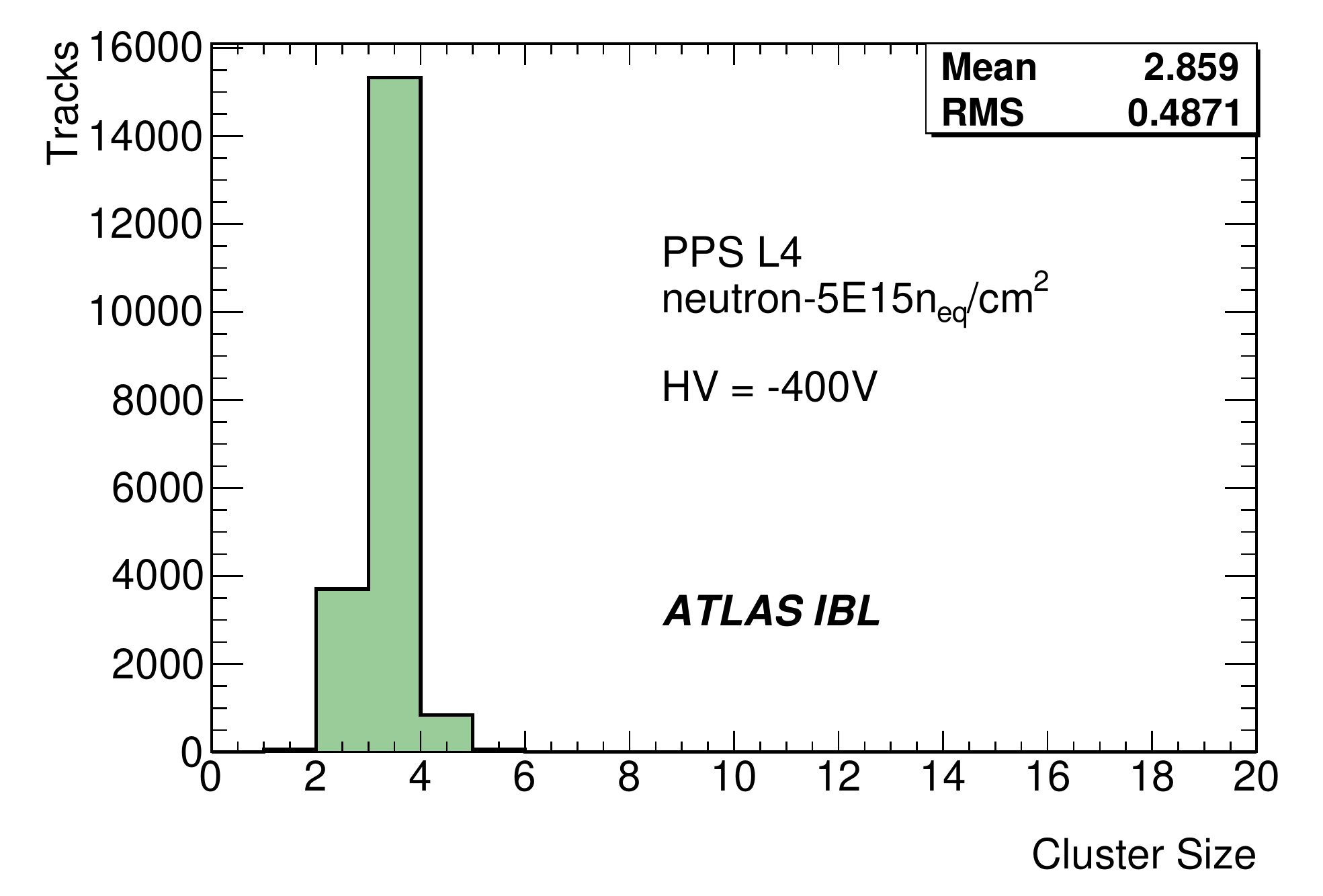}}
\subfigure[b][]{\includegraphics[width=0.49\textwidth] {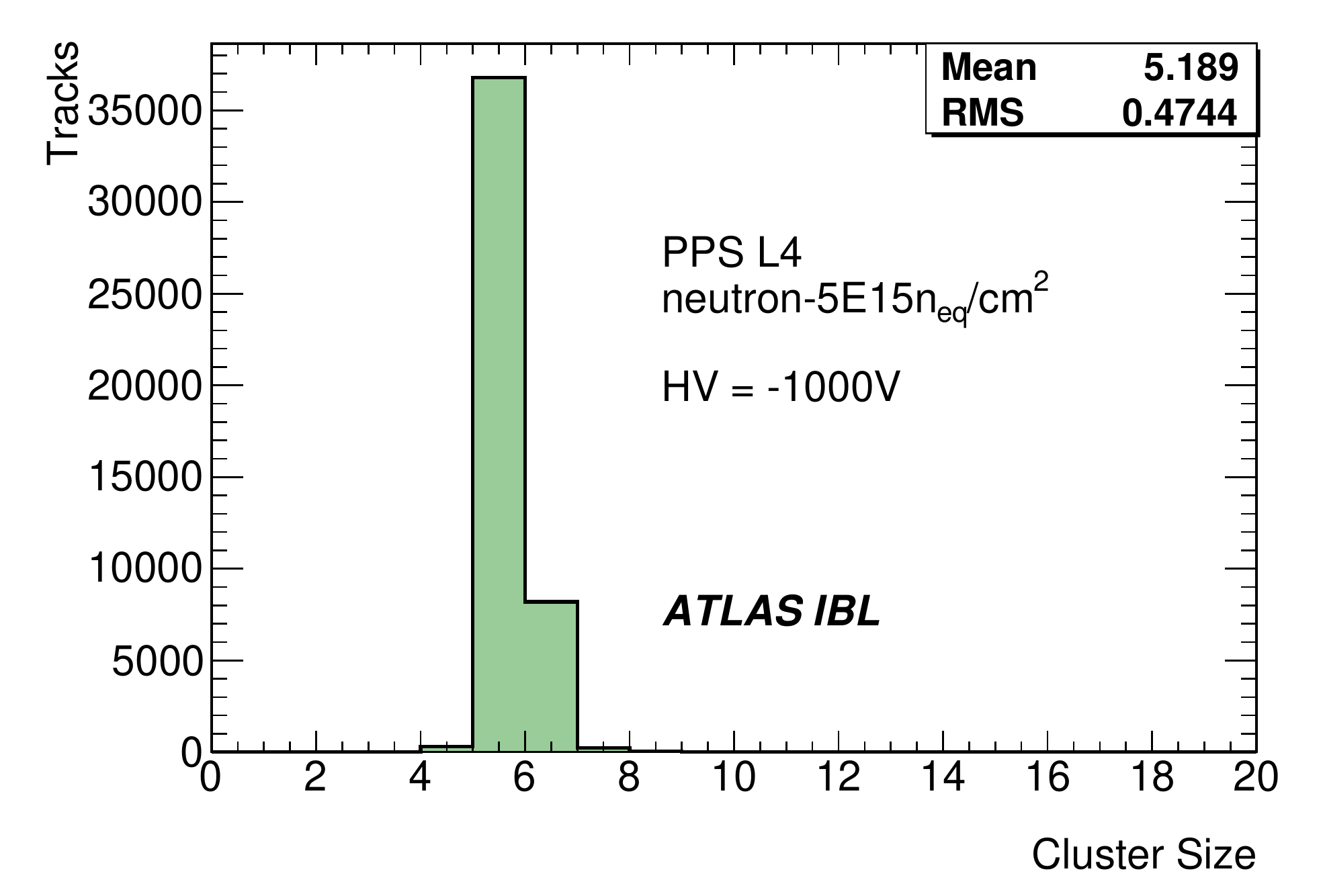}}
\caption{Cluster size distribution for the PPS L4 module, following proton irradiation to a fluence of  $5\times10^{15} \ \mathrm{n_{eq}/cm^2}$,  for tracks at $80^{\circ}$ incident track angle with respect to normal incidence. The data are collected at bias voltage of (a)  V$\rm{_{b}}$ = -400~V and (b) V$\rm{_{b}}$ = -1000~V. In each case,  the nominal threshold setting was 1600 e$^{-}$.}
\label{fig:fig30_higheta_PPS}
\end{figure}

\section{Conclusions {\label{section:conclusions}} }

The first prototype pixel modules designed for the ATLAS Insertable B-Layer (IBL) have been fabricated and fully characterized. A selection of the fabricated modules have been irradiated using a 25 MeV proton beam, or using reactor neutrons, to the fluences expected during the IBL operational lifetime. A selection of both irradiated and non-irradiated modules have been characterized in a CERN test beam. 

The single-chip (SCS) modules have used either conventional planar single chip sensors (PPS), or two types of 3D single chip sensor (CNM or FBK) bump-bonded to a single FE-I4A front-end readout IC that was developed by the collaboration for high radiation and high luminosity applications.

The FE-I4A IC performance is very satisfactory using all sensor types. Following irradiation, the measured threshold dispersion remains stable and virtually unchanged, while stable low mean threshold settings of $1000-2000$ e$^{-}$ are achievable following irradiation. The intrinsic noise increase is only $20-25$~\% in comparison with pre-irradiation values. Following high levels of proton (as opposed to neutron) irradiation during initial studies, a significant fraction of pixels became digitally unresponsive. This was traced to a non-optimized internal DAC current setting that can be adjusted. With minor improvements, the FE-I4B IC will be used for production IBL modules.

Single-chip PPS modules operate satisfactorily, both before and after irradiation. Depletion voltages of $30-40$~V are measured before irradiation, with a typical noise of $115-150$~ENC when biased at  $\rm{|V_{b}|}= 80$~V. Following irradiation at a level of $2-5\times10^{15} \ \mathrm{n_{eq}/cm^2}$, stable operation with adequate charge collection is possible at bias voltages of $\rm{|V_{b}}| = 800- 900$~V.  At these operating conditions, the measured noise is typically 160 ENC for a mean threshold setting of 1600~e$^{-}$. Using test beam data, hit efficiencies of around 98\% are typically measured, with spatial resolutions that are not significantly deteriorated from charge sharing. Beyond a bias of $\rm{|V_{b}}| = 900$~V, there is some evidence of charge multiplication. The cell response is very uniform within any pixel cell, and the 50\% active edge efficiency is approximately 200~$\rm{\mu}$m from the cutting edge after irradiation.  

The single-chip CNM and FBK modules also behave satisfactorily, with performances that are equivalent to the PPS modules, but with much lower operating voltage. Because of their geometry, the pre-irradiation depletion voltage is at the level of $15-20$~V for FBK modules, and $20-25$~V for CNM modules before irradiation, and noise values of order 150 ENC are achieved with a mean threshold setting of approximately 1600~e$^{-}$. That allows initial operating voltages $\rm{|V_{b}}|\lesssim$30~V. The ENC is only slightly deteriorated following irradiation at a level of approximately $5\times10^{15} \ \mathrm{n_{eq}/cm^2}$, and hit efficiencies exceeding 97.5\% are achieved for bias voltages of $\rm{|V_{b}|}=160-170$~V. Excepting small geometric effects due to the 3D electrodes, the cell uniformity and edge cell efficiency are excellent for modules having 3D sensors, and the track hit spatial resolution is close to that of modules having planar sensors. Although the inactive edge of the 3D sensors is also 200~$\rm{\mu}$m in the present design, this could be reduced to close to 0~$\rm{\mu}$m in the case of an active edge design, or less than 100~$\rm{\mu}$m with a more aggressive fence design.

An extensive test beam campaign is underway to fully understand the module behavior (all types) as a function of the radiation dose. A key step is to calibrate and characterize the ToT measurement from each module type. Nevertheless, the initial bench-top and test beam measurements described in this publication demonstrate the satisfactory performance of all three module types (PPS, CNM, FBK), both before and after the irradiation excepted for the IBL detector. 

\acknowledgments

We acknowledge the support of NSERC, NRC and CFI, Canada; CERN; MSMT CR, MPO CR and VSC CR, Czech Republic; IN2P3-CNRS, France; BMBF, DFG, HGF, MPG and AvH
Foundation, Germany; INFN, Italy; MEXT and JSPS, Japan; FOM and NWO,
Netherlands; RCN, Norway; ARRS and MVZT, Slovenia; MICINN, Spain; SER, SNSF and Cantons of
Bern and Geneva, Switzerland; NSC, Taiwan; STFC, the Royal
Society and Leverhulme Trust, United Kingdom; DOE and NSF, United States of
America.
Irradiation studies at the KIT facility in Karlsruhe and the TRIGA reactor in Slovenia, as well as travel support for test beam operations at CERN and DESY, were partially funded by the European Commission under the FP7 Research Infrastructures project AIDA, grant agreement no. 262025. We acknowledge the support from the CERN IC group and in particular K. Kloukinas. 




\clearpage
\begin{flushleft}
{\Large The IBL ATLAS Collaboration}

\bigskip

J.~Albert$^{ \rm50}$,
M.~Alex$^{ \rm15}$,
G.~Alimonti$^{ \rm29(a)}$,
P.~Allport$^{ \rm24}$,
S.~Altenheiner$^{ \rm15}$,
L.S.~Ancu$^{ \rm9}$,
A.~Andreazza$^{ \rm29(a)}$$^{, \rm29(b)}$,
J.~Arguin$^{ \rm5}$,
D.~Arutinov $^{ \rm11}$,
M.~Backhaus$^{ \rm11}$
% ...

\bigskip

% ...
$^{5}$ Physics Division, Lawrence Berkeley National Laboratory and University of California, Berkeley CA, United States of America

% ...
$^{9}$ Albert Einstein Center for Fundamental Physics and Laboratory for High Energy Physics, University of Bern, Bern, Switzerland

% ...
$^{11}$ Physikalisches Institut, University of Bonn, Bonn, Germany

% ...
$^{15}$ Institut f\"{u}r Experimentelle Physik IV, Technische Universit\"{a}t Dortmund, Dortmund, Germany

% ...
$^{24}$ Oliver Lodge Laboratory, University of Liverpool, Liverpool, United Kingdom

% ...
$^{29}$ $^{(a)}$INFN Sezione di Milano; $^{(b)}$Dipartimento di Fisica, Universit\`a di Milano, Milano, Italy

% ...
$^{50}$ Department of Physics and Astronomy, University of Victoria, Victoria BC, Canada

% ...

\end{flushleft}

\end{document}